\documentclass{book}
\usepackage{graphicx,amsmath,psfrag,bbm}
\newcounter{footnote_temp}
\begin{document}
\normalsize
\thispagestyle{empty}
\Large
\begin{center}
ALBERT-LUDWIGS-UNIVERSIT\"AT, FREIBURG IM BREISGAU\\
FAKULT\"AT F\"UR PHYSIK\\
\vspace*{2cm}
DIPLOMARBEIT / DIPLOMA THESIS\\
$ $\\
\vspace*{1cm}
\bf\huge The calibration of the vector\\
\vspace*{.2cm} polarimeter POLIS\rm\Large\\
\vspace*{2cm}
CHRISTIAN BECK\\
\vspace*{2cm}
\resizebox{6cm}{!}{\includegraphics{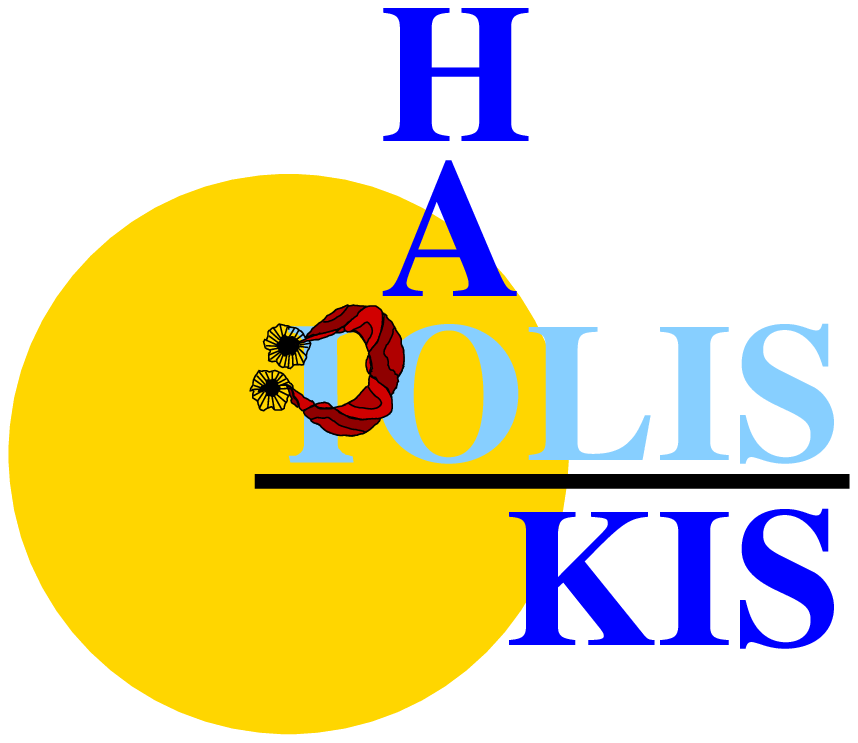}}\\
\vspace*{2cm}
angefertigt am\\
Kiepenheuer-Institut f\"ur Sonnenphysik, Freiburg\\
April 2002
\end{center}
\newpage
\normalsize
\bf\Huge Abstract\\
$ $\\
$ $\\
\rm\normalsize
\thispagestyle{empty}
In this diploma thesis, the calibration of the new vector polarimeter
POLIS will be described. The instrument is built by the
Kiepenheuer-Institut f\"ur Sonnenphysik (KIS, Freiburg, Germany) in
cooperation with the High Altitude Observatory (HAO, Boulder, USA). It
will be operated at the German Vacuum Tower Telescope (VTT) in Tenerife.\\

The instrument yields simultaneously the polarization state of light in
two spectral ranges at 396\thinspace nm and 630\thinspace nm. The
measurement is performed with a rotating retarder, which modulates the incident
polarization. The modulation is transformed into a varying intensity
through polarizing beamsplitters. The demodulation uses a weighted
integration scheme to obtain the full Stokes vector of the radiation.\\

To achieve a sufficient polarimetric accuracy of 0.1\thinspace $\%$ of
the continuum intensity, it is necessary to calibrate the polarimeter and
remove the instrumental polarization due to the telescope.\\

The calibration of the polarimeter is performed through the evaluation
of the calibration data set. This data is produced with a calibration unit
consisting of a linear polarizer and a retarder. Both elements are
placed in rotatable mounts and can be steered by remote control. The
calibration unit will be placed inside the vacuum tank at the VTT Tenerife.

The polarimeter response function or \bf X\rm-matrix can be determined
from a comparison between created input and measured output. The
calibration data images have to be corrected for the detector properties
before, which can be achieved with additional flatfield and dark current
data. The application of the inverse matrix \bf X\rm$^{-1}$ removes the
properties of the polarimeter from measured data.\\

The instrumental polarization of the telescope, which changes the Stokes
vector incident from the sun, will be removed through
the usage of a model of the polarimetric properties of the telescope. The
optical elements in the telescope are modelled by appropriate
Mueller matrices with free parameters. To calculate the total optical
train, it is necessary to consider the specific geometry of the light
path at each moment of time. For this purpose a numerical approach was
developed that can be applied to other optical setups as well, if the
beam path is known. The free parameters in the telescope model will be
derived from a least-square-fit to telescope calibration
data. Similar to the polarimeter calibration, known input states can be
created with an array of polarizing sheets, which can be placed either
on top of the first mirror or on top of the entrance window of the
vacuum tank. The corresponding measurements should allow to obtain
the required parameter values, after the polarimeter properties have been
removed by the application of \bf X\rm$^{-1}$.\\

Due to the delayed setup of POLIS, data from the Advanced Stokes
Polarimeter (ASP) is used to display the calibration steps and probable
results. The last chapter presents a preliminary evaluation of two data
sets from the ASP.
\tableofcontents
\begin{chapter}{Introduction}
\begin{minipage}{12cm}
''\it These results leave no doubt in my mind that the doublets and
triplets in the sun-spot spectrum are actually due to a magnetic
field.\rm'' G.E.Hale (1908)\\
\end{minipage}\\
$ $\\
The sun as the nearest and for us most important star is known for
centuries to show activites of periodic and random character, the
most prominent being the sun spots. The key feature of solar physics was
already well known at the beginning of the last century, but many of the details still are far from being understood.

The only sources of information available for ground observations on earth
are some of the particles in the solar wind, and a limited range of the solar
spectrum, which can penetrate the atmosphere of the earth. From
spectroscopic examinations of the sunlight a number of discoveries were
made, for example the detection of helium through its absorption
lines. The combination of spectroscopy with polarimetric
measurements permitted G.E.Hale to prove the existence of magnetic
fields on the solar surface and obtain the value of the field strength. 

The polarimeter POLIS\footnote{POlarimetric Littrow Spectrometer, built
in cooperation with the High Altitude Observatory (HAO) at the
Kiepenheuer Institut f\"ur Sonnenphysik, Freiburg (KIS). POLIS is intended to be used at the german Vacuum Tower Telescope
(VTT) in Tenerife.} could be regarded as a refined version of these
first measurements. The basic physics of the polarization of light, the
Zeeman effect and its interpretation has changed only little. Mayor
technical improvements are increased spectral and spatial resolution,
enhanced sensitivity of detectors, faster data acquisition and storage,
and a more complicated calibration procedure to remove most polarization
effects of non-solar origin.

The main difference to the observations at the beginning of the
20$^{\mbox{\small th}}$ century is the vector polarimetry, which
allows the calculation of the vector magnetic field, i.e. field strength
and direction. This technique has been succesfully applied in a number
of instruments, for example the Advanced Stokes Polarimeter (ASP). But most of
these instruments have a certain drawback: the needed spectral
resolution requires the restriction on one spectral line. The magnetic
field configuration can then only be established in a limited height in
the solar atmosphere, where this absorption line is formed.

This is the point, that makes POLIS an improvement to existing high
resolution instruments. It has been designed for the simultaneous
polarimetry in two spectral lines, which originate in different heights
in the solar atmosphere, a FeI-line from the photosphere and a
CaII-line from the chromosphere. This offers the opportunity to reconstruct the
magnetic field at two heights over the same region at the same
time. Moving the region of measurement on the solar image in the focal
plane a data set is obtained, which contains information from all three
spatial dimensions.

The additional dimension of the data, the height information, allows to
build a consistent model of the magnetic field lines from the
photosphere up to the chromosphere. Simultaneous observations with the
Tenerife Infrared Polarimeter (TIP) will be possible in the future,
adding information from the lower photosphere. 
But the calculation of the vector magnetic field has one requirement: the
observed polarization signal has to be of only solar origin.

To achieve a sufficient polarimetric accuracy two main points are of 
importance. The proper polarimeter has to be considered, which is
calibrated to check its response to polarization. Secondly, the instrumental
polarization due to the telescope, at which the polarimeter is used, has
to be removed.

This thesis will start with a short summary on solar magnetic phenomena,
and their influence on the polarization of light, in chapter
\ref{chap2}. The Stokes formalism is introduced to describe the
properties of polarized light. In connection with the Mueller matrix
calculus it is the theoretical base of the  measurement and the
evaluation of data. Chapter \ref{vecpola} explains the method of the
measurement of the polarization state, which is used for POLIS. The
instrument and its predecessor, the ASP, are described in detail in
chapter \ref{chap4}. The scientific goals of POLIS will be formulated in
the context of theory and instrumental design. Chapter \ref{polcal}
develops the calibration of the polarimeter, and the model for the polarization
properties of the VTT Tenerife, where POLIS will be installed. The
thesis finishes with an examination of observations with the ASP in
chapter \ref{polarimetric}, which are similar to a part of the data
POLIS will hopefully make available.\\

During the thesis it got clear that the main difficulty was -and will
be- the telescope model for the VTT. Unfortunately this problem can
not be rigidly discussed without actual measurements with POLIS at the
VTT.\\

To state it at least once, most of the work executed by the author was
to translate the theoretical concepts and methods of the calibration
into a set of program routines. The routines are supposed to almost
automatically perform the calibration from the respective data
sets. The routines must allow an external observer with no additional
knowledge on the instrument to obtain calibrated measurement data in
about half an hour after the observation. If that goal was achieved will
also have to be tested in Tenerife.
\end{chapter}
\begin{chapter}{Solar physics $\&$ polarized light\label{chap2}}
\begin{minipage}{12cm}
''\it Evershed's recent spectroheliographic results indicate that there
is an outward flow, parallel to the photosphere, from the center of
sunspots at the iron level,...'' \rm G.E.Hale (1910)\\
\end{minipage}\\
$ $\\
The following description of solar magnetic phenomena is far from
complete. It especially concentrates on topics, which can be displayed
from the examination of polarimetric data performed by the author
himself in chapter \ref{polarimetric}. For words in \it italics \rm a
short explanation can be found in the glossary on page \pageref{glossary}.
\begin{section}{Solar magnetism\label{solmag}}
\begin{figure}[ht]
\psfrag{G}{\large 15 $\% \cdot I_{max}$}
\psfrag{H}{\large \hspace*{-2.4cm} 100 $\% \cdot I_{max}$}
\psfrag{C}{\large -22 $\% \cdot I_{mean}$}
\psfrag{T}{\large \hspace*{-1.6cm}8 $\% \cdot I_{mean}$}

\begin{minipage}[b]{7.5cm}
\hspace*{.3cm}
\includegraphics[width=7.2cm,height=8.4cm]{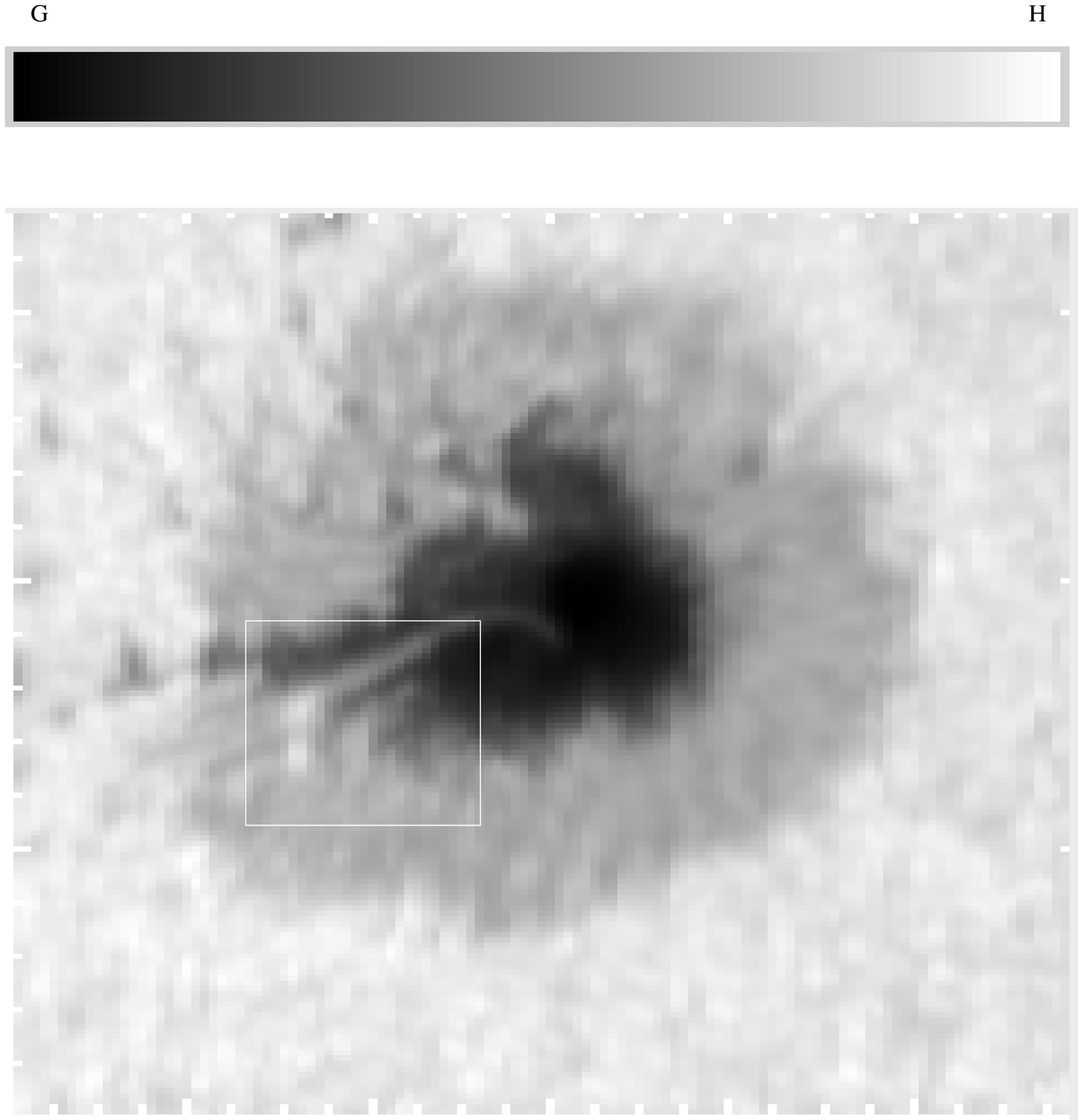}
\caption[Sunspot near disc center]{Sunspot near disc center, continuum
intensity. The image area is 62'' x 55'', ASP data from 10.2.2000
(courtesy M.Sigwarth). The suppression of convection by the magnetic
field causes a decrease in temperature and emitted intensity. Indside
the spot the dark umbra can be distinguished from the brighter
penumbra. The rectangle marks an area with a radial oriented
substructure in the penumbra of the spot, extending into the umbra as
well.\label{sunspot}}
\end{minipage}
\hfill
\begin{minipage}[b]{7.5cm}
\hspace*{.3cm}
\includegraphics[width=7.2cm,height=8.4cm]{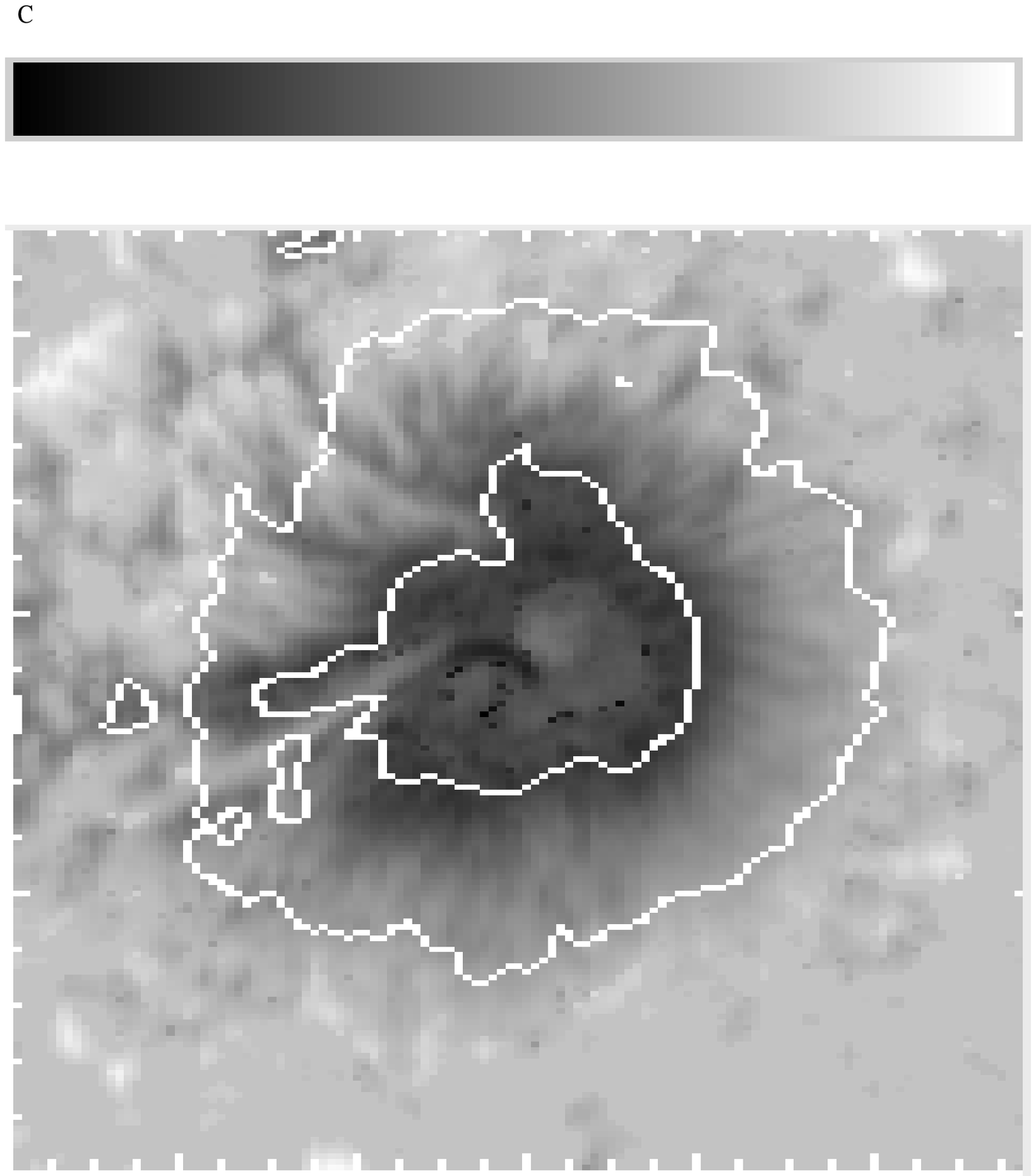}
\caption{The signed Stokes-V amplitude with superimposed borders of
the umbra and penbumbra from the continuum image.\newline  The Stokes-V
amplitude is proportional to the line-of-sight component of the magnetic
field, while the sign indicates the direction of the field to (+) or
away (-) from the observer. The magnetogram is constructed from the
amplitudes in the Stokes V signal, see section \ref{stokesveval} for
details.\label{stokesvspot}\newline}
\end{minipage}
\end{figure}
The activities on and of the sun include a number\footnote{For a more
detailed overview see for example M.Stix, \it The sun \rm, \cite{stix}.} of
phenomena like:
\begin{itemize}
\item sun spots
\item flares
\item coronal mass ejections
\item the sun spot cycle
\item the solar wind
\item total solar luminosity.
\end{itemize}
Some of them are of no or little concern to us, while
others like solar storms or variations of the total solar luminosity can
influence the existence of life on our planet. Today most of these features are
supposed to result from an interaction between the material
motions of a \it gas plasma \rm of electrons and ionized atoms and \it
magnetic fields\rm.\\

The basic feature most suited to display this interaction are 
also the most prominent structures on the sun, the sun
spots. Fig. \ref{sunspot} displays a high resolution intensity image of
a typical spot near the disc center.

Sun spots have been reported as early as in the 17th century after the
invention of  the telescope. They appear as darker areas on a bright
background. The first and simplest explanation is that they are
cooler than their surroundings. This has been proven by the observation of
absorption lines from molecules, which dissociate at higher
temperatures, in the spectra of spots. This leads to the next question,
namely why are they 'cool' ?

A hint to the answer were the observations made by G.E.Hale in
1908\footnote{G.E.Hale, \it On the probable existence of a
magnetic field in sun spots \rm, \cite{hale}, ApJ \bf 28\rm, 1908},
where the Zeeman effect was used to measure the magnetic
field strength in sun spots. The existence of these fields was
suspected, because photographs in the H$_\alpha$ spectral
line ''suggest that all sun spots are vortices''\footnote{\cite{hale},
p. 315}. It is a consequence of Maxwell's laws, that these moving
charges, or in other words, a current, induced a magnetic field. Even as 
 it was proven later by velocity measurements with the Doppler shift
that the supposed whirling motion did not exist at all, the results from
Hale's observations were non-ambiguous: in sun spots existed magnetic fields
of about 2900 Gauss\footnote{\cite{hale}, p. 325} (or 0.29 Tesla\footnote{1
Tesla = 1 Vs/m$^2$ = 10000 Gauss} in SI units). Fig. \ref{stokesvspot}
displays the existence of the magnetic field, as the shown Stokes-V
amplitude is proportional to the line-of-sight (LOS) component of the
magnetic field.

The decrease of the temperature in sun spots is no direct result of the
magnetic field, only in connection with the conditions prevailing on
the sun can this be achieved. The upper layers of the sun's atmosphere
transport energy mainly by \it convection\rm, in difference to the \it
radiative \rm core. The convection can be seen as \it granulation
cells \rm on the surface of the sun. In sun spots this convection 
is suppressed, because the plasma can only move along the
magnetic field lines and not normal to them.

Like the sun spots most other solar features are coupled to magnetic
fields influencing material motions or vice versa:
\begin{itemize}
\item The magnetic field appearing later in sunspots is supposed to be
caused by the \it solar dynamo\rm, which generates it from
convective motion and the \it differential rotation \rm of the sun.
A comparison with other active stars shows a strong dependence on
the velocities of the material motions\footnote{see K.G. Strassmeier,
\it Aktive Sterne \rm, \cite{strassmeier}}.
\item The structure of the \it corona \rm is dominated by field lines,
building magnetic \it loops \rm or \it filaments\rm. Regions with open
field lines are the origin of the \it solar wind\rm.
\item Flares or coronal mass ejections consist of a massive energy
release in a short time. The energy is assumed to result partly from \it
magnetic reconnection\rm. These outbursts lead to an acceleration of
ionized material, which may eventually hit the atmosphere of the earth,
causing polar lights.
\item Sun spots can be stable for some weeks, while their surrounding is
permanently changing. In this case the strong magnetic field dominates
over other dynamical influences.
\item Sun spots exhibit internal structure like the division into umbra
and penumbra. The penumbra has a substructure with varying magnetic
fields. On the static background field of the sun spot moving \it
penumbral grains \rm hint to mass motions.
\item The \it Evershed flow \rm as a global phenomen in sun spots also
indicates a systematic mass flow, approximately radially outwards from
the spot center.
\end{itemize}
To understand the features appearing on the sun two different
environment parameters have to be known: the velocity and direction of mass
motions, which can be measured at least in the line-of-sight (LOS) with
the \it Doppler shift\rm, and the magnetic field. For the first part one
can use a spectrograph to obtain resolved spectral lines, the second
part is the task of \it polarimetry\rm. As the magnetic field influences
the shape of the intensity spectrum and the polarization of the
radiation, it is possible to invert this process and reconstruct the
magnetic field from polarimetric data measured by instruments like POLIS.
\end{section}
\begin{section}{The Zeeman effect}
G.E.Hale made use of the splitting of spectral lines due of
the Zeeman effect to prove the existence of magnetic fields on the
sun. This method works without taking into account any polarization
effects, because in fields of sufficient strength the splitting
can be seen in the spectrum of the intensity\footnote{Hale's
measurement method was the comparison with a spectral line produced with
a spark and a electromagnet of known field strength, \cite{hale},
p. 325f.}. The required field strength depends magnetic sensitivity of
the spectral line used. One has to use additional polarimetric
information to obtain the field direction, or for weaker fields, due to
the line broadening by collisions and random motions.\\

The Zeeman effect describes the change of the energy of atomic
levels in the presence of a magnetic field\footnote{For a detailed
discussion see for example Mayer-Kuckuck, \it Atomphysik \rm,
\cite{mayer}, or any textbook on quantum mechanics.}. Its main
feature is to break the energy degeneracy of atomic levels with
different quantum number $m$, which is the case for an undisturbed
atom. The actual fields on the sun are 'weak' in the terminology of
atomic physics. The so called Russel-Saunders coupling
of the angular momentum\footnote{Both vectors
and matrices will be printed bold, the context should exclude
confusions.} \bf L \rm and the electron spin \bf S \rm to the total
angular moment \bf J \rm can be assumed. If the magnetic field \bf B \rm
is taken to be (0, 0, $B_z$)\footnote{This is assumed everywhere in the
following text, if not especially mentioned otherwise.}, the problem
can be solved in the following way:
\begin{itemize}
\item \bf J \rm = \bf L \rm + \bf S\rm, where $|$ L - S $|$ $\leq $  J
$\leq $ $|$  L  +  S $|$
\item the quantum number $m_j$ can have the (2J+1) values -J,-J+1,...,J-1,J
for a fixed value of J
\item the energy level corresponding to $m_J$ is corrected by a value
$\Delta E_{B,m_j}$, for which perturbation theory yields
\begin{eqnarray}
\Delta E_{B,m_j} = \mu_B\;g_j\;m_j\;B 
\end{eqnarray}
where $\mu_B = \frac{e \hbar}{2 m_e c}$ is Bohr's magneton and the Land\'e
factor $g_j$ is calculated by
\begin{eqnarray}
g_j = 1 + \frac{J(J+1) + S(S+1) - L(L+1)}{2 J(J+1)}\;\;.
\end{eqnarray}
\end{itemize}

This result can best be visualized with an example like
Fig. \ref{zeemanbild}, using one of the spectral lines observed by
POLIS. The transition is one of neutral iron (FeI) with a rest
wavelength around 630.25 nm.
\begin{figure}
\psfrag{A}{\huge $^5D_0$}
\psfrag{B}{\huge $g_j = 0$}
\psfrag{C}{\huge $^5P_1$}
\psfrag{D}{\huge $g_j = 2.5$}
\psfrag{E}{\huge -1}
\psfrag{F}{\huge 0}
\psfrag{G}{\huge +1}
\psfrag{H}{\huge $\Delta M$}
\psfrag{I}{\huge $m_j$}
\psfrag{J}{\huge $\sigma^-$}
\psfrag{K}{\huge $\pi$}
\psfrag{L}{\huge $\sigma^+$}
\psfrag{M}{\huge no \bf B \rm field}
\psfrag{N}{\huge external field}
\resizebox{9cm}{!}{\includegraphics[height =10cm,width
=16cm]{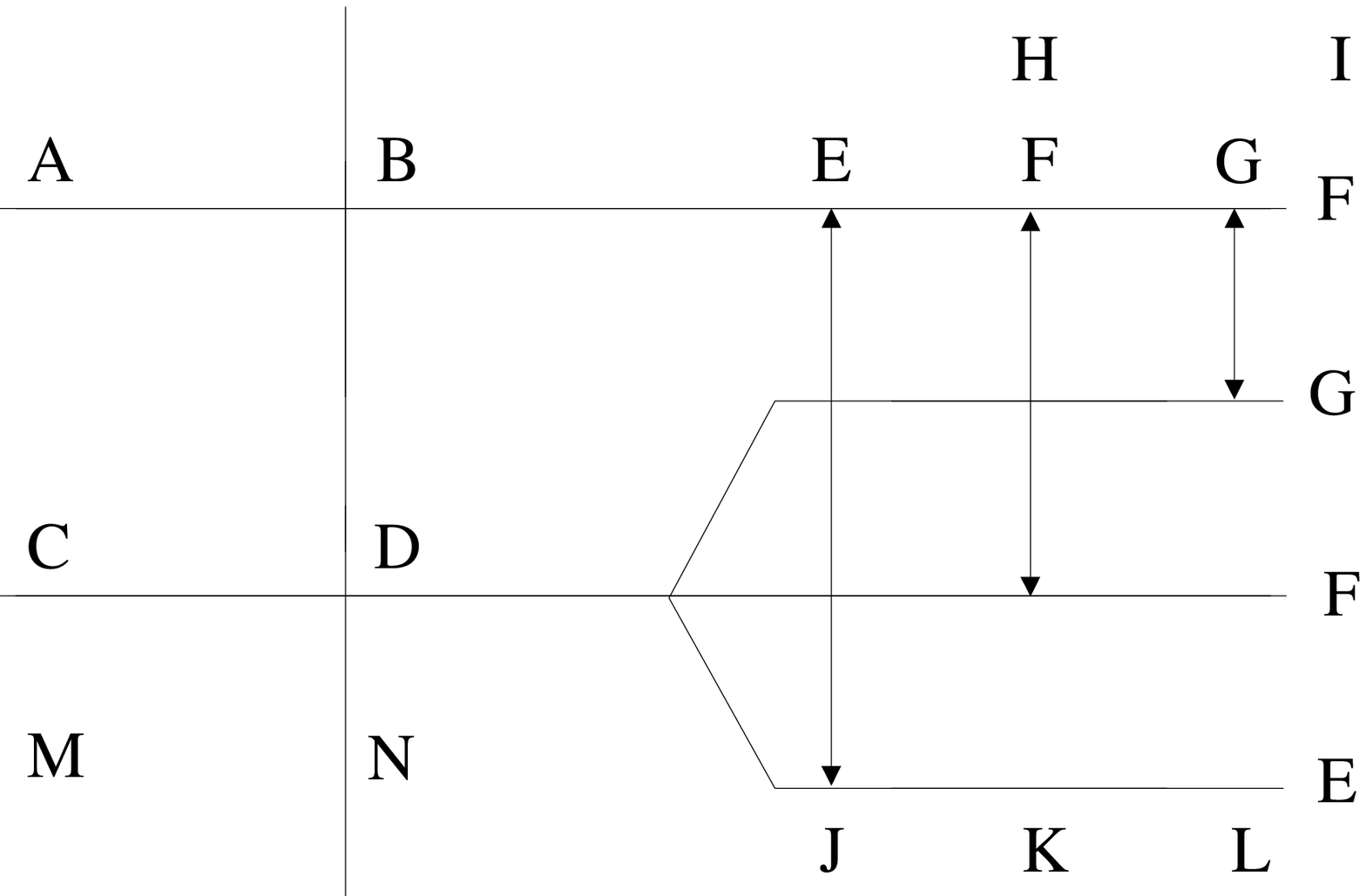}}
\hfill
\begin{minipage}[t]{6cm}
\psfrag{A}{\large $\sigma^-$}
\psfrag{B}{\large $\pi $}
\psfrag{C}{\large $\sigma^+$}
\psfrag{D}{\large $\lambda$}
\includegraphics[height=5cm,width=6cm]{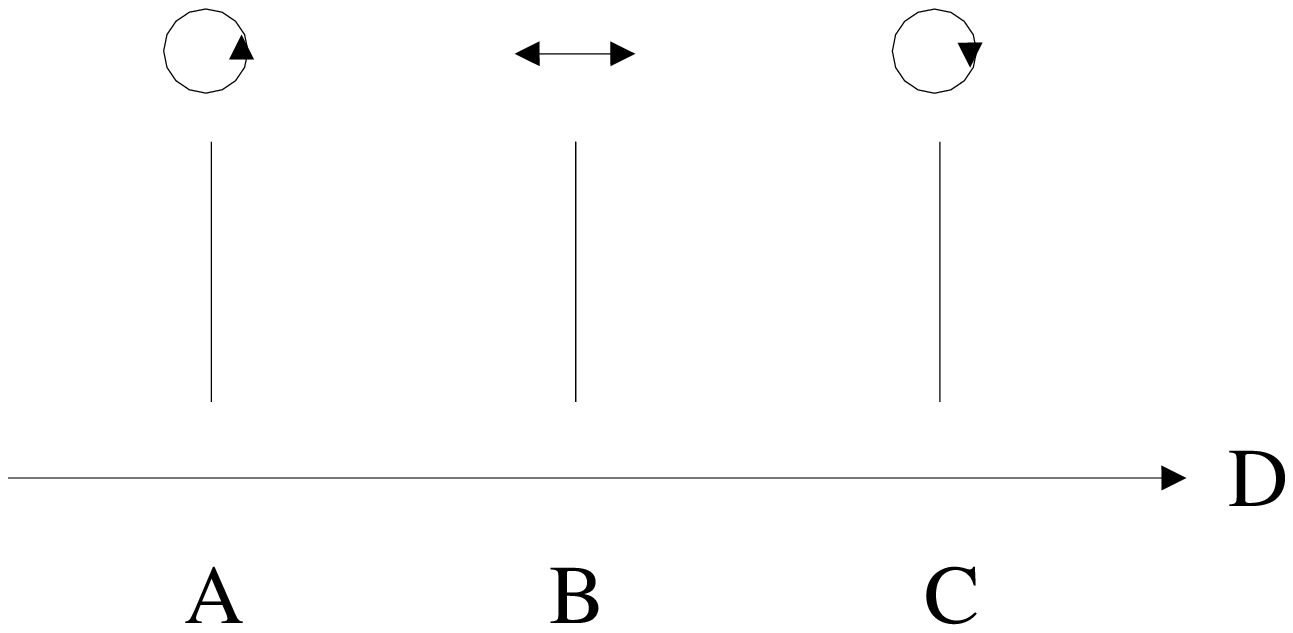}
\end{minipage}
\caption[Zeeman splitting]{(\it left\rm) Zeeman splitting of the FeI
line at 630.25 nm. The notation is $^{2S+1}L_J$. (\it right\rm) Resulting
spectrum in the external field.\newline
The upper level remains unaffected by the magnetic field, as \bf J \rm =
0, and also the Land\'e factor $g_j$. The lower splits into three
different levels. The possible 
dipole transitions $\pi, \sigma^+$ and $\sigma^-$ are coupled to a fixed
polarization each, corresponding to linear and left or right circular
polarized light, depending on the value of $\Delta M$. The $\pi$-component
lies at the wavelength of the undisturbed transition.\label{zeemanbild}}
\end{figure}
The lower level $^5P_1$ splits into three energetically different
equidistant levels, while the upper level cannot split up (\bf J \rm$=0
\rightarrow m_J=0,\; g_J=0$). The three possible dipole transitions
$\pi,\sigma^+$ and $\sigma^-$ differ in wavelength\footnote{$E_\gamma = h \cdot
\nu = h \cdot c\; /\; \lambda$, and $E_\gamma = E_0\; (\rightarrow \pi)$
respectively $E_0 \pm \Delta E_{B,m_j}\; (\rightarrow \sigma^\pm)$.} as 
well as in the polarization state. A transition with $\Delta M$ =
0 corresponds to linear and $\Delta M = \pm1$ to right respectively left
circular polarized light. The resulting spectrum consists of three
different polarized spectral lines instead of a single unpolarized one.

The case just discussed is the simpliest that can
happen. If the Land\'e factors are not zero for both levels, one has to
calculate all energy levels, and use the rules for dipole
transitions\footnote{See for example W. Demtr\"oder, Band 3 : \it
Atome, Molek\"ule und Festk\"orper \rm, \cite{demtroeder}, p. 205f.} to
otain the allowed transitions. Usually a spectral distribution results,
in which lines of different polarization are mixed. Fortunately it is often
possible to use an \it effective \rm Land\'e factor to group lines
with the same polarization to reach an analogon to the situation
depicted.\\

The importance of the observation of the polarization does not lie
in the calculation of the absolute field strength. This would for
example be possible from the distance between the two circular polarized
components, if they are measured separately. The circular
components would lie at different wavelengths in the spectrum, even if the
splitting can not be resolved in the intensity profile due to the line
broadening. A classical treatment of the Zeeman effect\footnote{Collett,
\cite{collett}, chapter 18.} reveals that in the strength of the
different polarization components the information on the direction of
the magnetic field is also included.
\begin{figure}[ht]
\psfrag{A}{\Huge \bf B \rm = (0,0,$B_z$)}
\psfrag{B}{\Huge z}
\psfrag{C}{\Huge x}
\psfrag{D}{\Huge y}
\psfrag{E}{\Huge $\omega_3$}
\psfrag{F}{\Huge $\omega_1$}
\psfrag{G}{\Huge $\omega_2$}
\resizebox{9cm}{!}{\includegraphics[height =15cm,width
=16cm]{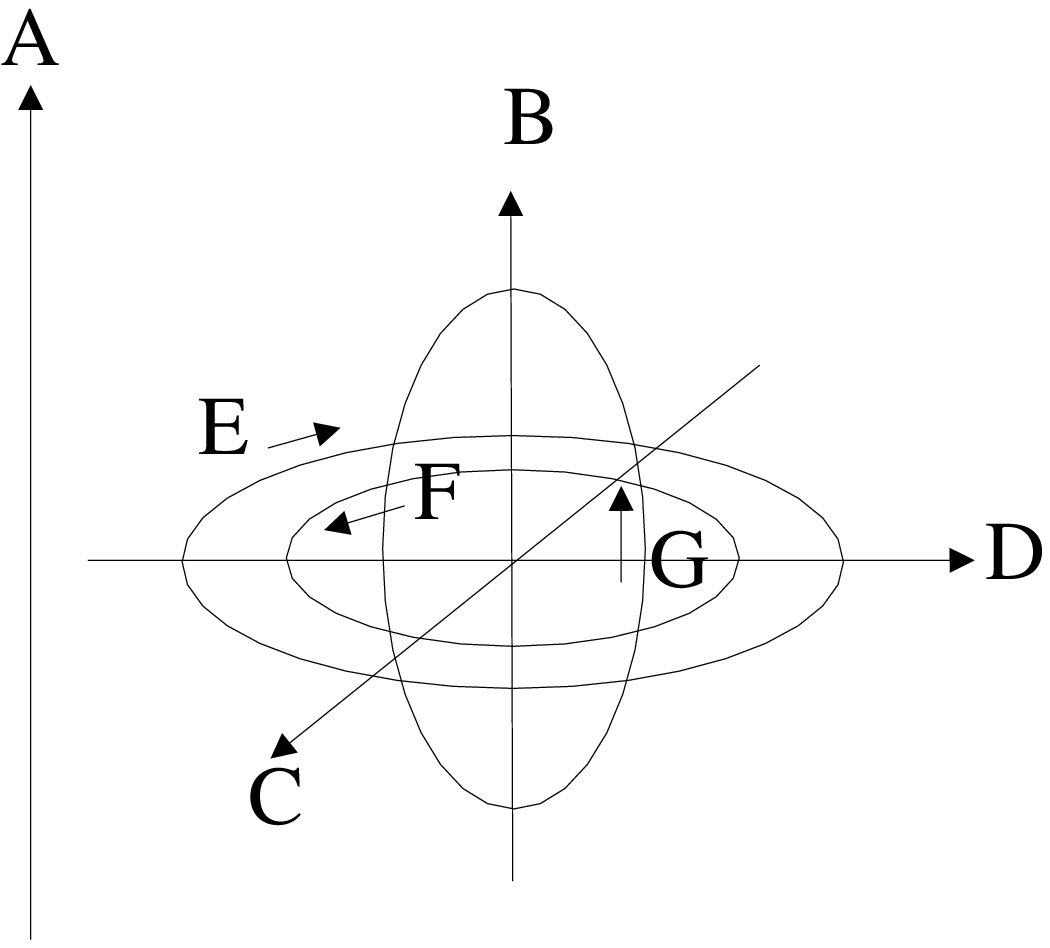}}
\hfill
\begin{minipage}[t]{6cm}
\psfrag{A}{\large $\omega_3$}
\psfrag{C}{\large $\omega_1$}
\psfrag{D}{\large $\lambda$}
\psfrag{E}{\large $\omega_3$}
\psfrag{F}{\large $\omega_2$}
\psfrag{G}{\large $\omega_1$}
\psfrag{H}{\large longitudinal Zeeman effect}
\psfrag{I}{\large transverse Zeeman effect}
\includegraphics[height=8cm,width=6cm]{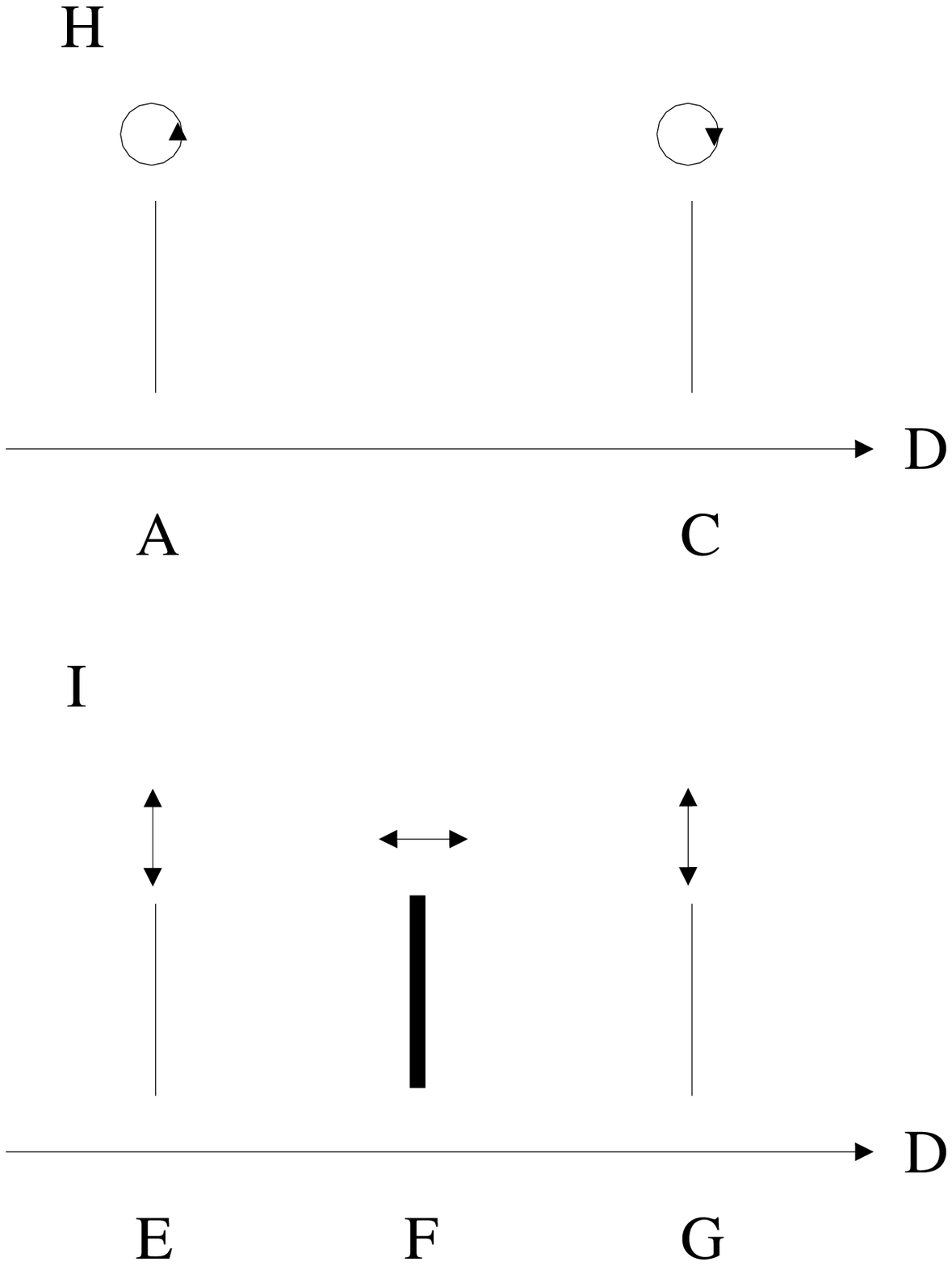}
\end{minipage}
\caption[Classical Zeeman effect]{(\it left\rm) A symbolic sketch of the
classical treatment of the Zeeman effect.(\it right\rm) Resulting
spectra for observation along and perpendicular to the magnetic
field.\newline
The rotation of an electron around the atomic nucleus in an external
field results in different rotation frequencies $\omega_i$ corresponding
to the three wavelengths. The frequencies depend on the sense of
direction of the rotation and the differences are caused by the Lorentz
force \bf v \rm x \bf B \rm,  with $\omega_1\; (\sigma^+) < \omega_2\;
(\pi) <\omega_3\; (\sigma^-)$.\newline
On observation parallel to the z-axis one can only see the two circular
components $\omega_1$ and $\omega_3$ because of the radiation
characteristic of a dipole (longitudinal Zeeman effect). On observation in
the x-y-plane normal to the magnetic field there are three linear
polarized oscillations with the frequencies $\omega_i$ (transverse
Zeeman effect). From the relative strength of the components one can, as
the direction of observation is known, retrieve information on the
direction of the magnetic field.\label{classzeeman}}
\end{figure}

The motion of an electron around an atomic nucleus is considered there as an
oscillation in three dimensions, with a central force given by the
coulomb attraction. This system is disturbed by the
presence of the magnetic field, which leads to a Lorentz force $\propto$
\bf v \rm x \bf B \rm. This results in three different oscillation
frequencies depending on the sense of revolution in the x-y-plane, while
movement parallel to the z-axis is not influenced. Figure
\ref{classzeeman} gives a symbolical sketch of the situation: on observation
parallel to the magnetic field, i.e. looking down on the x-y-plane, only
the two circular polarized components $\omega_1$ and $\omega_3$ can be
seen, as the radiation characteristic of an electric dipole
forbids emission in the oscillation direction (\it longitudinal \rm
Zeeman effect). If one observes in the x-y-plane normal to the magnetic
field lines, one sees three distinct linear polarized lines at the
frequencies $\omega_i$ with i=1,2,3 (\it transversal \rm Zeeman effect).\\ 
From the strength of the single polarization components relative to
each other and their spectral position, i.e. the displacement from
the zero level of undisturbed ground state transitions, one has enough
information to otain absolute strength as well as the direction of the magnetic
field. The intensity spectrum in Fig. \ref{stokesprof} from actual
obseraations shows the discussed necessity of polarization measurements:
even if the spectral line with the greatest Land\'e factor splits
visibly, the position and intensity of the components can not be
established accurate enough due to the doppler broadening of the line.

The usual way to obtain the magnetic field from observational data is an
inversion, i.e. the comparison of observed with synthetic line profiles from a
model atmosphere. The model atmosphere includes open parameters like
field strength and field direction, which are adjusted until the
synthetic profiles match the observation.\\

The effects discussed here concerning an \it emission \rm line can be also
applied on the actual measured \it absorption \rm line.
\end{section}
\begin{section}{The theoretical description of polarized light}
The Zeeman effect produces linear or circular polarized light. In
combination with the geometrical aspects of observation direction and
inclination of the magnetic field at last an arbitrary polarization is
usually obtained. The description to be used must therefore be capable of
including all possible states of polarization\footnote{Also totally unpolarized
light, which proves to be difficult to describe, and even more difficult
to artificially produce.}.
\begin{subsection}{The Stokes formalism}
The Stokes formalism was formulated around 1850 by Sir G.E.Stokes. The
derivation given here follows Collett, \cite{collett}, Chapter 2-4. 

The starting point is the wave character of light, which was the only 
existing view at the times of Stokes. The electric and
magnetic vector fields are supposed to be the solutions of a wave
equation, which was only later really proven by Maxwell's laws. This
empirical approach was motivated by many interference experiments, which found
their natural explanation in a wave character. Most of the basic optical
effects can be understood by using a wave equation taken from
mechanics, namely
\begin{eqnarray}
 \Delta \bf u\rm(\bf r\rm , t ) = \frac{1}{\nu^2}\; \frac{d^2}{dt^2}\; \bf
u\rm( \bf r\rm , t ) \label{schwingl} \; ,
\end{eqnarray}
where $\Delta = \frac{d^2}{dx^2} +\frac{d^2}{dy^2} + \frac{d^2}{dz^2}$,
$\nu$ is the oscillation frequency, \bf r \rm = (x,y,z), and the
components u$_x$, u$_y$ and u$_z$ are three displacements dependent on
place and time.\\

The interference experiments of Fresnel and Arago (1816) showed,
that light can only perform \it transversal \rm oscillations and no \it
longitudinal \rm ones\footnote{C.Beck, \it Kalibration und Auswertung von
ASP-Daten \rm, \cite{beck}, Appendix C.1, or Collett, \cite{collett},
Chapter 12.}, which can also be derived from Maxwell's laws. If the
z-axis always is parallel to the direction of the light propagation,
only two field components are left.\\

A simple solution of equation (\ref{schwingl}) can be given by
\begin{eqnarray}
E_x ( z, t ) & = & E_{x,0}\cdot \cos\;(\; \omega t - k z + \delta_x\; ) \label{efield}\\
E_y ( z, t ) & = & E_{y,0}\cdot \cos\;(\; \omega t - k z + \delta_y\; )
\label{ey}\nonumber \;,
\end{eqnarray}
where the displacement \bf u \rm has been substituted by the electrical
field \bf E\rm. The longitudinal component $E_z$ can be left
out, and the solution is characterized by the values of the angular
velocity, $\omega = 2\pi\cdot\nu$, the wave number, $k = 2\pi /
\lambda$, the field amplitudes, $E_{i,0}$, and an abritrary phase, $\delta_i$.

After squaring the equations in (\ref{efield}) the equation of the polarization
ellipse can be derived:
\begin{eqnarray}
\fbox{$\displaystyle
\frac{E^2_x}{E_{x,0}^2} + \frac{E^2_y}{E_{y,0}^2} - 2\cdot
\frac{E_x}{E_{x,0}}\cdot \frac{E_y}{E_{y,0}}\cdot \cos \delta  =
\sin^2 \delta \label{polellip}$} \;,
\end{eqnarray}
with the phase difference, $\delta = \delta_x - \delta_y$. The
argument (z,t) has been omitted.\\
Equation (\ref{polellip}) gives the locus of points ($E_x,E_y$), which are
taken in the course of one oscillation, forming a general ellipse. The
specific shape depends on the amplitudes $E_{x,0}$ and $E_{y,0}$, as well
as on the relative phase $\delta$. These three parameters encode the
polarization state of the light beam. For example,
$\delta\;=\;\frac{\pi}{2}$, and $E_{x,0}\;=\;E_{y,0}$, leads to the
equation of a circle, corresponding to circular polarized
light\footnote{For more  details compare Beck, \cite{beck}, Section
2.2.2, or Collett, \cite{collett}, Chapter 3.}.\\
With the polarization ellipse of the instantaneous field vector it is
only possible to describe fully polarized light. Furthermore, it can not be
used as measurement reading, as in the visible range the oscillation 
frequency amounts to about $10^{15}$ Hz, far beyond the time resolution
of most instruments. The special merit of Stokes' work was the 
transformation of all information needed to characterize a polarization
state into measurable quantities.

Without the ability to resolve the oscillation, it seems
natural to imitate the measurement process possible, i.e. to average
over a great number of oscillations. The time average is given by
\begin{eqnarray}
\langle\; E_i(t) , E_j(t)\; \rangle = \lim_{T \rightarrow 0} \frac{1}{T}
\int\limits^T_0 E_i(t)E_j(t)\; dt \;\; .
\end{eqnarray}
If one now substitutes the time dependent values ($E_x\;=\;E_x (z,t)$)
with their time averages in Eq. (\ref{polellip}), one obtains after a
short computation\footnote{One can assume z=0. the calculation can be
found in Collett, \cite{collett}, p. 36.} the equation
\begin{eqnarray}
\left( E_{x,0}^2 + E_{y,0}^2 \right)^2 = \left( E_{x,0}^2-E_{y,0}^2 \right)^2+\left( 2\cdot E_{x,0}\cdot E_{y,0}\cdot
\cos\, \delta \right)^2+\left( 2\cdot E_{x,0}\cdot E_{y,0}\cdot
\sin \delta \right)^2 \;. \label{equationtostokes}
\end{eqnarray}
With the terms in Eq. (\ref{equationtostokes}) the following definition
of the Stokes parameters $S_i$, usually arranged in a
vector\footnote{Even though addition and a norm are defined, the Stokes
vectors form no vector space in the strict mathematical sense.} for
convenience, can be made: 
\begin{eqnarray}
\fbox{\fbox{\bf Stokes parameters 
\bf S \rm$ = \begin{pmatrix} S_0 \cr S_1 \cr S_2 \cr S_3 \cr
\end{pmatrix} = \begin{pmatrix} E_{x,0}^2 + E_{y,0}^2 \cr
E_{x,0}^2-E_{y,0}^2 \cr 2\cdot E_{x,0}\cdot E_{y,0}\cdot \cos \delta
\cr 2\cdot E_{x,0}\cdot E_{y,0}\cdot \sin \delta\cr \end{pmatrix}
\label{stokesdef}$}} \;\; .
\end{eqnarray}

With different values for the phase difference ($\delta = 0,\; \pm
\pi/2,\; \pi$) and the amplitudes ($E_{x,0}, E_{y,0}$), it can be seen
that the Stokes parameters encompass the information on the polarization
state of the light in the following way:
\begin{itemize}
\item $S_0$ is the total intensity of polarized and unpolarized light.
\item $S_1$ is the difference of the intensity of \it linear horizontal
\rm and \it vertical \rm polarized light.
\item $S_2$ is the difference of the intensity of $+ 45^\circ$-\it
linear\rm\footnote{To the vertical.} and $- 45^\circ$-\it linear \rm
polarized light.
\item $S_3$ is the difference of the intensity of \it right \rm and \it
left circular \rm polarized light.
\end{itemize}
The following Stokes vectors, which contain only one polarization state
each, can be used as a base for all polarized light:
\begin{eqnarray} \bf S\rm_{unpol} = \begin{pmatrix} 1 \cr 0 \cr 0 \cr
0\cr \end{pmatrix},\;\; \bf S\rm_{lin,hor/ver} = \begin{pmatrix} 1 \cr
\pm 1 \cr 0 \cr 0\cr \end{pmatrix},\;\; \bf S\rm_{lin,\pm 45^\circ} =
\begin{pmatrix} 1 \cr 0 \cr \pm 1 \cr 0\cr \end{pmatrix},\; \bf
S\rm_{circ,right/left} = \begin{pmatrix} 1 \cr 0\cr 0 \cr \pm 1 \cr \end{pmatrix}\;.\label{mostimpstokes}
\end{eqnarray}
In contrast to the polarization ellipse, unpolarized as well as
partially polarized light is possible by linear combinations of the
vectors above. The requirements on a physical sensible vector are:
\begin{eqnarray}
S_0^2 \ge S_1^2 + S_2^2 + S_3^2 \;,\label{physsens}
\end{eqnarray}
i.e. the polarization degree $P = \frac{\sqrt{\rm S_1^2 + S_2^2 +
S_3^2}}{\rm S_0}$ can not exceed 100 $\%$, and of course $S_0 > 0$.\\

To establish the polarization state of a light beam one has to
measure, with a quantitative intensity detector like a photographic emulsion
or a CCD chip\footnote{A point which hindered the success of Stokes
theory at first was the absence of such a detector at his
time.}, the contribution of each of the polarization states in
Eq. (\ref{mostimpstokes}). This can be done for:
\begin{itemize}
\item The total intensity without any additional device.
\item The linear polarizations with a polarizer under different angles,
resulting in $S_1$ and $S_2$.
\item The circular polarization with a combination of polarizer and
retarder\footnote{Also called wave plate.}, resulting in $S_3$.
\end{itemize}

In newer publications the notation is choosen as \bf S \rm =
(I,Q,U,V). This will also be adopted here in the following.
\end{subsection}
\begin{subsection}{The Mueller matrix calculus}
Optical elements that influence the polarization state in a linear way, can
be included easily into the Stokes formalism. The element is
represented by a matrix \bf M \rm $\in \mathbbm{R}^{4 \times 4}$, which
expresses the modifications of the base vectors performed by the
device. The resulting Stokes vector $\bf S\rm^\prime$ after the
transmission or reflection is given by $\bf S\rm^\prime = \bf M \rm \cdot
\bf S \rm $. The matrices for the commonly used optical devices can be
found in Appendix \ref{muellermat}. The matrix of the rotator
(\ref{mrot}) can be used to describe elements with an arbitrary position
angle. The matrices have to be also physical sensible, i.e. they must
not lead to a violation of eq. (\ref{physsens}) on application, either
by the introduction of negative intensities or hyper-polarization.\\

The calculations are simplified by the fact that most detectors can only
record intensity regardless of its polarization state. A complicated
optical train can be expressed by the matrix multiplication of its
elements in the correct order, i.e. their place in the light path, and
the selection of the intensity entry $S_0^\prime$ as measurement value.
\end{subsection}
\end{section}
\end{chapter}
\begin{chapter}{Vector polarimetry\label{vecpola}}
\begin{minipage}{12cm}
''\it The observer lay on his back while making the observations in the
following way...\rm'' C.E.St.John (1909)\\
\end{minipage}\\
$ $\\
Solar magnetometry has been performed for decades now. The instruments
used at the beginning differ in some respects to a vector polarimeter,
as they mostly only measure the Stokes V signal. A more detailed
decription of such \it magnetographs \rm can be found in M.Stix, \it The
sun \rm, \cite{stix}. This thesis also abstains from the description of
the various ways to built optical elements like retarders or polarizers
for the sake of shortness. The book of Shurcliff (\cite{shurcliff}) can
here be used as reference. Besides POLIS only one instrument shall be
introduced in more detail in section \ref{predecessor}, the vector
polarimeter ASP.\\

To establish the full Stokes vector of a light beam always at least four
intensity measurements are needed. For the 'classical' method this is
the intensity produced by a linear polarizer and a retarder at four fixed angle
positions. There are also various measurement methods with a single
element and a continuously changing position angle\footnote{compare
\cite{collett}, Chapter 6}. An extensive summary of possible
methods can be found in E. Landi Degl'Innocenti, \cite{landi}, (1992).\\

A general measurement scheme consists of two parts:
\begin{itemize}
\item[(1)] The transformation of a polarized input into a modulated
intensity signal by one or more optical elements. The intensity after
transmission has to be proportional to all polarization content in some way.
\item[(2)] The demodulation, the decomposition of the modulated intensity
into separate parts. Each part has to be related to
\begin{itemize}
\item a single polarization state
\item a linear combination of polarization states with known
coefficients.
\end{itemize}
\end{itemize}
To give only one example for an unuseable scheme: a continously rotating
polarizer would produce more then the minimal four intensity
measurements, but is opaque to circular polarization, which contradicts
the condition in (1). 
\begin{section}{A dual beam polarimeter with rotating
retarder\label{dualbeamtheory}}
\begin{figure}
\psfrag{A}{\Large retarder}
\psfrag{B}{\Large incoming beam}
\psfrag{C}{\Large beamsplitter}
\psfrag{D}{\Large $I^+$}
\psfrag{E}{\Large $I^-$}
\psfrag{F}{\Large detector}
\psfrag{G}{\Large angle $\theta$, ret. $\delta$}
\psfrag{H}{\Large s. appendix \ref{beamsplit}}
\psfrag{I}{\Large $I^\pm (\theta,\delta)$}
\includegraphics{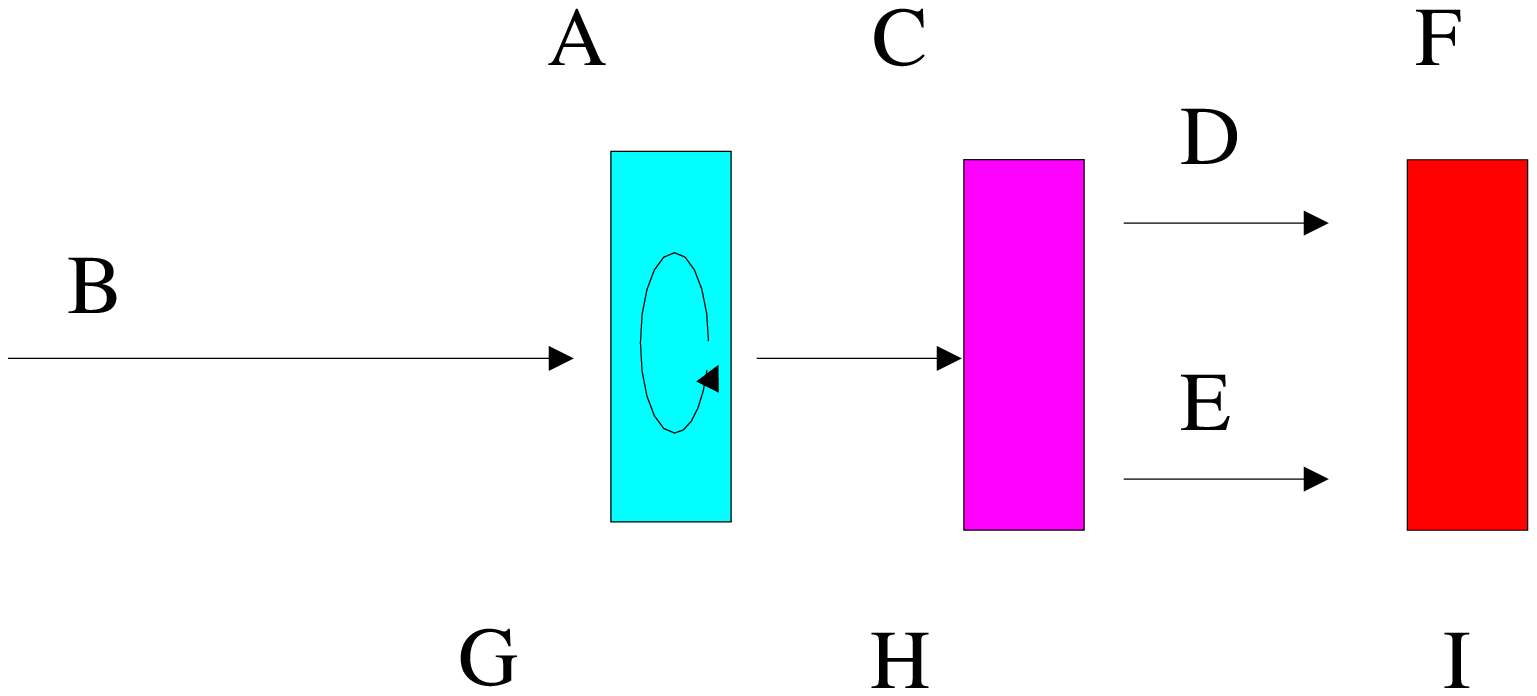}
\caption[Schematical design of a dual beam polarimeter]{Schematical
design of a dual beam polarimeter with rotating retarder.\newline
The incoming beam from the light source hits the
retarder, which is rotated continuously around the propagation
direction, and modulates the polarization state. The polarizing beam splitter
produces the two beams $I^+$ and $I^-$, which are registrated on spatially
separated regions of the detector or on two detectors. Their
respective intensity $I^\pm(\theta,\delta)$ depends on the retardance,
the angular position and the polarization of the incoming
light. \label{schemadual}}
\end{figure}
POLIS\footnote{As the ASP is similar in many respects, it will not be 
especially mentioned in the following.} uses a rotating retarder as
variable optical element. It is followed by a polarizing beamsplitter,
which helps to reduce intensity loss and the influence of
detector noise. The schematical design of a dual beam polarimeter with
this modulation procedure is given in Fig. \ref{schemadual}. The needed
devices are:
\begin{itemize}
\item A retarder of known retardance, $\delta$, which can be rotated
around the direction of light propagation with the position angle of its
fast axis, $\theta$.
\item A polarizing beamsplitter to produce two spatially separated
beams $I^\pm (\theta,\delta)$.
\item A suited set of detectors to record these intensities.
\end{itemize}

In the Mueller calculus the optical train is described by:
\begin{eqnarray}
\bf M\rm^\pm(\theta,\delta) = \bf M\rm^\pm_{beamsplitter}\cdot\bf M\rm_{rot}(-2
\theta)\cdot \bf M\rm_{ret}(\delta)\cdot\bf M\rm_{rot}(2 \theta)\;.
\end{eqnarray}
The resulting intensity for the beams $I^\pm(\theta,\delta)$ can be
calculated by:
\begin{eqnarray}
I^\pm(\theta,\delta) = ( \bf S\rm^{out} )_0^\pm = (\bf
M\rm^\pm(\theta,\delta) \cdot \bf S\rm^{in} )_0 \;.\label{opttrain}
\end{eqnarray}
\end{section}
\begin{section}{The measurement process}
Executing the calculations in Eq. (\ref{opttrain}) the intensities
$I^\pm$ of the two beams in dependence of an incoming Stokes vector \bf
S\rm$^{\rm in}$ = (I,Q,U,V) are obtained: 
\begin{eqnarray}
I^\pm (\theta,\delta) = \frac{r^\pm}{2} \left\{\rm I \pm Q  \cdot (\it
c\rm ^2_{2\;\theta} + \it s\rm ^2_{2\;\theta}\;\it c\rm _\delta) \pm U \cdot
\it s\rm _{2\;\theta}\; \it c\rm _{2\;\theta}(1-\it c\rm _\delta) \mp\rm
V \it\cdot s\rm _{2\;\theta}\;\it s\rm _\delta\right\}\label{iplusminus} ,
\end{eqnarray}
where $c_j = \cos j$, $s_k = \sin k$, $r^\pm$ is the transmission
coefficient for the respective beam\footnote{Set equal to 1 in the
following.}, and $\delta$ is the constant retardance.

Eq. (\ref{iplusminus}) shows a constant contribution by the total
intensity I, with an added modulation by the other Stokes parameters,
Q,U, and V. The problem of demodulation is to isolate the individual
contributions from the superposition of harmonic functions. A
transformation of the single terms gives their dependency on frequency :
\begin{itemize}
\item  Q is proportional to $\cos \,4\, \theta +$const.
\item  U is proportional to $\sin 4\, \theta$
\item  V is proportional to $ - \sin 2\,\theta $.
\end{itemize}
Fig. \ref{stokesq} displays this for the difference, $I^+ - I^-$,
and separate input of only one polarization state, i.e. +Q, +U,
and +V\footnote{The image was created with a simulation of a dual
beam polarimeter. The retardance was set to $\delta = 150^\circ$, the
actual value of the retarder used at the ASP.}.
\begin{figure}
\psfrag{A}{\Large \shortstack{\vspace*{-.3cm} $\theta \;\; [$deg$]$}}
\psfrag{B}{\large \hspace*{-.5cm}$I^+ - I^-$}
\psfrag{C}{1\hspace{.5cm} 2\hspace{.5cm} 3\hspace{.5cm} 4\hspace{.5cm} 5\hspace{.5cm} 6\hspace{.5cm} 7\hspace{.5cm} 8\hspace{.5cm} 9\hspace{.34cm} 10\hspace{.4cm} 11\hspace{.3cm} 12\hspace{.34cm} 13\hspace{.3cm} 14\hspace{.34cm} 15\hspace{.3cm} 16}
\includegraphics[height=8cm,width=16cm]{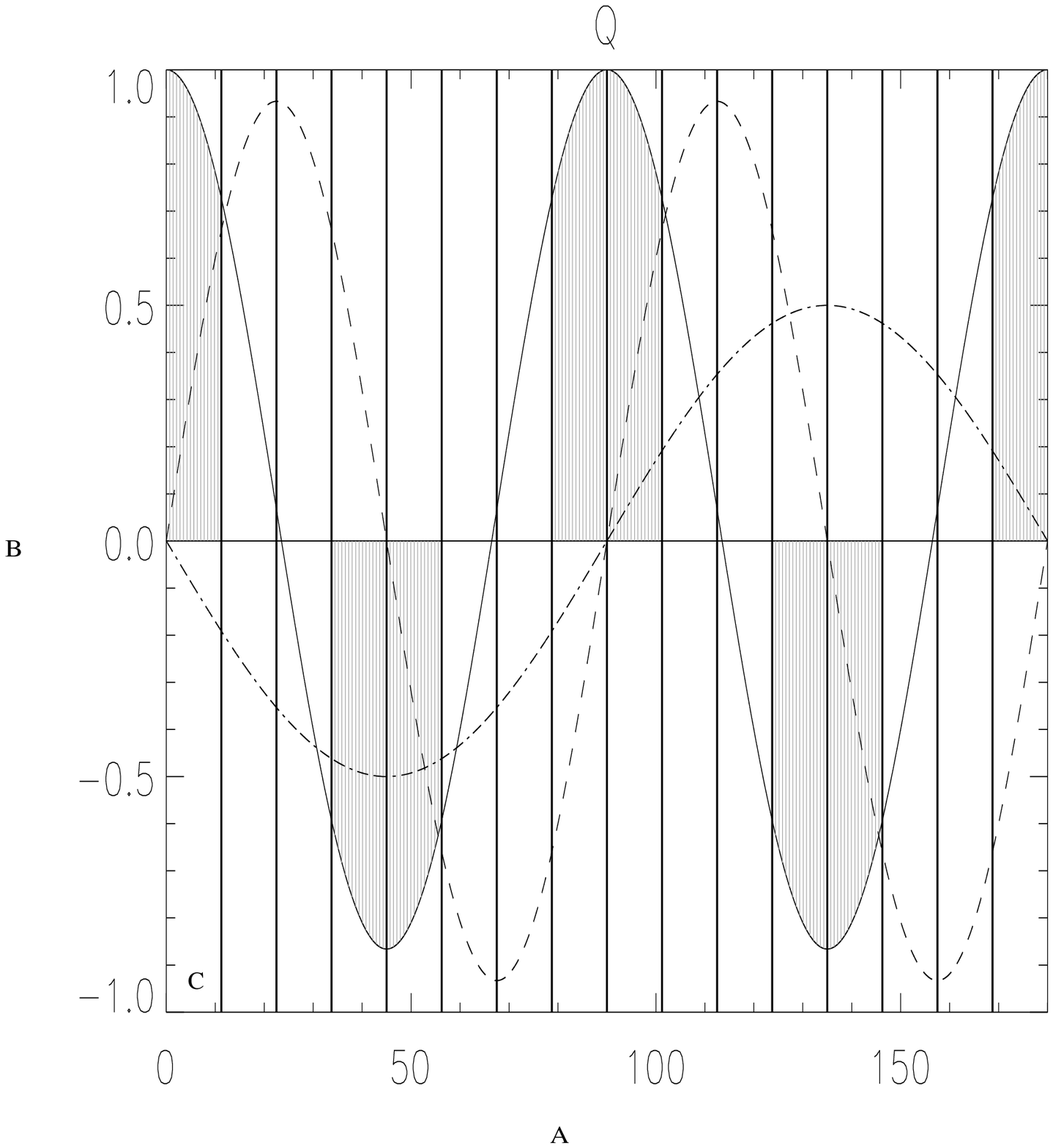}
\caption[Difference of beams $I^+ - I^-$ for pure input]{The difference of
the beams $I^+ - I^-$ for an input of pure Stokes Q (\it solid\rm), U
(\it dashed\rm), and V (\it dash-dotted\rm).\newline
The displayed course corresponds to the retarder position angle $\theta \in
[ 0, 180^\circ ] $. The vertical lines mark the integration steps. A slight
offset of the oscillation from zero can be seen in Q, the modulation
frequency is $4\; \theta$ for Q and U. Only the marked areas contribute to
the value of Stokes Q in the demodulation scheme (see the text for more
details).\label{stokesq}}
\end{figure}

There are some methods available to separate the individual
contributions. One possibility would be a Fourier analysis, which gives
a decomposition into single frequencies. This would need a continously
read out of the detector chips, which is impossible for CCD
detectors. The method of choice for POLIS uses a integration scheme,
which takes advantage of the specific shape of the modulation for each
polarization component.

Using Fig. \ref{stokesq} the procedure can quickly be seen. The
intensity curve for the position angle of the retarder from 0 to
180$^\circ$ is divided into 16 intervalls, over which the intensity is
integrated. The integration is done automatically by reading out the CCD
chips at the intervall limits. To obtain the value of a single polarization
component one has to add up the integration values by a demodulation
scheme with specific choosen signs.\\

For example the contribution from Stokes Q is calculated by taking the first
and last, the 8th and 9th integration value with a + sign, the 4th and
5th and also 12th and 13th value with a - sign. By comparison with the
curves for Stokes U and Stokes V it can be seen, that the integration
contributions from these polarization components nullify for that choice,
i.e. the value of the integration is proportional only to Stokes Q.

For each polarization a signed addition scheme can be found, in which
the other components cancel out. The total intensity I can be derived by
an addition of the beams, as can be seen from equation
(\ref{iplusminus}).

The order of demodulation and subtraction or addition of the beams is
not fixed. In the simulation used for the creation of Fig. \ref{stokesq}
the beams are first subtracted and then demodulated. Opposite to that
the data stored to file for POLIS or the ASP consists of the
demodulated beams $I^+$ and $I^-$, which have to be added or subtracted
afterwards. This allows the removal of individual pixel properties,  and
reduces the size of information to be saved.
\end{section}
\end{chapter}
\begin{chapter}{The vector polarimeter POLIS\label{chap4}}
\begin{minipage}{12cm}
''\it The instrument is presently being built at the KIS and the HAO
and will be completed by the end of the year 2000.\rm'' (Schmidt et al.,
undated poster)\\
\end{minipage}\\
$ $\\
POLIS stems from a cooperation between the KIS and the High Altitude
Observtory (HAO) in the USA. The design of POLIS drew strongly on the
heritage of the ASP\footnote{Operated by the HAO.}. Due to the delayed
setup of POLIS for the testing phase this thesis will have to use ASP data
to demonstrate the calibration procedure and expected results. It is
therefore necessary to introduce this instrument in more detail.
\begin{section}{The predecessor of POLIS: The Advanced Stokes Polarimeter\label{predecessor}}
The ASP at the Dunn VTT in New Mexico has been in operation for ten
years now. Its operation proved that it is possible to achieve a high
polarimetric accuracy, even at a telescope with considerable
instrumental polarization, by the usage of a telescope model.\\

The following description refers to Fig. \ref{asp_design}, displaying the
optical setup of the ASP with the inclusion of the telescope tower.

The Dunn VTT differs from the german VTT in Tenerife by two things.
The coelostat is mounted in an alt-azimuth configuration, leading to
fixed incidence angles of 45$^\circ$ on the coelostat mirrors. Further, all
mirrors before the polarimeter calibration unit are inside the vacuum
tank. This leads to a simplification in the telescope model, as the
mirror properties can be assumed to be identical.
\begin{figure}
\psfrag{A}{\huge \hspace*{-.1cm}$\downarrow$}
\psfrag{I}{\huge \hspace*{-.8cm} modulator}
\psfrag{B}{\huge \hspace*{-.8cm} $\leftarrow$ cal. polarizer}
\psfrag{C}{\huge \hspace*{-.8cm} $\leftarrow$ cal. retarder}
\psfrag{D}{\huge \hspace*{-.35cm} $\uparrow$}
\psfrag{J}{\huge \hspace*{-1.4cm} grating}
\psfrag{F}{\huge \hspace*{-.8cm} $\leftarrow$ beamsplitter}
\psfrag{E}{\huge \hspace*{-.4cm}$\leftarrow$ slit}
\psfrag{G}{\huge $I^+$}
\psfrag{H}{\huge $I^-$}
\includegraphics[height=14cm,width =18cm]{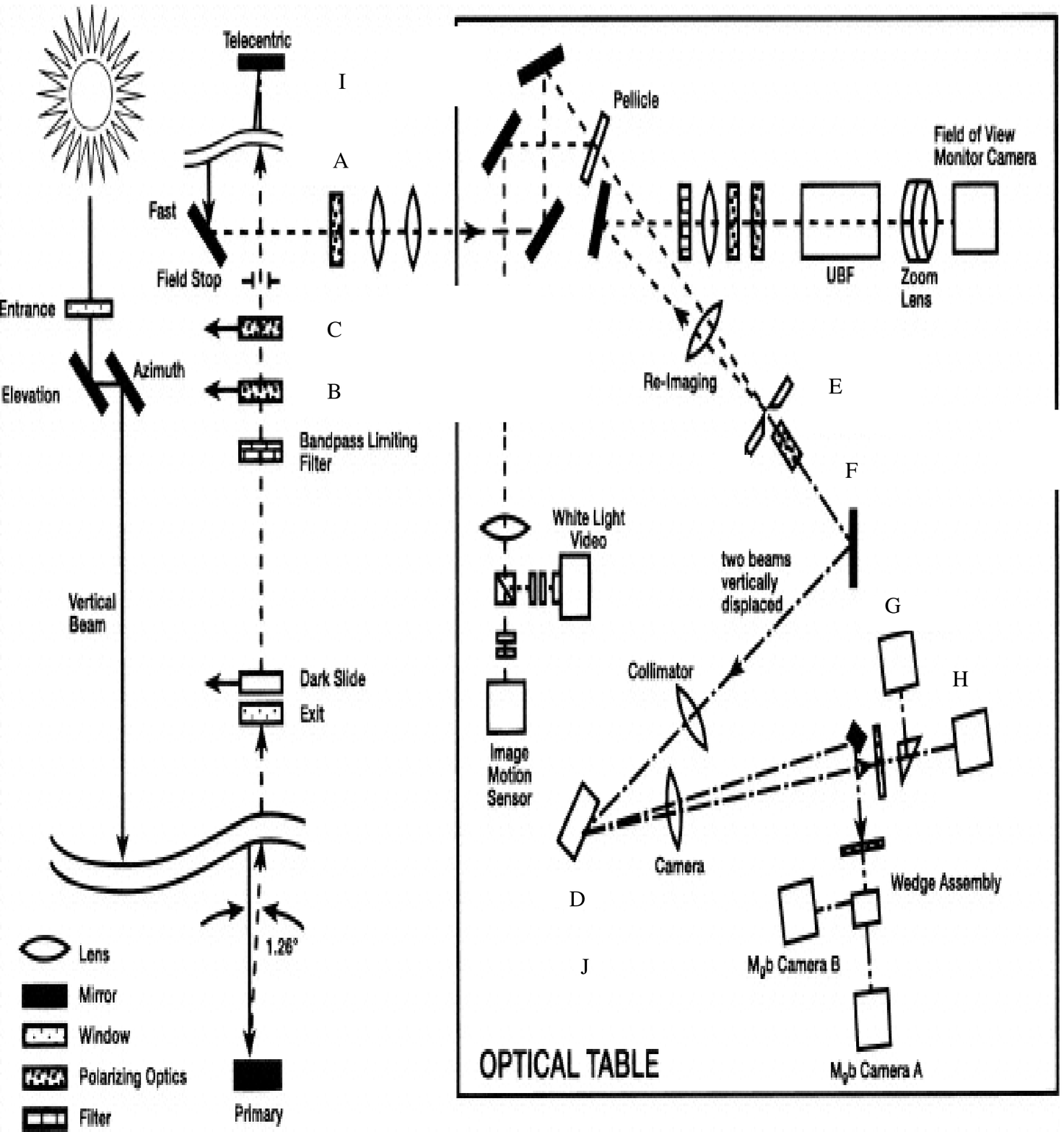}
\caption{Vertical section of the ASP including the light path in the
telescope. Original graph taken from \cite{lites1}, Fig. 2, slightly
modified. The more important optical devices have been given enlarged
labelling.\newline
The optical elements described in section \ref{dualbeamtheory} are the
modulator (=measurement retarder), the beamsplitter, and two CCD-cameras
denoted by $I^\pm$. The slit and the grating are necessary for spatially and
spectrally resolved data: the grating disperses the modulated signal,
while the slit restricts the area of the solar image under
observation. The slit, the following spectrograph, and the detectors are
on a separate optical bench (not drawn). To scan an region of interest
on the solar image it is necessary to move the whole bench by a motor
driven mandrel. The mirror labeled 'Fast', directly before the modulator,
compensates the image motion caused by the wedge on the modulator
(cp. section \ref{appwobble}). The calibration unit consisting of a
linear polarizer and a retarder is used for a calibration of all optics
between the cameras and its position. \label{asp_design}}
\end{figure}

The optical elements discussed in section \ref{dualbeamtheory}, the
modulator or measurement retarder, the polarizing beamsplitter, and the
intensity detectors are necessary for the measurement of the Stokes
parameters. Additionally now devices for spectrally and spatially
resolved information are present:
\begin{itemize}
\item The grating disperses the modulated polarization spectrally to obtain the
variation of the content with wavelength. In the images taken by the
detectors the dispersion direction corresponds to columns
(cp. Fig. \ref{gaintab_img}). The grating can be moved from the
wavelength range of observation to a spectral range without strong
absorption lines for flat field data (cp. section \ref{flatfield}). 

\item The slit restricts the area of the solar image under observation. The
height in the slight corresponds to rows in the images (also
cp. Fig. \ref{gaintab_img}). A scan of a region of interest on the solar
surface is performed by moving the complete spectrograph unit,
including slit and detectors. These elements are all mounted on a separate
optical bench, which can be shifted by a motor driven mandrel. 
\end{itemize}

The polarimeter calibration unit after the exit window of the vacuum
tank consists of a linear polarizer and a retarder. It is used for the
calibration of all optical elements afterwards, exactly like the one of
POLIS.

The polarization measurement of the ASP is performed in the same
wavelength range around 630 nm as for one of the spectral regions used at
POLIS. Therefore the properties of the data, the calibration procedure,
and to some extent the results of observations will be comparable.

The main disadvantage of the ASP is its non-permanent installation. The
setup on the optical table in the right part of Fig. \ref{asp_design}
has to be dismantled after an observation campaign. An improved version
of the ASP is planned to be installed permanently at the Dunn VTT.

For more technical details the reader should refer to Elmore et
al. (1992, \cite{lites2}), while the calibration procedure is discussed
in Skumanich et al. (1997, \cite{lites1}). Table \ref{asp_prop}
summarizes the instrumental characteristics of the ASP and the Dunn VTT.
\begin{table}
\begin{minipage}{8cm}
\begin{tabular}{| c | c |} \hline
\multicolumn{2}{|c|}{\rule[-2mm]{0mm}{6mm} ASP} \cr \hline
wavelength range & 630.1-630.3 nm \cr
spectral resolution & 225000 \cr
dispersion & 1.19 pm / detector pixel \cr
scan stepwidth & $\sim$ 0.4'' \cr
slit width & 0.6''\cr
integration time & 2.1 s \cr \hline
\multicolumn{2}{|c|}{\rule[-2mm]{0mm}{6mm} Dunn VTT} \cr \hline
coelostat mounting & alt-azimuthal \cr
main mirror diam. & 152 cm \cr
window diam. (entr./exit) & 76.2 / 20 cm \cr \hline
\end{tabular}
\end{minipage}\hfill
\begin{minipage}{7cm}
\caption{Instrumental characteristics of the ASP and the Dunn VTT. The
telescope uses a coelostat in alt-azimuth mounting inside the vacuum
tank. The wavelength range of observation is identical to one of the
spectral regions observed at POLIS. The corresponding values for POLIS
can be found in table \ref{polis_prop}. \label{asp_prop}}
\end{minipage}
\end{table}
\end{section}
\newpage
\begin{section}{The design of POLIS \label{polisdesign}}
POLIS was originally intended for a balloon experiment for the 1 m space
telescope Solar Lite. This design required to include an increased
number -in comparison with the ASP- of components on a L-shaped optical
bench of 180 x 50 x 50 cm.
\begin{subsection}{Optical layout}
\begin{figure}
\resizebox{16cm}{!}{\includegraphics{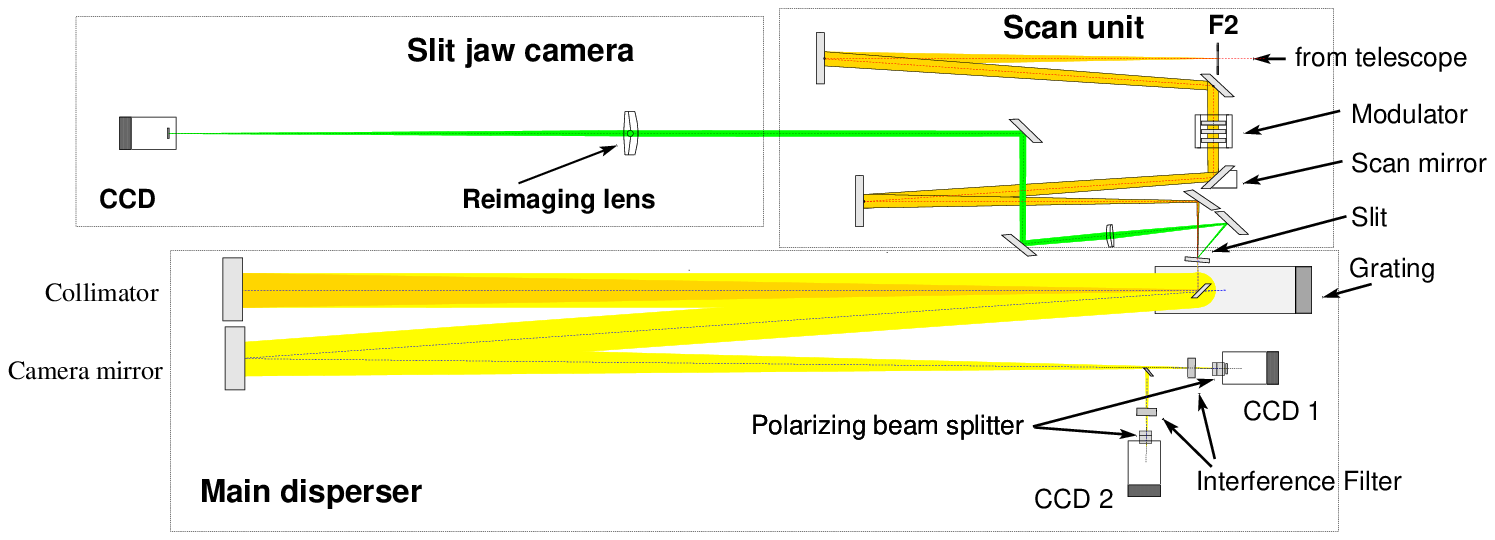}}
\caption{Different sections of POLIS. (\it top\rm) Slit jaw camera and
scanner unit, side view, (\it bottom\rm) main disperser, top
view. The images are combined to give the complete beam path.\newline
As for the ASP (cp. Fig. \ref{asp_design}) the modulator, beamsplitter
and detector are necessary for the measurement of the Stokes
parameters. The main differences to the ASP are:\newline
- the design of the scan mirror (see section \ref{appscanner})\newline
- the 'wobble compensation' included in the modulator device (see
section \ref{appwobble}), and most important,\newline
- the simultaneous observation in two spectral ranges.\newline
The beamsplitters are situated  after the grating directly in front of
the detectors CCD 1/2. Each of these detectors measures the two beams
$I^+$ and $I^-$ for a different wavelength range and such corresponds to
both ASP cameras. See the text for a discussion of all
elements present.\label{polis_section}}
\end{figure}

Fig. \ref{polis_section} displays the complete beam path in the
instrument. It can be separated in three mayor parts, the scan unit (\it
right top\rm) , the slit-jaw camera (\it left top\rm), and the main
disperser (\it bottom\rm). 
\begin{itemize}
\item[(1)] The scan unit contains:
\begin{itemize}
\item A mounting for targets at the focal plane F2 to establish image
scaling.
\item The modulator, consisting of a retarder and two additional glass
wedges. It modulates the polarization content of the incoming beam
according to the Mueller matrix in eq. (\ref{retrot}). The device is
described in detail in section \ref{appwobble}.
\item The scan mirror. This is a moveable mirror to select the image
area of observation in one spatial direction, which was necessary for
the balloon experiment. For the ground based version at the VTT it
allows to independently point the instrument without changing the
telescope orientation. Section \ref{appscanner} explains the design and
operation of this device.
\end{itemize}
\item[(2)] The slit-jaw camera: The back side of the slit is metallized,
to reflect the light outside the area covered by the slit. The reflected
light is focused on the slit-jaw camera, passing a filter (not
drawn). The images obtained in the wavelength selected display the 
orientation of the slit on the solar image. They are useful for the
alignment of POLIS data to simultaneous data from other instruments.
\item[(3)] The main disperser consists of:
\begin{itemize} 
\item The slit. It transmits the light of only a small region of the
solar image. Its metallized back side reflects light to the slit-jaw
camera. Across the slit horizontal hairlines can be placed to allow the
correct alignment of the images of the two observed wavelengths.
\item The grating. Its optical design is a compact combination of an
Echelle and reflective Littrow configuration. It disperses the modulated
polarization information spectrally. More details can be found
in Appendix \ref{grating}.
\item Interference filters. They prevent light from wavelengths and
spectral orders other then the observation ranges to reach the detectors.
\item The polarizing beam splitters. At POLIS they are placed directly
before the CCD-cameras. They transform the modulated, spectrally
dispersed polarization state into varying intensity on the single
detector pixels according to eq. (\ref{iplusminus}).
\item The CCD detectors. On CCD 1 the intensities of the two beams $I^+$
and $I^-$ for the visible wavelength range at 630 nm are measured on spatially
separated detector areas. CCD 2 is used to obtain the polarimetric
information of the Ca-line at 400 nm. More details can be found in table
\ref{instrutable}.
\end{itemize}
\end{itemize}
\begin{table}
\begin{minipage}{8cm}
\begin{tabular}{| c | c | c|} \hline
\multicolumn{3}{|c|}{\rule[-2mm]{0mm}{6mm} POLIS} \cr \hline
wavelength range & 630.1-630.3 nm & 396.7-396.9 nm\cr
spectral resolution & 145000 & 220000 \cr
dispersion & 0.96 pm / pixel & 1.5 pm / pixel \cr\hline
scan stepwidth & \multicolumn{2}{|c|}{\rule[-2mm]{0mm}{6mm} 0.1''} \cr
slit width & \multicolumn{2}{|c|}{\rule[-2mm]{0mm}{6mm} 0.1''} \cr
integration time & \multicolumn{2}{|c|}{\rule[-2mm]{0mm}{6mm} ??}
\cr \hline 
\multicolumn{3}{|c|}{\rule[-2mm]{0mm}{6mm} VTT Tenerife} \cr \hline
coelostat mounting & \multicolumn{2}{|c|}{\rule[-2mm]{0mm}{6mm} standard} \cr
main mirror diam. & \multicolumn{2}{|c|}{\rule[-2mm]{0mm}{6mm} 70 cm} \cr
window diam. (entr./exit) & \multicolumn{2}{|c|}{\rule[-2mm]{0mm}{6mm}
70 / 10 cm} \cr  \hline
\end{tabular}
\end{minipage}\hfill
\begin{minipage}{5cm}
\caption{Instrumental characteristics of POLIS and the VTT Tenerife. The
telescope uses a standard coelostat on top of the solar tower, outside
the vacuum tank. The wavelength range in the visible is identical to the
spectral region of the ASP. The corresponding values for the ASP 
can be found in table \ref{asp_prop}. \label{polis_prop}}
\end{minipage}
\end{table}

The performance tests of the scan mirror and the adjustment of the
'wobble compensation' included in the modulator unit were executed by
the author and shall be presented more elongated here. For other more technical
details refer to Appendix \ref{instruchar}.
\end{subsection}
\begin{subsection}{The modulator unit\label{appwobble}}
\begin{figure}
\begin{center}
\psfrag{A}{\large retarder}
\psfrag{B}{\large \hspace*{-.7cm}compensator}
\psfrag{C}{\large \hspace*{-.2cm}wedge}
\psfrag{D}{\large \hspace*{-.4cm}$+60^\circ$}
\psfrag{E}{\large \hspace*{-.8cm}$-60^\circ$}
\includegraphics[height=8cm,width = 18cm]{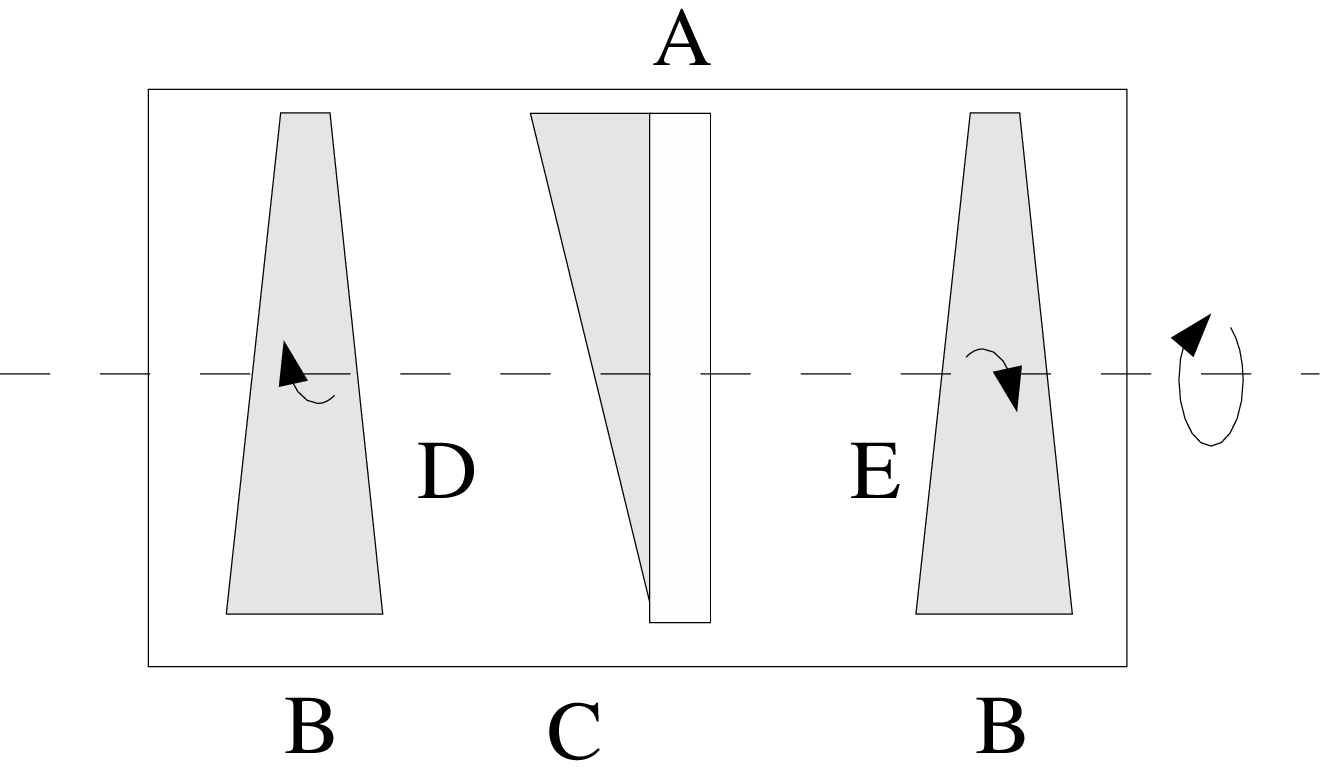}\\
\end{center}
\caption[Design of the wobble compensation]{Design of the 'wobble
compensation' in the modulator unit. The whole unit rotates around
the propagation direction.\newline
On one side of the measurement retarder a glass wedge has
been cemented. This defeats interferences effects due to the higly
polished parallel surfaces of the retarder. The introduced circular
image motion caused by the rotating wedge is compensated by two
additional wedges above and below. The compensator wedges have to be
twisted by $\pm 60^\circ$ relative to the wedge on the retarder to give
an effective wedge of exactly oppposite inclination. No parallel
surfaces are present in the final configuration.\label{wobbledesign}} 
\end{figure}
\begin{figure}
\begin{minipage}{8cm}
\psfrag{A}{\huge marker}
\psfrag{H}{\huge wedge}
\psfrag{B}{\huge 20$^\circ$}
\psfrag{C}{\huge 60$^\circ$}
\psfrag{D}{\huge guide slot}
\psfrag{0}{}
\resizebox{7.5cm}{!}{\includegraphics{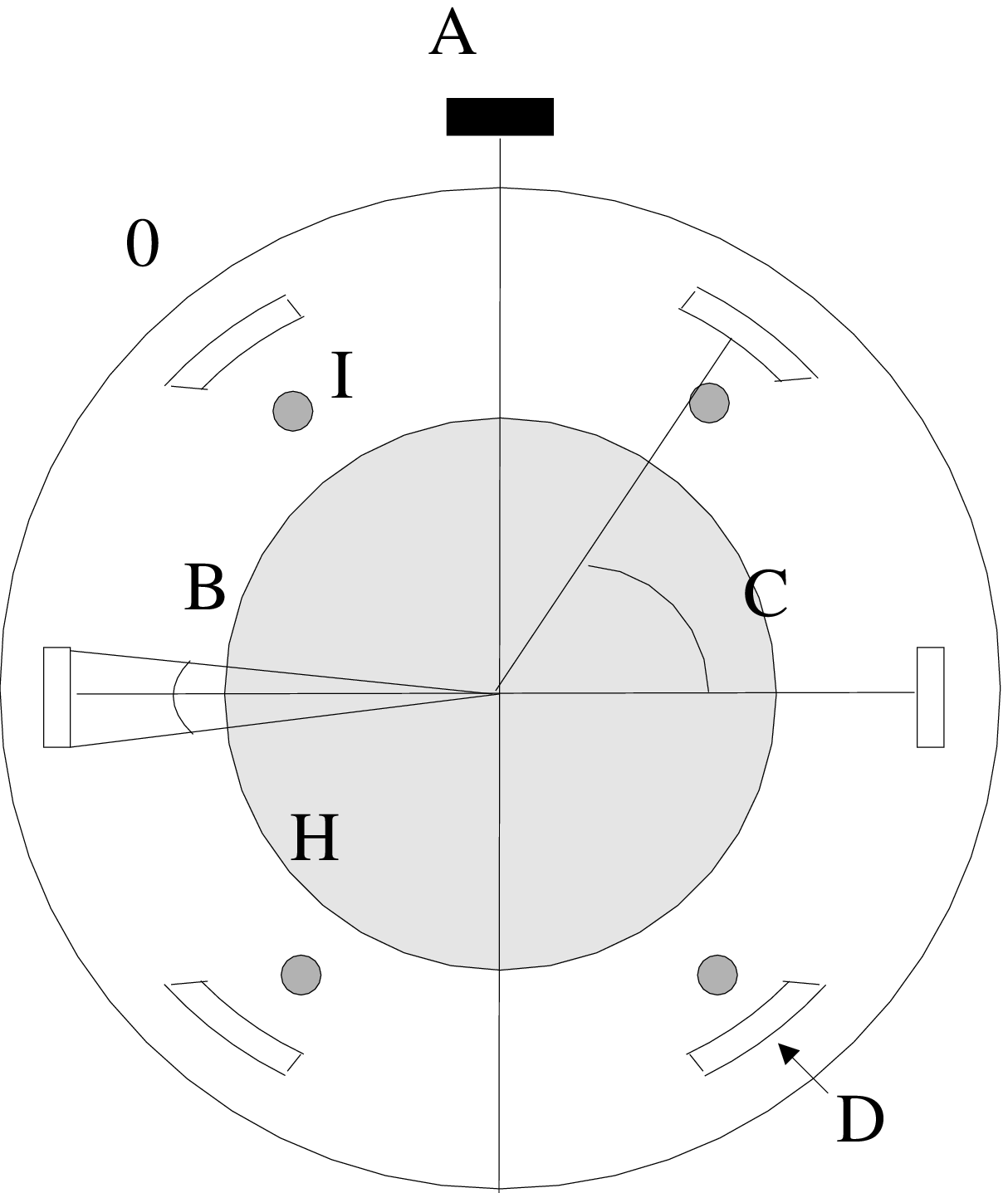}}
\end{minipage}
\begin{minipage}{8cm}
\caption[Vertical section of a wedge]{Vertical section of one
compensator wedge with guide slots. The marker was placed at an
arbitrary zero position.\newline
The screwholes are placed with $60^\circ$ angle separation. The
guide slots allow to additionallly change the angle by $\pm 10
^\circ$. The possibility to rotate the proper wedge in its own mounting
by $90^\circ$ was not needed, but extends the possible angle positions
to almost all values.\label{compmount}}
\end{minipage}
\end{figure}
The measurement retarder consists of a zero-order waveplate of
mica/quartz/marmor between two plates of BK7. It has two highly polished
parallel surfaces to ensure the same retardance on the full useable
area. These parallel surfaces act as a Fabry-Perot-Interferometer,
causing a wavelength dependent transmittance (cp. Fig. \ref{gaintab_img}
for the size of fringes in ASP data). To prevent this effect, a wedge of
4$^\circ$ inclination has been cemented onto one side of the
retarder. As the whole ensemble rotates, also the normal to the wedge
changes. This causes the image to move in a circle on the detector, the
so called 'beam wobble', because the incoming beam is bent into
different directions during one revolution.

To remove this effect at POLIS\footnote{At the ASP also a wedge is cemented
to the modulator. There a moveable mirror (the one labelled 'fast' in
Fig. \ref{asp_design}) is used to remove the image motion again,
requiring an active compensation.} a statically scheme was designed that
compensates the retarder wedge by two additional wedges. These are
mounted above and below the retarder, and are shaped to resemble a wedge
of opposite inclination, if they are twisted to $\pm 60^\circ$ relative
to to the retarder wedge (see Fig. \ref{wobbledesign}). No parallel
surfaces are present in the final configuration\footnote{In difference
to the use of only one compensator.}. The design is currently subject to
a patent application.

A vertical section of the compensator wedges is displayed in
Fig. \ref{compmount}. The angle position of the compensator wedges can
assume almost all values: the fastening screws are set in steps of
60$^\circ$, the guiding slots give another range of $\pm 10^\circ$, and
the wedge can be rotated inside its own mounting by $90^\circ$. It was
possible to reach a optimum compensation by using only the different
screw holes and guide slots.

The adjustment was performed in a reduced setup of POLIS, without
the cameras CCD 1/2. The displacements of a target, i.e. a glass plate
with a regular pattern of grooved lines, placed in F2 were
established with the slit-jaw camera mounted after the scan mirror.

To reach the compensation settings it was first necessary to establish
the inclination directions of each wedge, which were not marked. To
achieve this the compensator wedges were separately inserted in the
mounting above the retarder. The resulting radius of image motion for
the six main positions, i.e. increasing the position angle of the marker by
60$^\circ$ each time, leads to a minimal value, where the wedges are
antiparallel.

As the wedge angles are very small, their actual value may deviate from
the design specifications. This means that also the angle of 60$^\circ$,
for which the compensators are supposed to be twisted, is only
approximative. The fine tuning of the position angle of both wedges
was only practicable by trial and error. The method used was
iteratively fixing one of the compensator wedges, and adjusting the
second one for minimal radius of image motion.

The last result and some preceeding steps are presented in figure
\ref{nocircle}. For the final 'best settings' there is no systematic 
circular motion in the data. The remaining image motion of about 0.02''
is below a) the resolution limit of the telescope, and b) the spectral
and spatial size of the detector pixels.
\begin{figure}
\psfrag{A}{\large \hspace*{-1.3cm} \shortstack[c]{$\Delta y$\\$[arcsec]$}}
\psfrag{B}{\large \hspace*{-1cm}$\Delta x\; [arcsec]$}
\includegraphics[height=8cm,width=8cm]{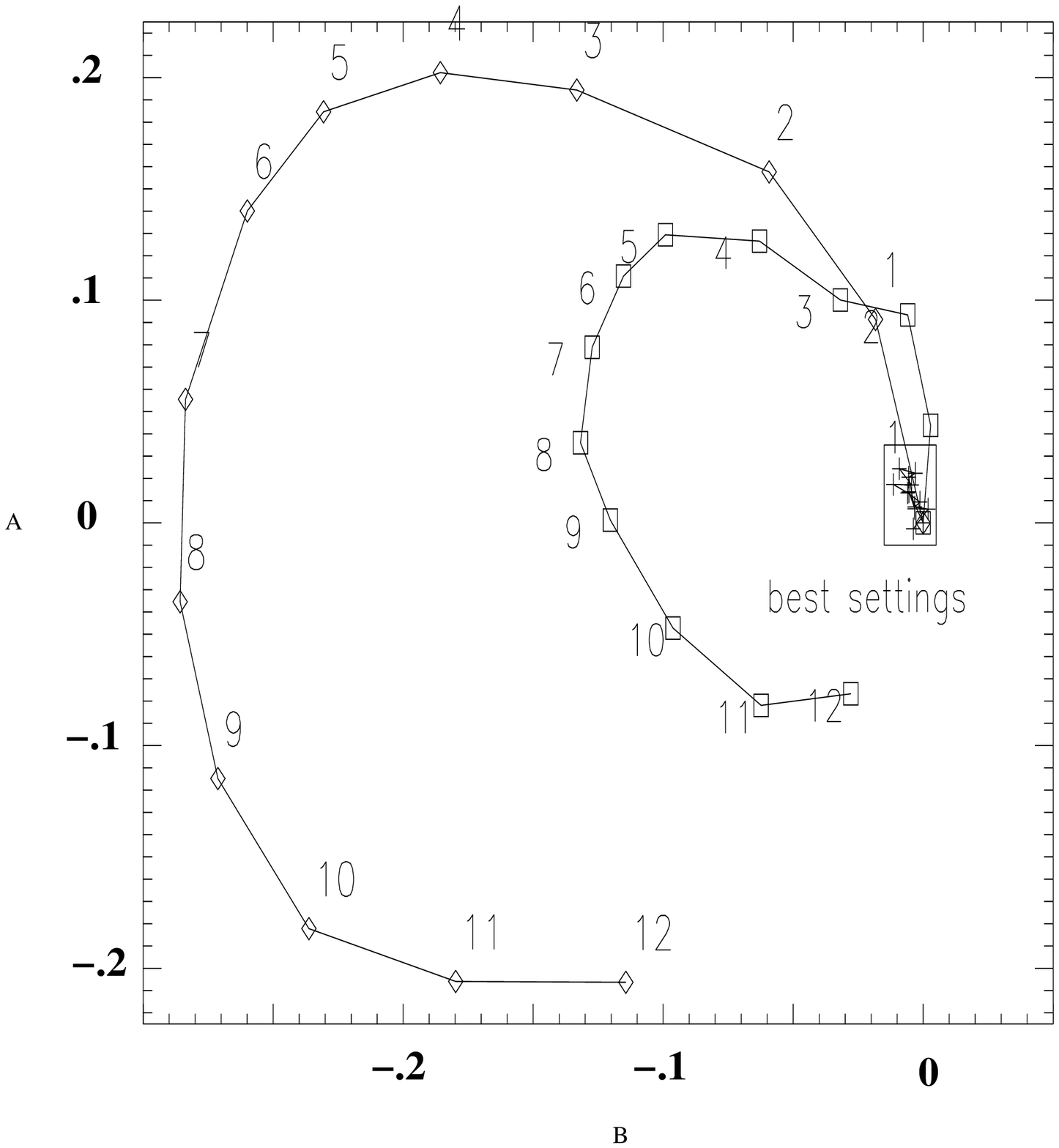}
\hfill
\includegraphics[height=8cm,width=8cm]{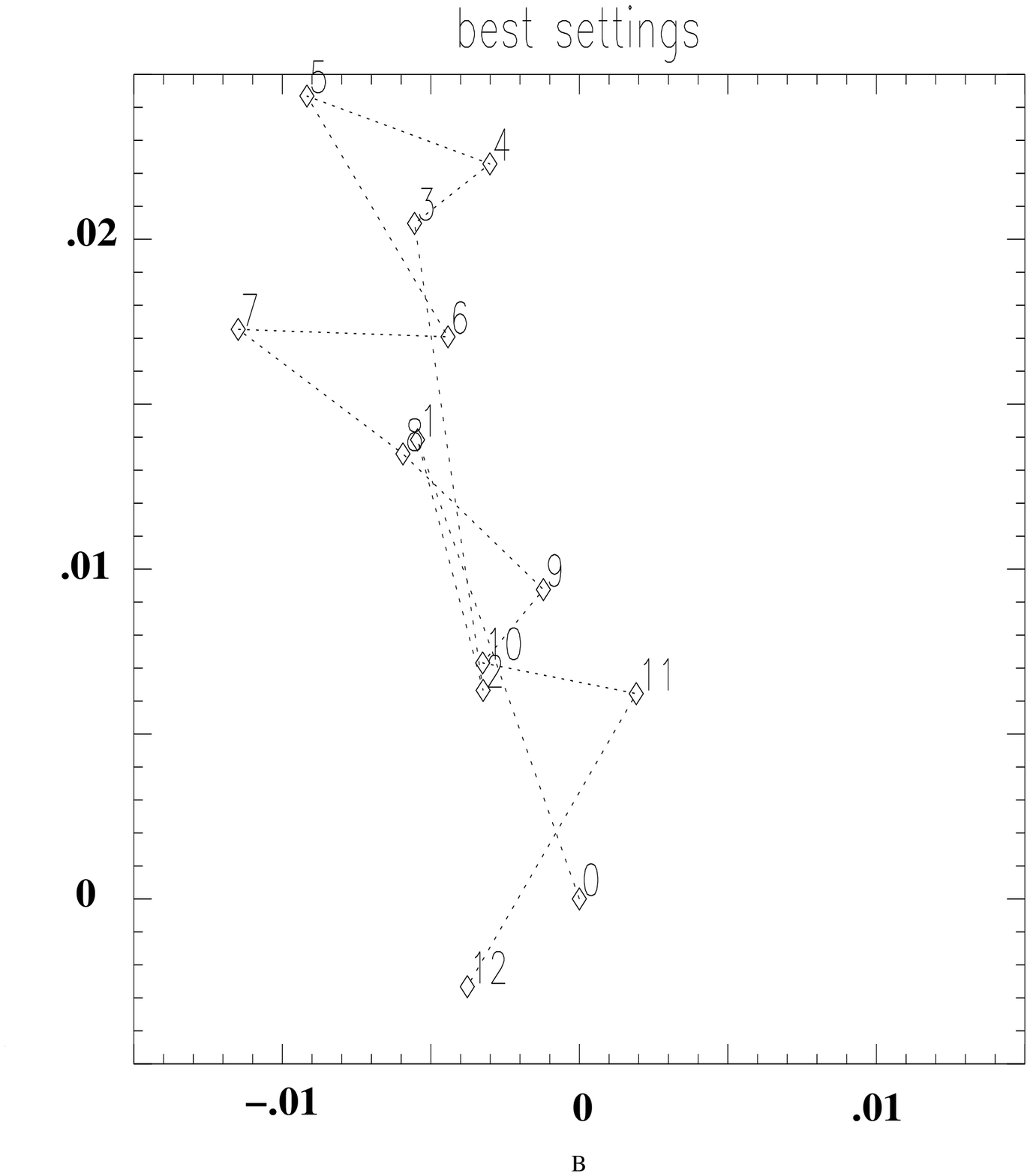}
\caption[Image movement with partial and full compensation]{The
displacement of the image N at the modulator position angle
N$\cdot 24^\circ$ relative to the first reference image at (0,0) in
arcseconds. (\it left\rm) The improvement through the iterative
adjustment of the compensator wedges led to (\it right\rm) the result of
the best setting. Not all intermediate steps are available.\newline
The resulting motion at the best setting is more like a random
walk, caused by the instability of the measurement setup, than a circle
due to a residual 'beam wobbling'. Note that the measurement setup is
estimated to be exact only to about 0.007''.\label{nocircle}}
\end{figure}
\end{subsection}
\begin{subsection}{The scan mirror\label{appscanner}}
The scan mirror is intented for positioning the slit on the desired
observation area on the solar image along one direction. A graphical
user interface\footnote{Written by Th. Kentischer.} gives access to the
different operation modes. The properties of the scan mirror were
established in the same reduced setup as in the preceeding section.
 
Figure \ref{scanschema} gives a side-view of the scan mirror. The proper
mirror sits at one end of a lever arm, which can be rotated around the
axis drawn. The other end of the lever is moved by a mandrel, whose
height can be changed with a servo-controlled DC-motor. The motor
position is given in single steps through a contactor. Due to the
magnification scale at the VTT 8 motor steps correspond to one scan step
of about 0.1'' on the sky.
\begin{figure}
\begin{minipage}{8cm}
\psfrag{A}{\huge Mirror}
\psfrag{B}{\huge Motor}
\psfrag{C}{\huge Mandrel}
\psfrag{D}{\huge Axis}
\psfrag{E}{\huge Lever arm}
\resizebox{8cm}{!}{\includegraphics{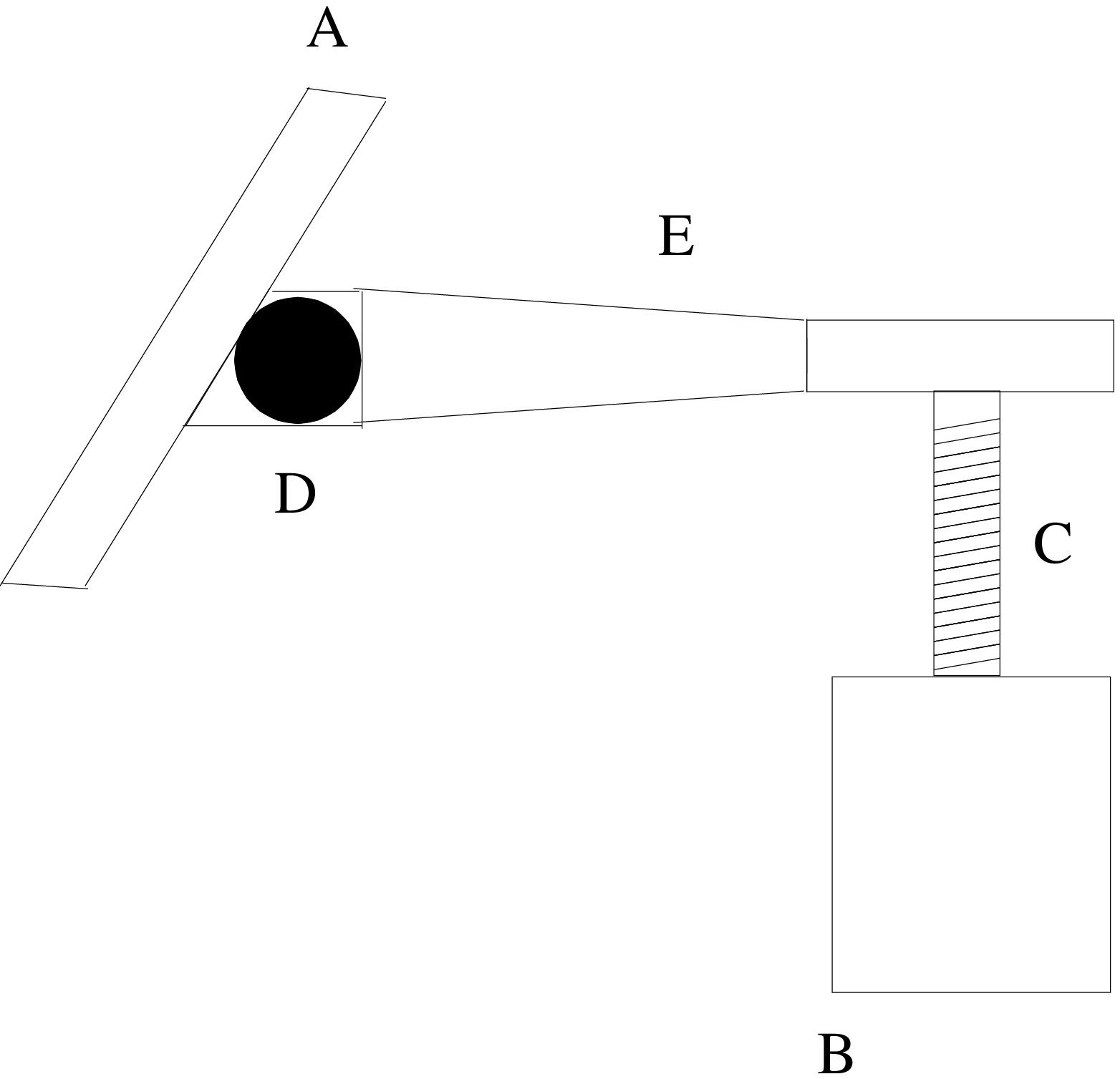}}
\end{minipage}\hfill
\begin{minipage}{7cm}
\caption[Schematic drawing of the scan mirror]{Schematic drawing of
the scan mirror, side view.\newline
The servo-controlled DC motor changes the height of the mandrel. This
tilts the mirror on the other end of the lever arm around the axis
drawn. The slit can be moved in one direction on the the solar image. In the
performance tests the actual mirror position was checked for various motor
positions.\label{scanschema}}
\end{minipage}
\end{figure}

The scan mirror device was checked for three design requirements:
\begin{itemize}
\item[(1)] variation of stepwidth for a single scan step of about 10 $\%$
\item[(2)] variation of the length of a scan of 100 steps of less then
1 step (=1 $\%$)
\item[(3)] 'homing' precision, i.e. the ability to hit a given start position
on repetition, of less then one pixel deviation in the data images.
\end{itemize}
The accuracy of the data acquisition was established through images of
the same constellation with no movement of the scan mirror. The routines
for the determination of the displacements were checked against an image
with a known displacement from a shifting routine. The total error
amounted to a value of app. 0.007''. The main contribution came from the
measurement setup, which was 'unstable'. The whole optical bench was
placed on a trolley and the camera itself was not fixed in the only
temporarily position.

15 area scans (in the following termed 'runs') of each 200 scan steps
with a length of 0.1'' over the same region were executed. The settings can be
found in table \ref{runtable}. This should mimic the later use of POLIS
with a repeated scan to obtain a timeseries. For coarser maps the
stepwidth can be increased to 0.2'' (=16 motor steps) or more.

The data was evaluated in two ways concerning point (1), the variation of
stepwidth.
Statistically, by the calculation of average value and standard
deviation of the 3000 measurement values, which resulted from the comparison of
subsequent images. At this straightforward method only 'trigger errors'
had to be additionally considered. The camera trigger to take the image
was sometimes sent out delayed or in advance, while the scan step was
not completed or had not even started. These were easy to identify in
the data and could be removed. The relative standard deviation of the
length of a step established was $\sigma_{8\; steps} \simeq 14.3 \%$.
\begin{table}
\begin{minipage}{8cm}
\begin{tabular}{|l|l|l|l|}\hline
stepwidth & start pos. & end pos. & number\cr\hline
8 steps & 8500 & 10100 & 200 \cr
0.1'' & 0'' & 20'' & \cr \hline
\end{tabular}
\end{minipage}\hfill
\begin{minipage}{7cm}
\caption[The settings of the 15 scanner test runs]{The settings of the
15 test runs of the scan mirror. The first line is in 'motor coordinates', the
second gives the corresponding values in arcseconds on the
sky. \label{runtable}}
\end{minipage}
\end{table}

The second method took advantage of the averaging effect in the
comparison of images with N scan steps inbetween.
The resulting variance obeys to the equation
\begin{eqnarray}
\sigma_{tot}^2 = \sigma_{offset=const.}^2 + \left( \frac{1}{\sqrt{N}} \cdot
\sigma_{8\; steps} \right)^2 \label{equat} \;.
\end{eqnarray}
The first part contains all contributions not due to the scan mirror,
the second is the variance of an average of N scan steps.

Values for two parameters $\sigma_{offset}$ and $\sigma_{8\; steps}$
were established by a least-square fit of eq. (\ref{equat}) to values of
the variance calculated from the data set with N=1,2,3,...,50. The
result was slightly greater than the statistical with $\sigma_{scanner}
\simeq 16 \%$ and $\sigma_{offset} \simeq 6 \%$. Summarizing the
consistent results the value of about 15 $\%$ variation is acceptable,
even if above the design reqirement.\\

Point (2), the full length of a run, showed only very little variations. The
fluctuations of scan stepwidth average out quickly. The constant
offset from above could partly be explained by a systematic
dependence of stepwidth on the motor position. This influences all
runs over the same range in the same way.

On average over the 15 runs a relative standard deviation of 0.16 $\%$ 
remained, which is well below the required 1 $\%$.\\

Point (3), the homing precision, was controlled through a statistical
evaluation of displacements between images of different runs at an
identical motor position. The mean value of about 0.5 pixel displacement
should permit an easy alignment of data of repeated scans through
sub-pixel shifts.
\end{subsection}
\begin{subsection}{Observed spectral lines}
POLIS is designed for the simultaneous observation of the polarization
state in two spectral ranges. The wavelength region from
630.082 nm to 630.318 nm is imaged on CCD 1. This includes four
prominent absorption lines. Two of them are telluric O$_2$-lines formed
in the atmosphere of the earth at their rest wavelength. The solar
photospheric absorption lines of neutral iron around 630 nm are the main
observation target. Table \ref{lines} gives the properties of the single lines,
their appearance in a spectrum can be seen in Fig. \ref{stokesprof}.

CCD 2 records the line core of a CaII line at 396.6
nm. This is a very broad chromospheric absorption line with
additional blends of photospheric iron lines. The left panel of
Fig. \ref{caline} shows a image of the spectral range with an
identification of different FeI lines. 
\begin{figure}[ht]
\begin{minipage}{8cm}
\resizebox{8cm}{!}{\includegraphics{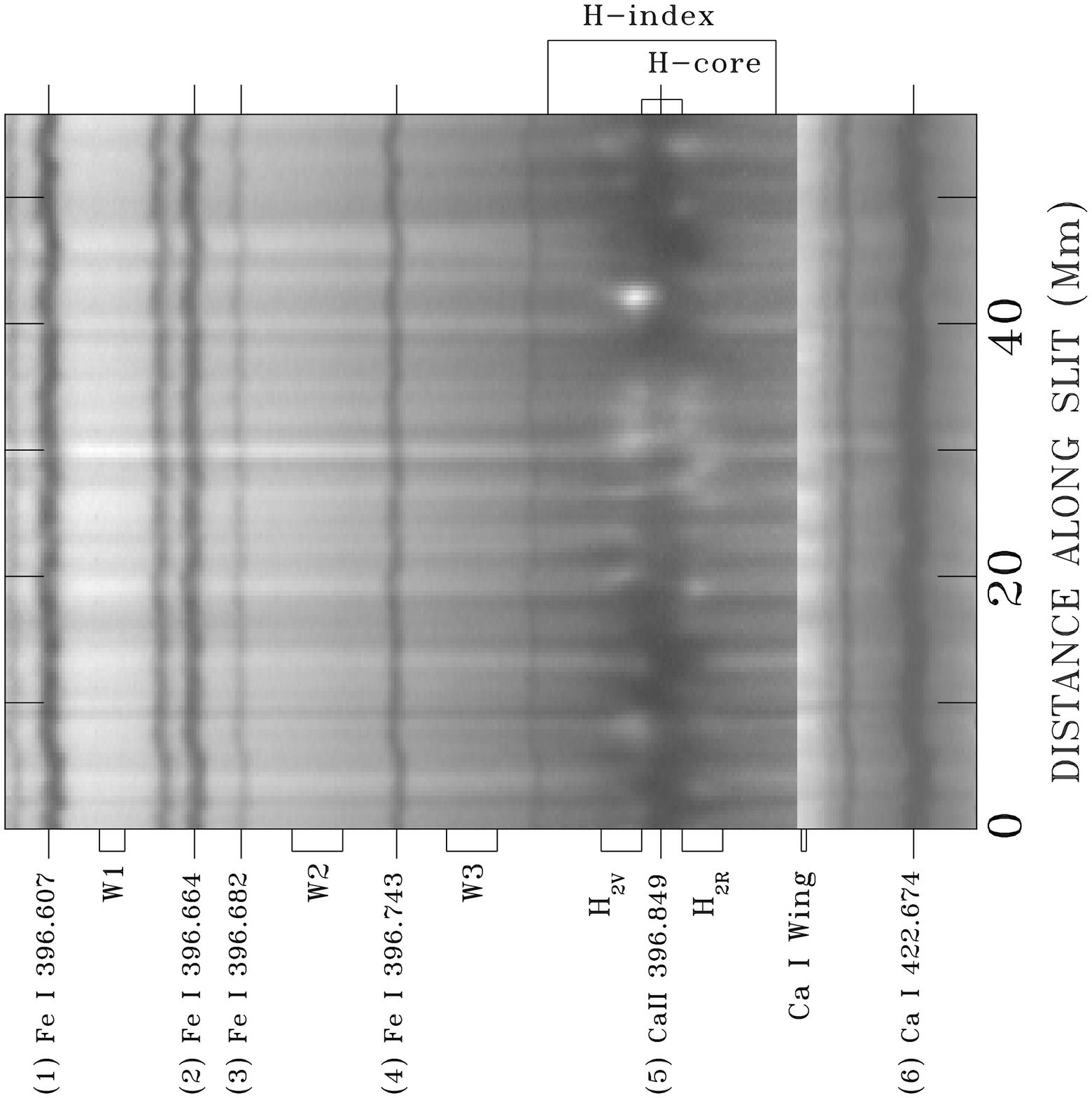}}
\end{minipage}\hfill
\begin{minipage}{8cm}
\resizebox{8cm}{!}{\includegraphics{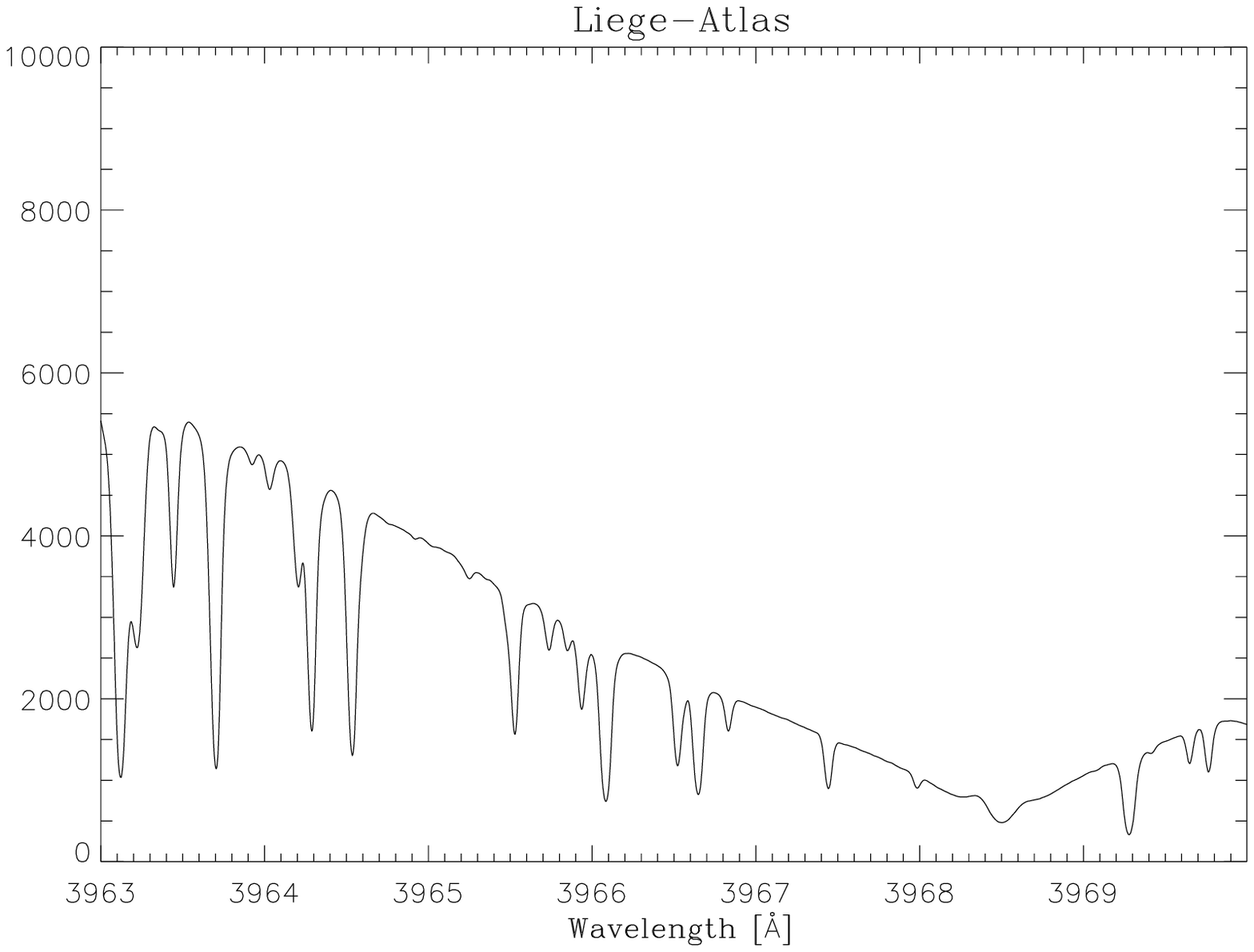}}
\end{minipage}
\caption{(\it left\rm) Spectrum of the wavelength range at 396 nm
(B.W. Lites, HAO). (\it right \rm) Corresponding section from the Liege
spectral atlas.\newline
The broad chromospheric Ca-line has additional blends of photospheric FeI
lines in its wing. The evaluation of the polarization signal is more
complicated then for a photospheric line. The contribution to a special 
wavelength stems from an extended height region in the solar
atmosphere. The shape of the spectrum will require more sophisticated
methods of data treatment, as no regular continuum or telluric reference
lines are available.\label{caline}}
\end{figure}
\begin{table}[hb]
\begin{minipage}{8cm}
\begin{tabular}{|c|c|c|c|c|c|} \hline
element & $\lambda$ & transition & $g_j$ & $\Delta h^c$ & $h_{max}^c$  \cr
 & $[$nm$]$ & $^{2S+1}L_{J}$ & & $[$km$]$ & $[$km$]$   \cr \hline
\multicolumn{6}{|c|}{\rule[-2mm]{0mm}{6mm} solar absorption lines, CCD 2} \cr \hline
CaII H & 396.649 & ? & ? & $>$ 500 & $>500$  \cr
FeI & 396.745 & ? & ? & ? & ?   \cr
FeI & 396.682 & ? & ? & ? & ?   \cr
FeI & 396.664 & ? & ? & ? & ?   \cr\hline
\multicolumn{6}{|c|}{\rule[-2mm]{0mm}{6mm} solar absorption lines, CCD 1} \cr \hline
FeI & 630.15091$^a$ & $^{5}P_{2} - ^5 D_{2} $& 1.67$^b$ & 300-450 & 390   \cr
FeI & 630.25017$^a$ & $^{5}P_{1} - ^5 D_{0} $& 2.50 & 260-410 & 310   \cr \hline
\multicolumn{6}{|c|}{\rule[-2mm]{0mm}{6mm} telluric absorption lines,
CCD 1}\cr \hline
O$_2$ & 630.20005$^a$  & - & - & earth atm. & -   \cr
O$_2$ & 630.27629$^a$  & - & - & earth atm. & -   \cr \hline
\end{tabular}
\end{minipage}\hfill
\begin{minipage}{5cm}
\caption{Summary of the properties of the observed spectral lines. \bf
a \rm wavelengths from Pierce $\&$ Breckinridge (1973), \bf b \rm
effective Land\'e factor from Solanki (1987), \bf c \rm height of line
core formation and maximum of contribution function above
$\tau_{\rm 500nm}=1$ (H.Schleicher, personal note). \label{lines}}
\end{minipage}
\end{table}
\end{subsection}
\end{section}
\begin{section}{Summary: Characteristics and intended usage of POLIS}
POLIS will yield the complete information of the polarization state of
light in two wavelength ranges simultaneously. The wavelength ranges
include spectral lines formed in different heights in the solar
atmosphere. The spatial accuracy of the instrument is of order of 0.1
arcseconds on the sky, to resolve structures of a few 100 km
diameter on the solar surface. The spectral resolution allows an
inversion of the polarimetric data to retrieve the vector magnetic
field. The intended time resolution for a scan step will be of order
of seconds. The calibration of the proper polarimeter will be accurate to
at least $0.1\; \% $ of the continuum intensity.\\

The following points outline some possibilities of the usage of POLIS:
\begin{itemize}
\item The construction of a consistent static model of magnetic field
lines from the photosphere to the corona. The strict simultaneity of the
polarimetric data is crucial for that.
\item Time series of a selected region. The additional information of
the chromospheric line allows the detection of vertical motions up- or
downwards in the atmosphere. This can help to solve the problem of the
heating of the upper solar atmosphere layers by propagating shocks or
acoustical waves.
\item Total magnetic flux difference. A comparison between the
total magnetic flux measured at the two heights can be used to establish
the amount of flux returning to the surface.
\item Co-observations with the TIP. As another polarimeter in the
infrared range is available at the VTT Tenerife, simultaneous
observations in a third different spectral line will be possible in the
future.
\item Studies of effects on small\footnote{'Small' in solar terms,
i.e. of order of some 100 km.} scales. The spatial accuracy will allow
to resolve many solar magnetic structures. For example, the substructure of the
penumbra of a sun spot, or the spatial dependence of the Evershed effect
on the magnetic field configuration can be examined in detail.
\item Relocation to the planned 1.5m-Gregory-Coud\'e-telescope. The
compact setup and the spatial accuracy allow an installation at the
greater telescope under construction.
\end{itemize}
To repeat the main problem of the examination of solar magnetic
phenomena with a polarimeter: the interpretation of the data rests on the
unspoken assumption that the polarization signal only stems from the
sun. As the following sections will display, POLIS as a polarimeter at
the VTT Tenerife will declare a polarization state to be Stokes U, when
it actually is Stokes Q, and initially has been Stokes V - before it
entered the telescope.
\end{section}
\end{chapter}
\begin{chapter}{The calibration of POLIS\label{polcal}}
\begin{minipage}{12cm}
''\it The beam of sunlight undergoes two reflections on the ... surfaces
of the coelostat ..., where elliptical polarization must again be
introduced.\rm'' G.E.Hale (1908)\\
\end{minipage}\\
$ $\\
The actual polarimeter output is influenced by a great variety of
factors. To give only some examples:
\begin{itemize}
\item The usage of the integration scheme requires an exact
synchronisation of read out timing and retarder position angle
\item The modulation efficiency of the retarder is not equal
for the different polarization components, and varies with wavelength.
\item The pixels of the two-dimensional detectors are not identical, so
their individual response has to be known.
\item Intensity variations due to the seeing at the observation site can
lead to spurious polarization signals because of the subtraction of
intensities\footnote{See B.W. Lites, \cite{lites3}.}.
\end{itemize}
The effects concerning the performance of the \it polarimeter \rm are taken
into account by so called \bf X\rm-matrix or polarimeter response
function. This $4\times 4$-matrix relates the output Stokes vector of
the polarimeter to the input vector at the position of the polarimeter
calibration unit.\\

The other main influence arises from the \it telescope \rm before the
proper polarimeter. Reflections under oblique angles change the incoming
polarization, which was already taken into consideration for the first
polarimetric solar measurements.
To quantify these effects a model of the polarization properties of the
telescope has to be used. Its result is the so called telescope or \bf
T\rm-matrix of all optical elements down to the polarimeter calibration unit.\\

The two matrices are a Mueller matrix of an optical train (\bf T\rm)
or an equivalent to describe the behaviour of the polarimeter
(\bf X\rm). They are needed, as the final polarimeter output is always
given by
\begin{eqnarray}
\fbox{$
\bf S\rm_{out} = \bf X\rm \cdot \bf T\rm \cdot \bf S\rm_{in,sun}$}\; ,
\end{eqnarray}
where one has to retrieve the Stokes vector of the incident light from
the output value.\\

The determination of these matrices is the crucial point for the
polarimetric accuracy, when the optical design of the polarimeter
components is decided.
\begin{section}{Polarimeter calibration\label{secpolcal}}
The calibration of the polarimeter is performed through the evaluation of the
polarimeter calibration data set. The 16 entries of the \bf X\rm-matrix
calculated in that way give the response of the measurement instrument
to the different polarization states. The \bf X\rm-matrix is assumed to be
constant for hours, it will be established on a daily\footnote{The
eventual need for an increased number of calibrations will have to be
tested at Tenerrife.} base during observation campaigns.
\begin{subsection}{Polarimeter calibration data}
The polarimeter calibration data set (in the following referred to as
'cal') has to be taken at the start and/or the end of the
measurements. The polarimeter calibration unit is used for the creation
of known input polarization states. This unit is located after the
deflection mirror inside the vacuum tank (see Fig. \ref{vactank}) and
consists of a linear polarizer and a retarder. They are placed inside
rotateable mounts and can be inserted into the light beam by remote
control\footnote{It is not possible to use them separately, always the
combination of first polarizer and then retarder is in the beam
path.}. The directions of the transmission axis of the polarizer and the
fast optical axis of the retarder have to be established before. The
accuracy of the position affects the calculated polarimeter response
function. It should be sufficient to fix the positions to $\pm 0.1^\circ$
(M.Collados, personal note). Appendix \ref{aligncalib} describes the
method to align the unit correctly. The optical parameters of the two
elements can be found in table \ref{instrutable}.\\ 

With the directions of the axes known the following equation between
created input and polarimeter output is valid:
\begin{eqnarray}
\bf S\rm_{out} = \bf X \cdot M\rm_{ret}(\theta_{ret})\bf \cdot
M\rm_{pol}(\theta_{pol})\bf \cdot T \rm \cdot 
\begin{pmatrix}\rm I_0 \cr 0 \cr 0 \cr 0 \cr \end{pmatrix} = \rm I\rm_0 \cdot \bf
X \cdot M\rm_{ret}(\theta_{ret})\bf \cdot M\rm_{pol}(\theta_{pol}) \cdot\rm
\begin{pmatrix} 1 \cr\rm T_{1,0} \cr\rm T_{2,0} \cr
\rm T_{3,0}  \cr \end{pmatrix}\; .\label{calibeq}
\end{eqnarray}
$\theta_{ret}$ and $\theta_{pol}$ are the angles to the zero positions
of the respective axes, and $\bf M\rm_i (\theta_i)$ the corresponding
Mueller matrices of rotated elements. The telescope matrix \bf T \rm  is
intensity normalized, i.e. T$_{0,0} \equiv 1$, and the initial sun light
is supposed to be unpolarized\footnote{The area on the sun during this
observation should be at disc center with no visible magnetic activity
like sun spots. Additionally the telescope pointing should be varied
randomly, to remove all spatial information.} with the intensity
$\rm I_0$.\\
\setcounter{footnote_temp}{\value{footnote}}
To obtain the 16 elements of the matrix the response to four
independent\footnote{'independent' means in that case, the 4 vectors
written in a matrix A have to give det(A)$\neq 0$, see \cite{lites1},
p. 362.} Stokes vectors has to be measured. The problem in choosing the
vectors is increased by the unknown values in eq. (\ref{calibeq}). These
are the \bf X\rm-matrix itself, the intensity and the telescsope matrix
entries. An elegant way to reduce the number of unknowns, and at the same time
decouple polarimeter and telescope, is the calibration data set used for
the TIP polarimeter, which will be adopted for POLIS. If the linear
polarizer is held on a fixed position, for example at $\theta_{pol}
\equiv 0^\circ$, eq. (\ref{calibeq}) transforms to
\begin{eqnarray}
\bf S\rm_{out} = \rm I\rm_0 \cdot \frac{1}{2}\;(1+T_{1,0})\cdot \bf X \cdot R
\rm(\theta_{ret})\bf\cdot \begin{pmatrix} 1 \cr 1 \cr 0 \cr 0  \cr
\end{pmatrix} \; \label{calibeq1} .
\end{eqnarray}
The intensity term $\rm I\rm_0 \cdot \frac{1}{2}\;(1+T_{1,0})$ can
assumed to be constant, if the variation of $\rm T_{1,0}$ during the
measurement of the cal data is neglegible\footnote{If seen to be
needed, a linear variation of intensity will be included.}. The
cal data set will therefore consist of a full revolution of the
retarder in 72 steps of $5^\circ$ for a fixed, but in principle arbitrary,
position of the polarizer. The evaluation of the data set is described
in section \ref{polresponse}.
\end{subsection}
\begin{subsection}{Flatfield $\&$ dark current data \label{flatfield}}
Before an evaluation of the polarimeter response the properties of the
individual detector pixels have to be removed. To this extent an
additional data set of flatfield images (in the following referred to as
'flat') has to be available$^{\arabic{footnote_temp}}$. Only the
intensity values of Stokes I are used. The flat data set will contain about
15 images.\\

The dark current images reflect the number of counts produced by stray
light, the electron noise, or read-out effects, when the light path to the
sun is blocked. The values have to be compared with the number of counts
in an actual observation to obtain the signal-to-noise ratio (S/N). At
the beginning of each file a small number of dark images will be taken to
ensure a close relationship between the actual and assumed noise level.
\end{subsection}
\begin{subsection}{Data reduction}
The following sections describe the procedure for one detector, for
example CCD 1. The data from the second camera will have to be treated
analogously, but also separately, in the same way.

The dual beam setup of the polarimeter results in the two beams $I^+$
and $I^-$ (see section \ref{dualbeamtheory}). They are measured on
different detector areas, so before the subtraction or addition of the
beams the data has to be corrected for the pixel properties. This
concerns all measurement data taken with POLIS, either cal or
actual solar measurements. All following images were created from ASP
data sets due to the lack of POLIS data. For the wavelength range around
630 nm this data is very similar, the wavelength range around 400 nm
will require slightly different methods.
\begin{subsubsection}{Gaintables}
The gaintables take into account the individual pixel responses. They
are constructed from the averaged flat field data in Stokes
I\footnote{Remark: the stored data is the already demodulated Stokes
vector, conisting of (I,Q,U,V) ($\lambda$).} after the subtraction of
the dark current. The ASP has two detectors with two gaintables, POLIS
will have two set of detectors with four different image areas,
i.e. four gaintables.

To remove the spectral information from the data the
line cores of one of the telluric\footnote{For the 630 nm wavelength
range, see the discussion for the co-alignment procedure below for other
possibilites.} lines along the slit are shifted to a fixed reference
position. The average of the shifted image along the spatial direction
gives the mean profile. The curvature (or linear variation) of the line
core along the slit is established. The gaintable results from the
division of the averaged flat by the mean profile, which is shifted in
$\lambda$ according to the curvature. Fig. \ref{gaintab_img} shows a gaintable
constructed for one of the ASP cameras with this procedure.

At first look the procedure may appear to be not satisfying. This is
mainly caused by the small number of flat field images (only 4) in an
ASP cal file. Flat field data for the ASP is created by rotating the
grating to a spectral range without strong absorption lines. As this
option is not available for POLIS, the intensity images from the ASP cal
file were choosen as most similar to flat field data of POLIS. The
residual spectral information is enhaced by the displayed range of the
gaintable. The visible spectral information corresponds only to app. 1$\%$
variation. The interference fringes present in ASP data can be
identified as periodic variations along the wavelength dimension. The
treatment of fringes for POLIS will depend on their shape and size in
actual data.
\begin{figure}
\begin{minipage}{8cm}
\psfrag{A}{\Huge 0.9}
\psfrag{B}{\Huge \hspace*{-1cm}1.1}
\psfrag{C}{\Huge \hspace*{-1cm} $\leftarrow \lambda$}
\psfrag{D}{\Huge \hspace*{-1.4cm} $\uparrow$ height in slit}
\resizebox{6.8cm}{!}{\includegraphics{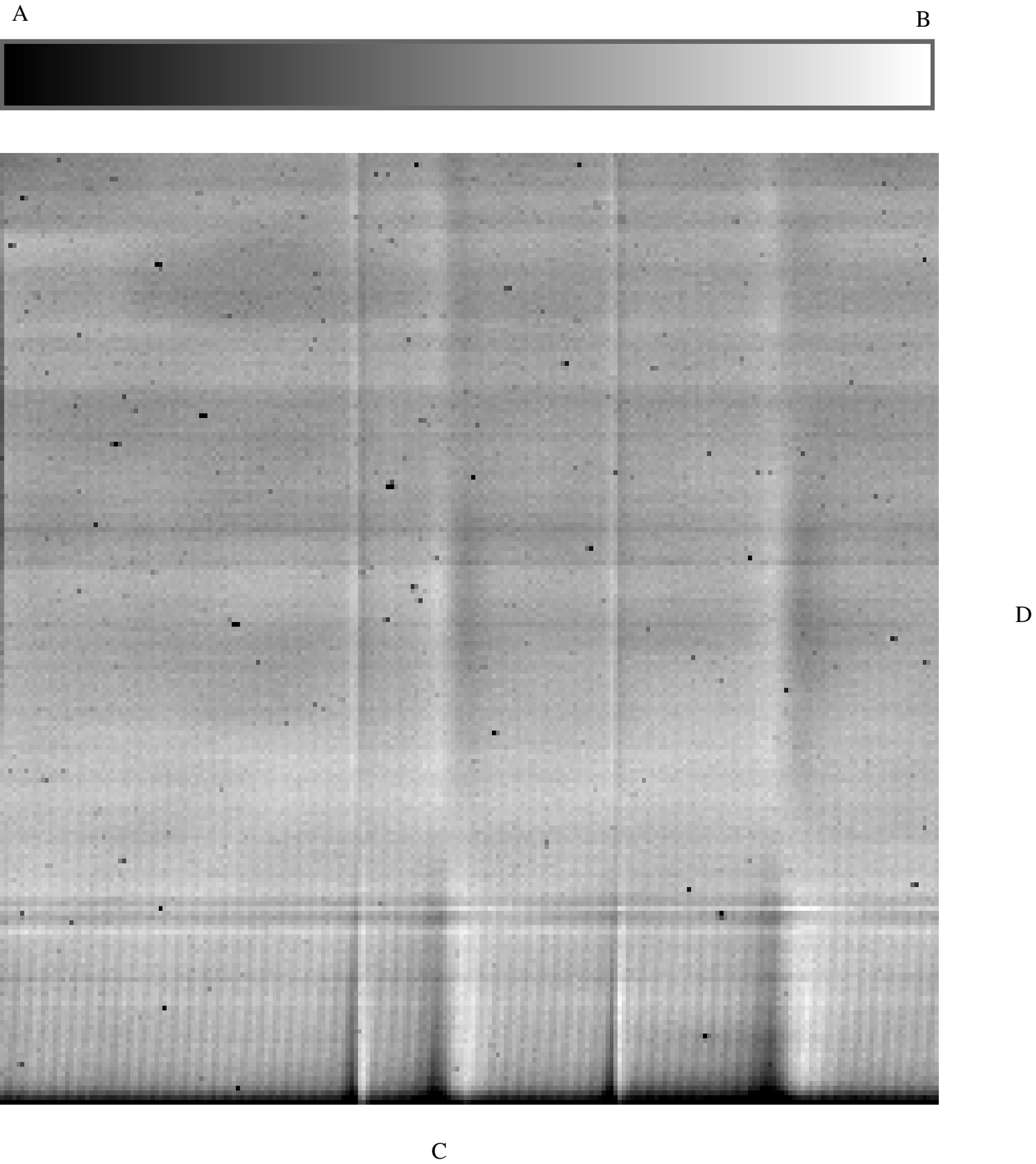}}
\end{minipage}
\begin{minipage}{8cm}
\caption[Gaintable for one of the ASP detectors]{Gaintable for one of
the ASP detectors (camera B), in the range of 0.9 to 1.1.\newline
The spectral information is not fully removed, as can be seen by the
residual spectral lines. The single black spots correspond to pixels
with greatly reduced intensity response. The periodic variation in the
image along the wavelength dimension are the interference
fringes present in ASP data. See text for a discussion of the quality of
the gaintable. The wavelength dimension $\lambda$ corresponds to columns
of the image, the height in the slit to rows.\label{gaintab_img}}
\end{minipage}
\end{figure}
\end{subsubsection}
\begin{subsubsection}{Balancing, coaligning and merging}
After the subtraction of the dark current and the application of the
gaintables the pixel properties of the detectors are not yet
fully removed. The gaintable can be interpreted as a comparison of the
intensity on single pixels to the mean intensity on the respective
detector area. This does not account for differences in the mean
detector intensities due to the different light path in the beam
splitter. To balance the detectors the mean intensity in Stokes I of an
abitrary choosen beam, say $I^+_{mean}$, is normalized to the intensity
$I^-_{mean}$.

The two images of Stokes I from $I^+$ and $I^-$ should then
contain the same spectral information, with identical intensities,
i.e. number of counts. In the next step they have to be coaligned in
spectral and spatial dimension. Again the position of the line core of
one telluric O$_2$-line can be used. One of the images is shifted row by
row, until the line cores in both images lie on the same column in each
row\footnote{cp. Fig. \ref{gaintab_img} for the definition of column and
rows of the image}. The alignment along the slit can be performed with
the horizontal hairlines inserted in the slit. These hairlines will be
especially useful for the correct alignment of images of the
chromospheric and photospheric lines.

The values of the intensity normalization, and the respective shifts in
spatial and spectral direction are established from Stokes I, but
have of course to be applied to Stokes Q,U and V as well.\\

For the coaligning of images a number of possibilites exist, the one
cited using line cores is only an example. The actual program code here
needs to consider mainly two things. First, quality of the calculation
of the shifts, second, stability under complicated conditions,
i.e. intensity profiles distorted by gradients or fully split
lines. Especially for the Ca-line a correlation method using an extended
image area may be the only stable option.

After balancing and co-aligning the data from the two beams the
components of the Stokes vector can be merged according to
eq. (\ref{iplusminus}) by:
\begin{eqnarray}
\bf S\rm (\lambda) = \begin{pmatrix} \rm I_+ + I_- \cr\rm  Q_+ - Q_- \cr
\rm  U_+ - U_- \cr\rm   V_+ - V_- \cr \end{pmatrix} (\lambda)\; ,
\end{eqnarray}
where the subscript + or - indicates the information from the beam $I^+$ or
$I^-$, and $\lambda$ is identical to the column number.\\
\end{subsubsection}
$ $\\
\fbox{\fbox{\parbox{16cm}{Summary:\newline
The polarimeter calibration data consists of three types, calibration,
flat field, and dark current images. The individual pixel response of
the detectors is established from dark and flat and stored in the
gaintables. The data images from $I^+$ and $I^-$ have to be
gain-corrected separately, balanced, and co-aligned before merging.}}}
\end{subsection}
\begin{subsection}{Determination of the polarimeter response
function\label{polresponse}}
After the calibration data has been treated in the way described in the
preceeding sections, a data set consisting of 73 images remains,
corresponding to the number of input vectors from the calibration unit. Along
the wavelength dimension the polarimetric value should be constant,
whereas along the slit it may vary\footnote{Along the slit spatial
inhomogeneties of the calibration unit or in the later beam path may
cause variations, while the grating only spectrally disperses the
incoming modulated polarization state.}. The polarimeter response function is
therefore established at four different heights in the slit. 
\begin{subsubsection}{Intensity normalization}
As the input intensity $\rm I\rm_0 \cdot
\frac{1}{2}\;(1+T_{1,0})$ from eq. (\ref{calibeq1}) is unknown, the data
set has to be normalized. If the slit heights are set to $r_i$
(i=1,2,3,4), the normalization is given by:
\begin{eqnarray}
\int\limits^{\Delta\lambda, \Delta x_i} \frac{\bf
S\rm_{out}(\lambda)}{\rm I(\lambda)}\; \rm d\lambda
dx \cdot \frac{\hspace*{-.2cm} \int \limits^{\Delta\lambda, \Delta x_i} I(\lambda)\;\rm d\lambda dx\;}{<
\int\limits_{\Delta\lambda, \Delta x_i} I(\lambda)\; \rm d\lambda dx >_{cal}} 
\equiv \frac{\int\limits^{\Delta\lambda, \Delta x_i} \bf S\rm_{out} (\lambda)\; d\lambda
dx}{< \int\limits_{\Delta\lambda, \Delta x_i}\rm I (\lambda)\; d\lambda dx
\;>_{cal}}\nonumber\; .
\end{eqnarray}
$\Delta \lambda$ (=columns) is the restriction on the
useable wavelegth range, excluding eventual image distortions at the
borders. The $\Delta x_i$ (=rows) are a decompostion of the slit height
into four non-overlapping intervalls, each centered around the value
$r_i$. $< .. >_{\rm cal}$ indicates the average over all images in the
calibration file.
Note that the normalization retains the information on the intensity
variations in the cal file, which are necessary to establish the first
row of the \bf X\rm-matrix.
\end{subsubsection}
\begin{subsubsection}{Matrix inversion}
For a single slit height $r_i$ the data is reduced to 73 output
Stokes vectors with this method. The corresponding input vectors
can be calculated from the position angles of the polarizer and retarder of the
calibration unit, using the right side of eq. (\ref{calibeq}). The
resulting linear problem can be written in the form
\begin{eqnarray}
(\bf S\rm_{out})_{cal} = (\bf S\rm_{in})_{cal} \cdot \bf X\rm ^T \; ,
\end{eqnarray}
where $(\bf S\rm_{out,in})_{cal}$ are $73\times 4$-matrices and \bf X\rm
$\in \mathbbm{R}^{4\times4}$. With the substitutions of $(\bf
S\rm_{out})_{cal} = \bf y\rm$ and $(\bf S\rm_{in})_{cal} = \bf M\rm $ 
the problem can be solved through
\begin{eqnarray}
\centerline{$\bf y\rm = \bf M\rm \cdot \bf X\rm^T$}\nonumber\\
\centerline{\mbox{\hspace*{.6cm}}$\bf M\rm ^T \cdot \bf y\rm = \bf M\rm ^T
\cdot \bf M\rm\cdot \bf X\rm ^T  =
\bf D \rm \cdot \bf X\rm  ^T \;\mbox{\hspace*{.6cm}by}$}\nonumber \\
\nonumber\\
\centerline{\fbox{$\bf X\rm ^T = \bf D\rm^{-1} \cdot \bf M\rm ^T \cdot
\bf y\rm$}}
\end{eqnarray}
$^{\rm T}$ denotes transposition, and $^{-1}$ the inverse of the matrix.
The errors can be calculated from the matrix $\bf A\rm = \bf D\rm^{-1}
\cdot \bf M\rm ^T$ in the following way:
\begin{eqnarray}
\sigma_{X_i}^2 &=& \sum_j A_{ij}^2 \sigma_{y_j}^2 = \bar{\sigma}^2 \sum_j
A_{ij}^2 \mbox{, where}\\
\bar{\sigma}^2 &=& \frac{1}{N} \cdot \sum_{cal} ( \bf y\rm - \bf M\rm
\cdot \bf X\rm ^T)^2 \; . \label{totalchi}
\end{eqnarray}
The errors of the single measurements, $\sigma _{y_i}$, are approximated by the
total deviation of the fit, $\bar{\sigma}$. As the same $A_{ij}$
(j=1,...,73) are used for the calculation of $\rm X_{\it ik}^T$ (k=0,1,2,3), the
procedure results in identical errors for each column of the polarimeter
response function \bf X\rm. The matrix is normalized in intensity by
division through $\rm X_{00}$ at the end of the calculation.
\end{subsubsection}
\begin{subsubsection}{Properties of the calibration unit}
The matrix inversion relates polarimeter input and measured output by the \bf
X\rm-matrix. To ensure the correct inclusion of the properties of the
calibration unit three additional parameters are introduced in the
construction of the input $(\bf S\rm_{in})_{cal}$. The polarizer is
assumed to be ideal\footnote{The transmission coefficient of light
polarized linear in the blocking direction is app. $10^{-5}$.} and
aligned correctly. For the wave plate it is useful to use the
retardance, $\delta_{ret}$, the dichroism, $b = r_x - r_y$, and a
position error of its fast axis, $\theta_{ret,pol}$, as free
parameters. The resulting matrix of the retarder is identical to
eq. (\ref{mirrmat}). The values of $\theta_{ret}, b,$ and
$\theta_{ret,pol}$ are established by minimizing $\bar{\sigma}^2$ in
eq. (\ref{totalchi}) with regard to the parameters through a gradient method.
\end{subsubsection}
\begin{subsubsection}{Final result and evaluation}
The determination of the polarimeter finally gives the following values
for each slit height $r_i$:
\begin{itemize}
\item \bf X\rm-matrix 
\item retardance $\delta_{ret}$, dichroism $b$, and position error
$\theta_{ret,pol}$.
\end{itemize}
The entries of the \bf X\rm-matrices are interpolated linearly to obtain
the \bf X\rm-matrix for an arbitrary slit height. The values describing
the properties of the wave plate can be used to control the consistency
of the evaluation. The variation should be rather small for one cal data
set.

The following values, which resulted from an application of the procedures on a
calibration data set of the ASP, can also serve as a good example for the
values to be expected for POLIS.

The \bf X\rm-matrix for $r_4$ = 200 and its error were calculated to be
\begin{eqnarray}
\bf X\rm = \begin{pmatrix} 1.0000 & -2.7\cdot 10^{-8} &   2.8\cdot 10^{-8}&
3.8\cdot 10^{-9}\cr
  -0.0004   &   0.1022   &   0.4108   &   0.1696\cr
   -0.0017   &  -0.4459   &  0.0975   &  0.0299\cr
    0.0011 &  -0.0088  &   -0.1624 &    0.3979\cr
\end{pmatrix}\; , \mbox{ error: } \pm
\begin{pmatrix} 0.0013 & 0.0037 & 0.0037& 0.0019 \cr
 0.0013 & 0.0037 & 0.0037& 0.0019 \cr
 0.0013 & 0.0037 & 0.0037& 0.0019 \cr
 0.0013 & 0.0037 & 0.0037& 0.0019 \cr \end{pmatrix} \;.
\end{eqnarray}
Here each image of the polarimeter output was normalized separately with
the average intensity of the image itself. This causes the information
about the first row to get lost. The values of the diagonal elements
clearly reflect the necessity of the polarimeter calibration. A measured
signal of Stokes Q is mainly caused by an input of U, and vice versa.
Only for Stokes V the matrix element $X_{33} \equiv$ V$\rightarrow$V is
the greatest contribution.  
Table \ref{rettable} displays the results for the properties of the
calibration retarder\footnote{The position error was not included in
this calculation.}, with the expected result of almost uniform
retardance.
\begin{table}
\begin{minipage}{5cm}
\begin{tabular}{|c|c|c|}  \hline
$r_i$ & $\delta_{ret}$ & $b$ \cr\hline
15  & 81.7352  &   0.0067  \cr
85  &82.0277 & 0.0034 \cr
155  &  82.1160 & -0.0003 \cr
200  & 81.7162 &  -0.0030  \cr\hline
\end{tabular}
\end{minipage}\hfill
\begin{minipage}{11cm}
\caption{The parameters of the retarder of the calibration unit for four
slit heights $r_i$. They were established with the procedures described
in this section from a data set of the ASP. The retardance $\delta$ is almost
constant along the slit, while the value of the dichroism $b$
systematically decreases.\label{rettable}}
\end{minipage}
\end{table}
\end{subsubsection}
\end{subsection}
\end{section}
\begin{section}{Telescope calibration}
The polarimeter response function takes into account all optical
elements behind the calibration unit, and the performance of the proper
measurement instrument itself. The application of the inverse
matrix $\bf X\rm^{-1}$ on measurement data results in the Stokes vector
of the light beam at the position of the calibration unit. This vector
is not identical to the polarization signal incident in the
telescope. 

The various elements in the beam path up to this point, the
coelostat mirrors, the windows of the vacuum tank, or the main and
deflection mirror change the  polarization to a certain degree. These
effects are usually summarized in the term 'instrumental polarization'
of the telescope. Some telescope designs are suited to minimize the
instrumental polarization due to their layout, for example a
Gregory-Coud\'e optic system\footnote{M.Stix, \cite{stix},
p. 76f}. Unfortunately a coelostat does not belong to that
category. Nonetheless, the ASP demonstrated how to successfully operate
a polarimeter at a telescope, which is in principle not suited for
polarimetry (cp. section \ref{asp_design}).

It is necessary to first introduce the telescope model, because the calibration
data set is adjusted its specific shape.
\begin{subsection}{The telescope model of the vacuum tower telescope in
Tenerife\label{telmodel_theory}}
The telescope model has to include all optical elements before the
calibration unit of the polarimeter to give effective values for the
polarimetric properties of the VTT. In the description of a mirror the
specific geometry of the reflection enters (section
\ref{mirrorrf}). This requires an explicit calculation of the beam path
for every moment of time, as the coelostat orientation and the beam path
are permanently changing (section \ref{beampath}). 
The main aim of the model is to give an accurate estimation of the
instrumental polarization from an only small number of
parameters. Opposite to the \bf X\rm-matrix no constancy in time can be
assumed, thus one has to deal with 16 variable entries in the \bf
T\rm-matrix.\\

Each mirror is represented in the model by a separate Mueller matrix,
whose entries depend on the physical properties of the mirror, and
geometrical factors of the beam path like the incidence angle. The
matrices are only valid for a specific set of reference frames (RFs) for each
mirror. It is necessary to additionally include rotation matrices to 
switch between the different RFs.\\

A semianalytical approach is used to calculate the incidence and
rotation angles in the telescope model. It uses analytical
solutions and some numerical procedures. One of the advantage of this
method is that one can control all actual directions, i.e. the different
reference frames, the mirror normals, the sun position and the position
of the second coelostat mirror, as they are given in a fixed
coordinate system as unit vectors. It is further possible to choose
different reference frames as input or intermediate systems. This is
important for the calibration of the telescope by inserting a sheet
polarizer at different places in the light path. The input created
depends on the orientation of the polarizer and has to be described in a
suited reference frame.

Another advantage is that the procedure developed is not restricted to
the specific design of the coelostat, but applicable on other
optical setups as well. If the beam path and the respective optical
elements are known, it is possible to calculate the polarimetric
properties according to the equations given in the following
sections\footnote{To take the worst possible case, for a randomly
oriented light beam no analytic solution is possible, but the numeric
method will work.}. This concerns for example an examination of a single
mirror through polarized light under different incidence angles.\\

The accuracy of the numerical part of the routines can be checked by the
comparison of axes, which are supposed to coincide after a rotation. The
numerical errors introduced corresponded to less then 6$\cdot 10^{-4}$
degrees in all cases.\\

The results for the telescope model are equivalent to the analytical set
of equations for a coelostat given by Cap. et al. (1989),
\cite{capitani}, which will be discussed in Appendix \ref{compltel} for
comparison. The model was developed in cooperation with
members\footnote{Thanks to M.Collados for his patience at this point.}
of the Instituto de Astrofisica de Canarias (IAC\footnote{The IAC
operates the Tenerife Infrared Polarimeter at the VTT Tenerife, and some
other telescopes at Iza\~na, Tenerife, and at the Roque de los
Muchachos, La Palma.}).
\begin{subsubsection}{Geometry of the Vacuum Tower Telescope\label{geometry}}
\begin{figure}
\begin{minipage}{11cm}
\psfrag{H}{\huge height}
\psfrag{J}{\huge sun}
\psfrag{A}{\huge celestial pole}
\psfrag{B}{\huge azimuth}
\psfrag{S}{\huge S}
\psfrag{E}{\huge E}
\psfrag{W}{\huge W}
\psfrag{N}{\huge N}
\psfrag{F}{\huge C1}
\psfrag{G}{\huge C2}
\psfrag{C}{\huge }
\resizebox{9.5 cm}{!}{\includegraphics{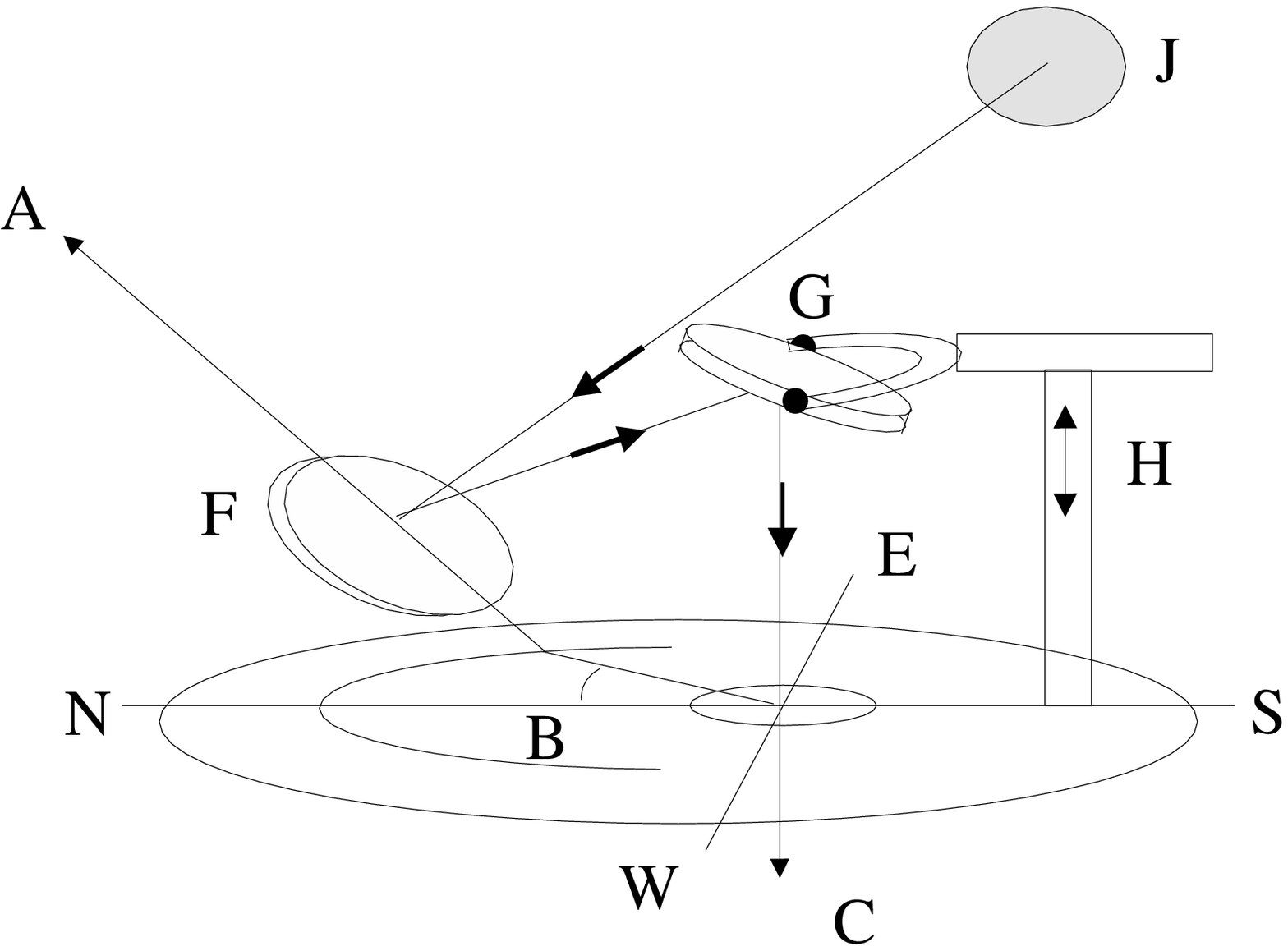}}
\end{minipage}
\hspace*{-1.3cm}
\begin{minipage}{8cm}
\caption[Scheme of the coelostat at the VTT Tenerife]{Scheme of the coelostat at the VTT Tenerife.\newline
The coelostat consists of two mirrors C1 and C2. The mirror C1 can be
moved on a circle of 1.55 m radius around the center of the entrance
window. The position on the circle is the mirror azimuth angle, which is
by convention positive from N to E. The rotation axis is inclined
28$^\circ$ (= geographical latitude of Tenerife) to the
horizontal plane and points to the celestial north. C2 is mounted
on a pillar with variable height and can be tilted along two axes to
deflect the incoming light vertically downwards to the main mirror in the
vacuum tank. \label{coelo}}  
\end{minipage}\\
\begin{minipage}{8cm}
\psfrag{E}{\Huge from coelostat}
\psfrag{A}{\Huge $0.84^\circ$}
\psfrag{B}{\Huge $0.84^\circ$}
\psfrag{C}{\Huge main mirror}
\psfrag{D}{\Huge deflection mirror}
\psfrag{F}{\Huge to focal plane}
\psfrag{G}{\Huge W}
\psfrag{I}{\Huge E}
\psfrag{H}{\Huge app. 19.7 m}
\psfrag{J}{\Huge 0.58 m}
\psfrag{K}{\Huge entrance window}
\psfrag{L}{\Huge exit window}
\psfrag{M}{\Huge calibration unit}
\resizebox{7.3cm}{!}{\includegraphics{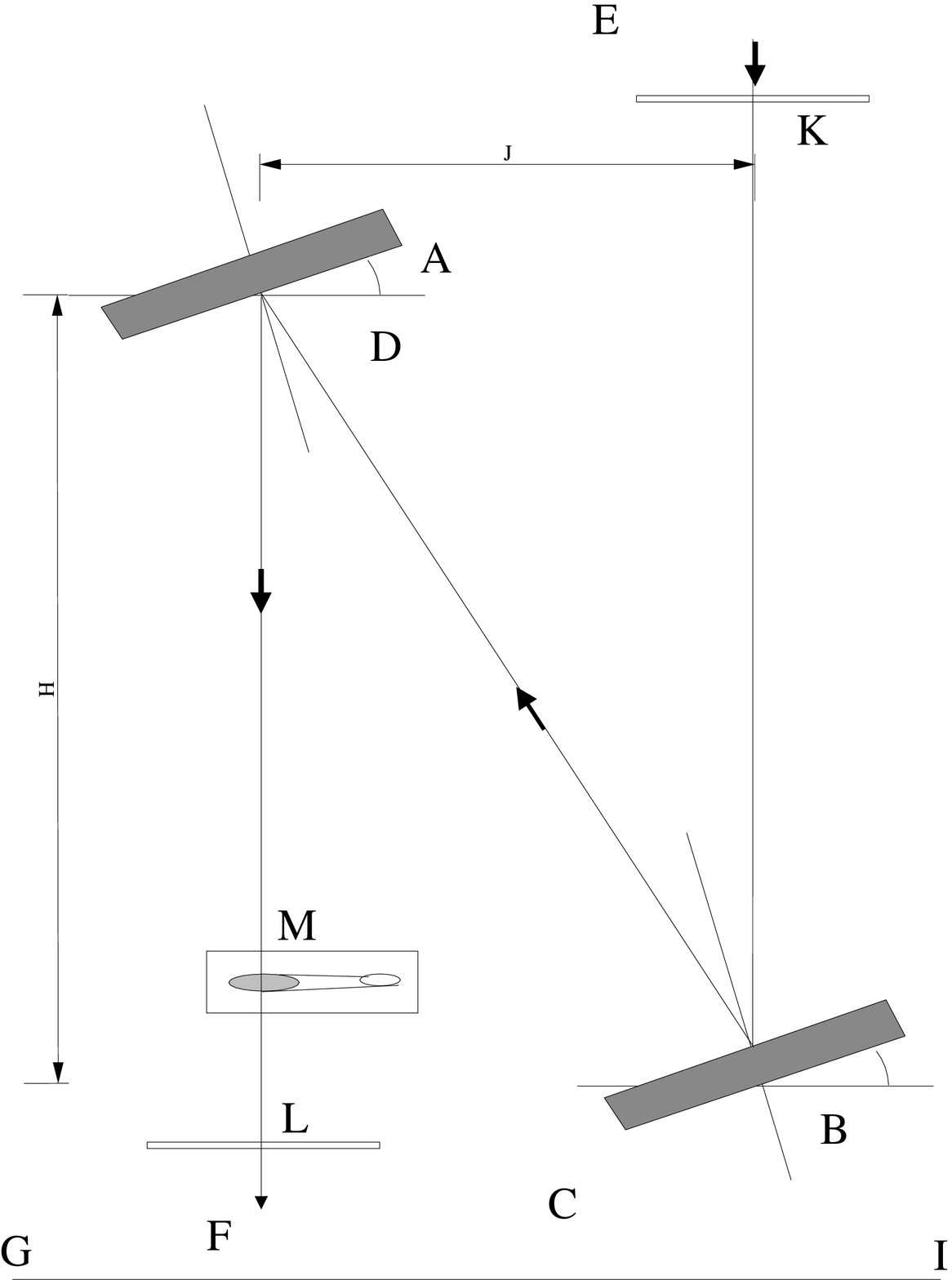}}
\end{minipage}\hfill
\hspace*{1.7cm}
\begin{minipage}{8cm}
\caption[The contents of the vacuum tank]{The contents of the vacuum tank.\label{vactank}\newline
The light enters vertically downwards from the second coelostat mirror
C2 through the entrance window. The main mirror is inclined by 
$0.84^\circ$ (drawing strongly exaggerated) to the W and reflects
the light to the deflection mirror app. 19.7 m above it. The deflection
mirror is inclined by the same amount and can be moved $\pm 1$ m
in height to adjust the position of the focal plane. The calibration unit
of the polarimeter is positioned before the exit window, if the correlation
tracker tank (not drawn) is also evacuated. After passing the
window the light reaches instrumentation like the spectrograph or POLIS.}
\end{minipage}
\end{figure}
The telescope is built primarily from two separate parts: the coelostat
on the top of the tower, which reflects the sunlight vertically
downwards, and the main and deflection mirror in the vacuum tank, which
produce an  image of the solar disc in the focal plane.\\

The coelostat of the german VTT at Tenerife consists of the two mirrors
labelled C1 and C2 (see Fig. \ref{coelo}). The primary coelostat mirror
C1 can has two degrees of freedom:
\begin{itemize}
\item The footpoint of its mounting can be moved on a circle of rails
with a radius of 1.55 m laid around the center of the entrance
window. The necessity of an adjustable footpoint arises to 
avoid C1 being shaded by C2 at noon, or to reduce the pillar height, when the
sun has a negative declination (cp. section \ref{beampath} for the
connection between these parameters). The position on the circle is
measured with the socalled mirror azimuth, by convention positive from N
to E.
\item C1 can be rotated around an axis pointing to the celestial north
pole to reflect the sun light on the second mirror C2. This feature defines the
optical design of a coelostat opposite to for example a
heliostat\footnote{see M.Stix, \cite{stix}, p. 69f}.
\end{itemize}
The vacuum tank contains the main and the deflection mirror (see
Fig. \ref{vactank}). These are fixed in their position with an
inclination angle of about 0.84$^\circ$ to the W direction. The
height of the deflection mirror above the main mirror can be changed by
$\pm 1$ m to adjust the postion of the focal plane, but this affects the
incidence angles important for the polarimetric properties only slightly. It is
therefore assumed that the default values can be used.\\

The combination of Fig. \ref{coelo} and \ref{vactank} gives the
complete beam path inside the telescope down to the polarimeter
calibration unit. If the correlation tracker (CT) tank (not shown) is
evacuated, the calibration unit sits before the exit window. If the CT
is not in vacuum, the window will be before the calibration
unit (cp. section \ref{windowmtx} for the implications).  
\end{subsubsection}
\begin{subsubsection}{Mueller matrices in the model}
There are two kinds of Mueller matrices used in the calculation. One are
matrices describing the polarizing properties of single mirrors or other
optical active media like the entrance window, the other are rotation
matrices between different reference frames.
\begin{paragraph}{Mirror matrices $\&$ reference
frames\label{mirrorrf}}
$ $\\

For the description of a mirror with a simple expression, a special set
of RFs has to be chosen explicitly. The Stokes vector of the incoming
and reflected beam are described in two different RFs. The following
definition of the RF is sometimes labelled
'canonical'\footnote{\cite{lites1}, p.361}.

The RF consists of the vector $\vec{\bf e}_3\rm$, parallel to the
propagation direction of the light beam. The vector $\vec{\bf e}_1\rm$
lies in the incidence plane defined by $\vec{\bf e}_3\rm$ and the mirror
normal. It is orthogonal to $\vec{\bf e}_3\rm$, and points from $\vec{\bf
e}_3\rm$ to the normal. $\vec{\bf e}_2\rm$ must be orthogonal on both to
form a right-handed RF. The output Stokes vector is given in the RF of
the reflected beam, which is constructed analogously from the
propagation direction and the mirror normal. This is depicted in
Fig. \ref{mirror} for the first coelostat mirror C1 and a positive
declination of the sun.\\
\begin{figure}
\begin{minipage}{8cm}
\psfrag{F}{\huge $\vec{\bf e}_2\rm$}
\psfrag{E}{\huge $\vec{\bf e}_1\rm$}
\psfrag{D}{\huge $\vec{\bf e}_3\rm$}
\psfrag{C}{\huge $\vec{\bf e}_2\hspace{0.05cm}^\prime\rm$}
\psfrag{B}{\huge $\vec{\bf e}_1\hspace{0.05cm}^\prime\rm$}
\psfrag{A}{\huge $\vec{\bf e}_3\hspace{0.05cm}^\prime\rm$}
\psfrag{G}{\huge polar axis}
\psfrag{I}{\huge $\vec{N1}$}
\psfrag{H}{\huge C1}
\psfrag{J}{\huge from the sun}
\resizebox{8cm}{!}{\includegraphics{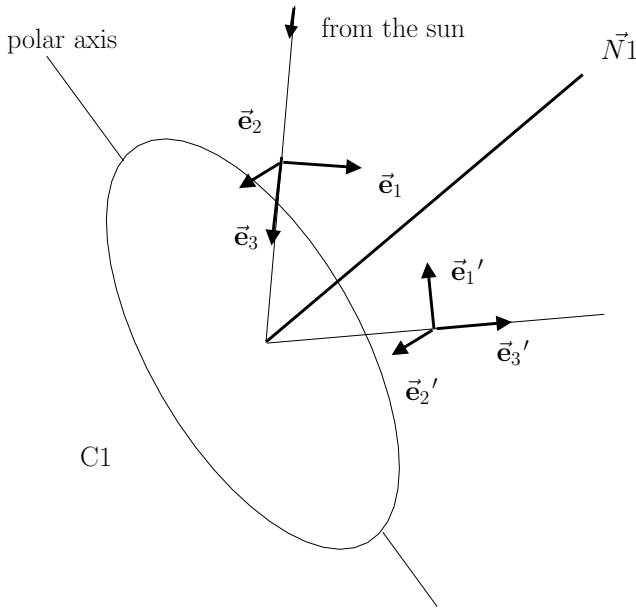}}
\end{minipage}
\hfill
\begin{minipage}{7.5cm}
\caption[Reference frames of a single mirror]{Reference frames of a
single mirror for the incoming and reflected beam (here for C1).\newline
The RF for the incoming beam ($\bf \vec{e}_1\rm, \vec{\bf e}_2\rm,
\vec{\bf e}_3\rm$) is defined by $\vec{\bf e}_3\rm$ in the propagation
direction, $\vec{\bf e}_1\rm$ in the incidence plane of mirror normal
$\vec{N1}$, orthogonal to $\vec{\bf e}_3\rm$, and pointing to the mirror
normal. $\vec{\bf e}_2\rm$ is orthogonal on both, forming a right-handed
RF. The RF of the reflected beam is constructed analogously from the
respective directions. Note that $\vec{\bf e}_2\rm$ and $\vec{\bf
e}_2\hspace{0.05cm}^\prime\rm$ are identical and parallel to the mirror
surface.\label{mirror}}
\end{minipage}
\end{figure}
The following discussion will concentrate on the more general
optical thin aluminium coatings\footnote{i.e., the value of the
thickness of the coating $d/\lambda \ll 1$}. The corresponding
formulation for optically thick coatings can be found in Appendix
\ref{compltel}.

The mirror matrix \bf M \rm in this case is given by\footnote{Jaeger $\&$
Oetken, \cite{jaeger}}
\begin{eqnarray} 
\bf M\rm_{mirror} = \bf M\rm(\bf \vec{e}_1, \vec{e}_2, \vec{e}_3\; | \;
\vec{e}_1\hspace{.051cm}^\prime, \vec{e}_2\hspace{0.05cm}^\prime, \vec{e}_3\hspace{.05cm}^\prime\rm) = \begin{pmatrix} 
(r_x+r_y)/2 & (r_x-r_y)/2& 0& 0\cr
(r_x-r_y)/2 & (r_x+r_y)/2& 0& 0\cr
0 & 0 & \;\;\sqrt{r_x r_y} c_\delta & \sqrt{r_x r_y} s_\delta \cr
0 & 0 & -\sqrt{r_x r_y} s_\delta & \sqrt{r_x r_y} c_\delta \cr
\end{pmatrix} \;. \label{mirrmat}
\end{eqnarray}
The conventions used are\footnote{Almost literally from Skumanich et
al., \cite{lites1}, p.379f.}:\\
(1) The electromagnetic wave is described by $\bf E\rm = \bf
E\rm_0 \cdot \exp ^{\bf i\rm (\bf k\cdot \bf r\rm - \omega \cdot t)}$.\\
(2) The components $E_i$ are defined in the canonical RF / RF$^\prime$
as given above.\\
(3) The reflectivities are defined by $E_i^{ref}$ = $R_i \cdot
E_i^{in}$.\\
(4) The given values are the incidence angle, $\eta$, the
refractive index, $n$, the absorption coefficient, $k$, the thickness of
the mirror coating in wavelength fractions, $d/\lambda$, and the
refractive index, $\nu$, of the substrate below the coating.\\
Under this conditions the matrix entries can be calculated by the following
equations\footnote{The equation were also taken from \cite{lites1},
p. 380. A derivation is possible from Born $\&$ Wolf, \it Principle of
optics\rm, \cite{born}, chapters 1.5, 13.2, and 13.4.}: 
\begin{eqnarray}
R_\| = \frac{r_\| + s_\| t}{1 + r_\| s_\| t}\;,\;\; R_\bot = \frac{r_\bot +
s_\bot t}{1 + r_\bot s_\bot t} \label{mirr1}
\end{eqnarray}
where
\begin{eqnarray}
r_\|&=& \frac{(n + ik)^2 \cos \eta - u - iv}{(n + ik)^2 \cos \eta + u +
iv}\;,\;\;\;\;\;\;\; r_\bot = \frac{(\cos \eta - u) -iv}{(\cos \eta + u) +iv}
\nonumber\\
s_\| &=& \frac{\nu (u + iv) - (n + ik)^2 \cos \alpha}{\nu (u + iv) + (n +
ik)^2 \cos \alpha}\;,\;\; s_\bot = \frac{(u + iv) - \nu \cos \alpha}{(u
+ iv) + \nu \cos \alpha} \nonumber\\
t &=& e^{4 \pi (d/\lambda)(u+iv)i} \label{teq}
\end{eqnarray}
and
\begin{eqnarray}
u &=& \sqrt{ \frac{1}{2} [n^2 - k^2 -\sin^2 \eta + \sqrt{n^2 - k^2 -\sin^2
\eta + 4 n^2 k^2\;}]\;}\nonumber \\
v &=& \sqrt{ \frac{1}{2} [-n^2 + k^2 +\sin^2 \eta + \sqrt{n^2 - k^2 -\sin^2
\eta + 4 n^2 k^2\;}]\;} \nonumber\\
\cos \alpha &=& \sqrt{1-\frac{\sin^2 \eta}{\nu^2}\;} \nonumber 
\end{eqnarray}
Taking $R_\| = R_x e^{i \epsilon_x}$ and $R_\bot =  R_y e^{i
\epsilon_y}$ with real $R_x, R_y$, the entries of the matrix are
\begin{eqnarray}
r_x &=&
R_x^2,\;\;\;\;\;\;\;\;\;\;\;\;\;\;\;\;\;\;\;\;\;\;\;\;\;\;\;\;\;\;\, r_y
= R_y^2 \label{mirr2}\\ 
\sqrt{r_x r_y} \cos \delta &=& Re(R_\| \cdot R_\bot^*),\;\;\;\sqrt{r_x
r_y} \sin \delta = - Im(R_\| \cdot R_\bot^*) \nonumber
\end{eqnarray}
For optical thick coatings the substrate below the coating can be
neglected, and the equations simplify (see Appendix \ref{compltel}).

It is assumed, that a sufficient polarimetric accuracy of the telescope model
can be achieved also with the simplier equations (M.Collados, personal
note). The resulting values for the refraction indices would in that
case have to be interpreted more as effective than as physical
values. Both sets of equations will be used to check the consistency. If
the coating thickness $d/\lambda$ is used as free parameter in a
least-square-fit to telescope calibration data, its value should allow
the decision, which equations are best suited to describe the telescope.\\
$ $\\
\fbox{\fbox{\parbox{16cm}{Summary:\newline
The Mueller matrices of the single mirrors have to be calculated for
each point in time with the momentaneous incidence angles, while the
other parameters entering are constant physical mirror properties.}}}
\end{paragraph}
\begin{paragraph}{Rotation matrices}
$ $\\

To transform between different RFs, in general a base
transformation has to be performed. This is simplified by the fact
that the beam directions of subsequent optical elements are identical, i.e. for
example the reflected beam of C1 ($\vec{\bf
e}_3\hspace{.05cm}^\prime\rm$ from Fig. \ref{mirror}) is the incoming
beam of C2 ($\vec{\bf e}_3\rm$ of C2).\\
Only the axes orthogonal to the beam direction can be different and the 
transformation can be established by a rotation around the $\vec{\bf
e}_3\rm$-axis.
The matrix of a rotation is given in Appendix \ref{mrot}.
The only thing to be considered additionally is the direction of the
rotation: the rotation angles are to be measured by convention
counter-clockwise from $\bf \vec{e}_1\rm$ to $\bf
\vec{e}_1\hspace{0.05cm}^\prime\rm$, when looking towards the light
source. The primed vector is from the RF to be transformed
to\footnote{Capitani et al., \cite{capitani}, p. 177}. Neglection of the
convention can easily lead to sign errors in the calculated angles,
which are very difficult to track down.
\end{paragraph}
\begin{paragraph}{Window matrix\label{windowmtx}}
$ $\\

The windows of the vacuum tank can exhibit stress induced
birefringence. The entrance window is modellized as pure retarder with
an arbitrary, but fixed, position angle of the fast and slow
axis\footnote{As in Skumanich et al., \cite{lites1}, p.362 .}. Its
properties can be included by the two parameters retardance, $\delta$,
and position angle, $\theta$. The exit window is assumed too small to
show polarizing effects at all. Values for the parameters result from the least
square fit to the telescope calibration data.\\

The paper of Owner-Petersen, \cite{owner}, gives a derivation for this
modellization. If the telescope model is seen to give an unsufficient
fit, a more rigid approach for the entrance window will be tried following the
considerations in the paper. Additionally the possibility to exclude the
window will be investigated. Then the main effect caused by the window, an
increased retardance, should be included in the effective physical values
of the main and deflection mirror. The way of treatment in this case is
still an open issue, depending on actual measurements at Tenerife.
\end{paragraph}
\end{subsubsection}
\begin{subsubsection}{Additional reference frames}
In addition to the RFs of the single mirrors, which are defined by the
beam path and the mirror normals, three other RFs are needed. The first is a
RF, which stays fixed on the sun. Into this RF the measurement result has to be
transformed at the end. The second one is the actual measurement RF of the
polarimeter, while the third one is the coordinate RF ('main RF') for all
following calculations of vectors.
\begin{itemize}
\item The main RF has its origin in the center of the first mirror C1 and is
defined by:
\begin{eqnarray}
\vec{x} = \begin{pmatrix} 1\cr 0\cr 0\cr \end{pmatrix}\;\mbox{points from
N to S,}\; \vec{y} = \begin{pmatrix} 0\cr 1\cr 0\cr \end{pmatrix}\; \mbox{
from W to E, and}\;  \vec{z} = \begin{pmatrix} 0\cr 0\cr 1\cr
\end{pmatrix}\; \mbox{from nadir to zenith}.\nonumber
\end{eqnarray}
Its orientation can be seen by moving the coordinate directions in
Fig. \ref{coelo} from the center of the entrance window to the center of
C1.
Four different ways are used later to describe a vector in the main RF :
\begin{itemize}
\item[(1)] Horizontal plane coordinates: the horizontal angular height,
$h$, equal to 90$^\circ$ - zenith distance, and the azimuth, $A$, given
by the angular distance between S and the projection of the vector to
the x-y-plane, increasing from S to W\footnote{H.-H. Voigt, \it Abriss
der Astronomie\rm, \cite{voigt}, p.1f}.
\item[(2)] Equatorial coordinates: declination, $\delta$, the angular distance
from the equatorial plane. This plane is inclined by
(90$^\circ$ - geographical latitude) to the horizontal plane. The second
coordinate is the hour angle, $t$, in the equatorial plane, which also is
increasing from S to W. This is not identical to the azimuth $A$ because of
the angle between these two planes\footnote{Voigt, \cite{voigt}, p.2}.
\item[(3)] Spherical coordinates: $\theta$, the angle to the $\vec{z}$-axis
(equal to zenith distance), and $\phi$, the angle in the
$\vec{x}$-$\vec{y}$-plane, this time increasing from $\vec{x}$ to
$\vec{y}$, i.e. from S to E, therefore equal to $-A$.
\item[(4)] Cartesian coordinates $x, y$ and $z$ as given by the definition of
the axes.
\end{itemize}
Transformations between these coordinate systems are:
\begin{itemize}
\item for $\delta, t \;\rightarrow h, A$\footnote{Voigt, \cite{voigt}, p.5}:
\begin{eqnarray}
\cos h \cdot \sin A& = &\cos \delta \cdot \sin t \label{eq1}\\
\sin h &=& \sin \phi \cdot \sin \delta + \cos \phi \cdot \cos \delta \cdot
\cos t \label{eq2}\\
- \cos h \cdot \cos A &=& \cos \phi \cdot \sin \delta - \sin \phi \cdot
\cos \delta \cdot \cos t \; .\label{eq3}
\end{eqnarray}
\item for $h, A \;\rightarrow \theta, \phi$ :
\begin{eqnarray}
\theta = 90^\circ - h,\;  \phi = - A \label{eq4}
\end{eqnarray}
\item for $\theta, \phi \;\rightarrow \vec{r} = (x,y,z)$ :
\begin{eqnarray}
\begin{pmatrix} x \cr y \cr z \cr \end{pmatrix} =
\begin{pmatrix}  \cos \phi \cdot \sin \theta \cr
\sin \phi \cdot \sin \theta \cr
\cos \theta \cr
\end{pmatrix} \;.\label{eq5}
\end{eqnarray}
\end{itemize}
\item For the solar RF it is sufficient to fix one axis. This reference
axis is the tangent to the equatorial plane at the hour angle of the
sun. If this vector is moved in angular height to the sun center, it
bisects the sun from W to E. This direction is given in the cartesian
main RF by
\begin{eqnarray}
\vec{x_0} = \begin{pmatrix}-\sin \phi \cdot \sin\; t_{sun}\cr
 -\cos\; t_{sun} \cr
-\cos \phi \cdot \sin\; t_{sun}\cr \end{pmatrix} \;,
\end{eqnarray}
where $\phi$ denotes the geographical latitude and $t_{sun}$ the hour angle.
\item The measurement RF of the polarimeter is defined through the zero
position of the transmission axis of the linear polarizer of the calibration
unit as $\vec{\bf e\rm}_1$. 
The rotation between telescope and polarimeter at the end of the optical
train will be used as free parameter, to compensate errors in the
initial positioning of the calibration unit.
\end{itemize}
\end{subsubsection}
\begin{subsubsection}{Calculation of the telescope matrix}
To calculate the polarimetric properties of the complete telescope,
consisting of the four mirrors C1, C2, main and deflection mirror, and
the entrance window of the vacuum tank, the following
parameters\footnote{Remark: for optically thin coatings.} are needed:
\begin{itemize}
\item mirror properties : refraction index, $n$, absorption coefficient,
$k$, thickness of the aluminium coating in wavelength fractions,
$d/\lambda$, refraction index of substrate, $\nu$
\item window properties : retardance, $\delta_{entrance}$, position angle, $\theta_{entrance}$,
\end{itemize}
which are constant properties of the system, and
\begin{itemize}
\item position angle of C1 on the circle of rails (azimuth)
\item sun position,
\end{itemize}
which define the beam path.\\
With the last two parameters the path of the light beam in the telescope
can be constructed, also the mirror normals and the RFs of the single
mirrors. This gives the values needed for the calculation of the Mueller 
matrices of the optical train, namely
\begin{itemize}
\item incidence angles on the mirrors : $i_1, i_2, i_3, i_4$
\item rotation angles between RFs : $\theta_1 \mbox{
(sun$\rightarrow$C1), } \theta_2 \mbox{ (C1$\rightarrow$C2), } \theta_3
\mbox{ (C2$\rightarrow$main), }\\ \theta_4 \mbox{
(defl. mirror$\rightarrow$polarimeter)}$.
\end{itemize}
The actually used input parameters of the routine are the constant
physical properties, which will be stored in a separate file, and the azimuth
of the first mirror, date, and time of the measurement.
\begin{paragraph}{The construction of the beam path \label{beampath}}
$ $\\

The beam path is defined by three direction vectors, corresponding to the bold
arrows in Fig. \ref{c2pos} : 
\begin{itemize}
\item The position of the sun as seen from C1. The program
r$\_$frame$\_$asp of the ASP-library returns the sun position for
Tenerife in equatorial coordinates, $\delta,t$. These can be transformed
to the cartesian vector $\vec{r}_{sun}$ with the equations
(\ref{eq1})-(\ref{eq5}).
\item The position of the second mirror C2 as seen from the center of C1.
\begin{figure}
\psfrag{A}{\huge sun}
\psfrag{B}{\huge $\vec{N1}$}
\psfrag{C}{\huge $A_{C2}$}
\psfrag{D}{\huge N}
\psfrag{E}{\huge E}
\psfrag{F}{\huge W}
\psfrag{G}{\huge S}
\psfrag{H}{\huge $\delta = 0$}
\psfrag{I}{\huge $\delta = \delta_{sun}$}
\psfrag{J}{\huge $\delta = \delta_{C2}$}
\psfrag{K}{\huge azimuth}
\psfrag{L}{\huge C2}
\psfrag{M}{\huge equ. plane}
\begin{center}
\resizebox{13cm}{!}{\includegraphics{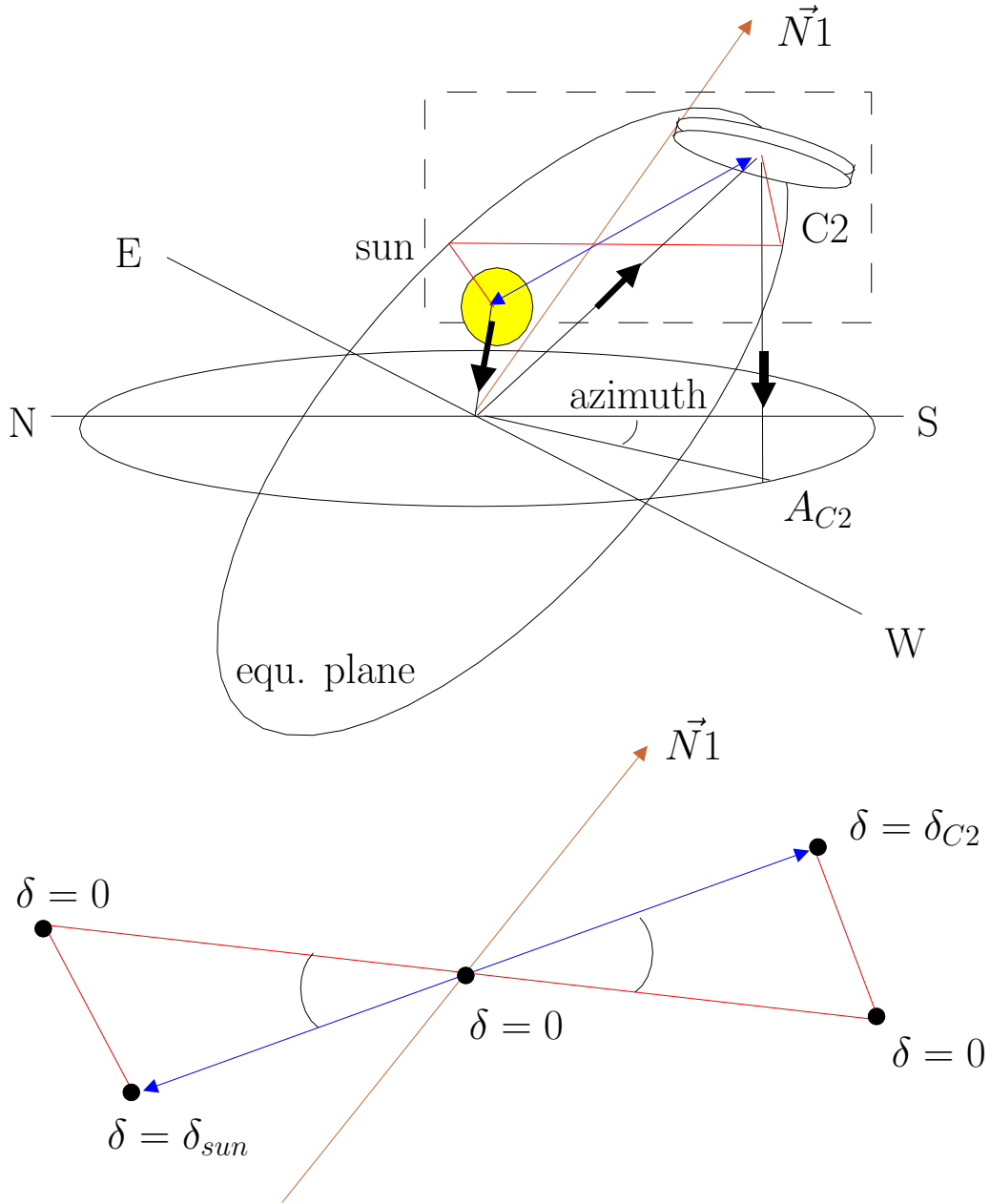}}
\end{center}
\caption[The position of the second mirror]{(\it top\rm) The position of
the second coelostat mirror C2 relative to C1. The origin lies in the center of
C1. The position angle $A_{C2}$ in the horizontal plane is fixed by the
azimuth of the first mirror. The mirror normal on C1, $\vec{N1}$,
is in the equatorial plane, which is inclined 90$^\circ -\phi$ to the
horizontal plane. Bold arrows denote the beam path. The declination of
the sun is negative. The entrance window would be vertically below C2.\newline
The declination of the second mirror, $\delta_{C2}$, is the negative of
the declination of the sun, $\delta_{sun}$, as the (\it bottom\rm) part
displays. There the line with the arrows connects the centers of the sun and
the second mirror. From the law of reflection it follows, that it is
exactly bisected by the mirror normal $\vec{N1}$, and is orthogonal to
it. The triangles built by the footpoints in the equatorial plane, the
intersection point and the positions of sun or C2 are identical,
therefore $\delta_{C2} = -\delta_{sun}$.\label{c2pos}}
\end{figure}
The coordinates of C2 are -unfortunately in two different coordinate
systems- given by
\begin{eqnarray}
\mbox{declination in equ. plane coordinates}\;&:&\;\; \delta_{C2} = -
\delta_{sun} \nonumber\\
\mbox{azimuth in hor. plane coordinates}\;&:&\;  A_{C2} = \mbox{azimuth} \;.\nonumber
\end{eqnarray}
The first stems from the fact that the mirror normal on the first
mirror, $\vec{N1}$, always is in the equatorial plane. Due to the law of
reflection C2 then must have the value opposite to the sun with regard
to the equatorial plane (see the lower part of Fig. \ref{c2pos}). The
second coordinate, $A_{C2}$, can easily be derived from the definition
of the azimuth in section \ref{geometry}.
 
The value, which needs to be established here, is the horizontal height,
$h_{C2}$. To solve the equations (\ref{eq1})-(\ref{eq3}) in this case a
numerical solution is used. Multiplication of Eq. (\ref{eq2}) with $\sin
\phi$ and of Eq. (\ref{eq3}) with $\cos \phi$, and addition of the
modified equations gives:
\begin{eqnarray}
\sin \phi \cdot \sin h - \cos A_{C2} \cdot \cos \phi \cdot \cos h = \sin
\delta_{C2} \;, \label{geth}
\end{eqnarray}
with the known values $\delta_{C2}, \phi$ and $A_{C2}$, the only
unknown parameter is the angular height $h$.\\

The left half of equation (\ref{geth}) is calculated for values of $h
\in [0, 90^\circ]$. The angular height $h_{C2}$ sought is the
intersection point of this curve with the constant value $\sin
\delta_{C2}$. With eq. (\ref{eq4}) and (\ref{eq5}) the vector
$\vec{r}_{C2}$ is obtained.
\item The vertical beam path after C2. This direction simply is (0,0,-1)
in the main RF.
\end{itemize}
\end{paragraph}
\begin{paragraph}{The mirror normals}
$ $\\

The direction of the mirror normal on C2 is easy to construct. The
incoming beam has the direction $\vec{r}_{C2}$, the reflected beam must
be vertically downwards. The mirror normal therefore has to be inbetween
-$\vec{r}_{C2}$ and (0,0,-1). As these directions have the same length
(both are normalized), the mirror normal $\vec{N2}$ is given
by\footnote{ A proof is simple, this follows from the properties of an
isosceles triangle.}:
\begin{eqnarray}
\vec{N2} = \frac{\;-\vec{r}_{C2} + (0,0,-1)\;}{| \vec{r}_{C2} + (0,0,-1)
|}\;. 
\end{eqnarray}
The mirror normal of C1, $\vec{N1}$, is constructed in the same way from
$\vec{r}_{C2}$ and $\vec{r}_{sun}$:
\begin{eqnarray}
\vec{N1} = \frac{\;\vec{r}_{C2} + \vec{r}_{sun}\;}{| \vec{r}_{C2} +
\vec{r}_{sun}|}\;.
\end{eqnarray}
\end{paragraph}
\begin{paragraph}{The incidence angles}
$ $\\

The incidence angles on the respective mirrors can be calculated from the
standard scalar product, $< \vec{a} , \vec{b} > = \sum_i a_i b_i =
|\vec{a}||\vec{b}| \cos(\alpha)$, between mirror normal and the
direction of the incoming or reflected beam. So each $i_j$ can be
calculated in two ways\footnote{For control one can calculate $i_2 = (\pi /2 -
h_{C2})/2$, which is also valid.}:
\begin{eqnarray}
i_1 &=& \mbox{acos } ( < \vec{r}_{sun}, \vec{N1} > )\;\; =
\mbox{acos } (<\vec{N1}, \vec{r}_{C2} >) \; ,\; \mbox{and}\\
i_2 &=& \mbox{acos } (< -\vec{r}_{C2}, \vec{N2} >) = \mbox{acos }
(< \vec{N2}, (0,0,-1) >) \;.
\end{eqnarray}
The incidence angles on main and deflection mirror are constant with
$i_3 = i_4 = 0.84^\circ$, which can be derived from Fig. \ref{vactank}.
\end{paragraph}
\begin{paragraph}{Construction of the reference frames}
$ $\\

The RFs of the single mirrors are determined by beam directions and
mirror normals. An example will be executed for an incoming beam (see
Fig. \ref{rfcon}).

The axis $\vec{z}$, equal to the inverse direction of the incoming beam,
the mirror normal $\vec{N}$ and the x-axis $\vec{x}$ lie in the same
plane (incidence plane). The two vectors $\vec{x}$ and $\vec{z}$ are an
ONB for every vector in the incidence plane, thus $\vec{N}$ can be
decomposed by
\begin{eqnarray}
\vec{N} = < \vec{z} ,\vec{N} > \cdot \;\vec{z} + \vec{\beta} \;,
\end{eqnarray}
where $< \vec{\beta} , \vec{z} > = 0$.\\
With
\begin{eqnarray}
\vec{x} = \frac{\vec{\beta}}{| \vec{\beta} |} = \frac{\vec{N} - <
\vec{z} ,\vec{N} > \cdot \;\vec{z}}{| \vec{N} - < \vec{z} , \vec{N} > \cdot \;\vec{z} |}
\end{eqnarray}
\begin{figure}
\begin{minipage}{8cm}
\psfrag{A}{\huge \hspace*{-.6cm}mirror}
\psfrag{B}{\huge $\vec{\beta}$}
\psfrag{C}{\huge \hspace*{-.5cm}$ < \vec{z} ,\vec{N} > \cdot
\;\vec{z}$}
\psfrag{D}{\huge \hspace*{.8cm}$i$}
\psfrag{N}{\huge $\vec{N}$}
\psfrag{E}{\huge reflected beam}
\psfrag{F}{\huge \hspace*{-4cm}incoming beam,$-\vec{z}$}
\resizebox{8cm}{!}{\includegraphics{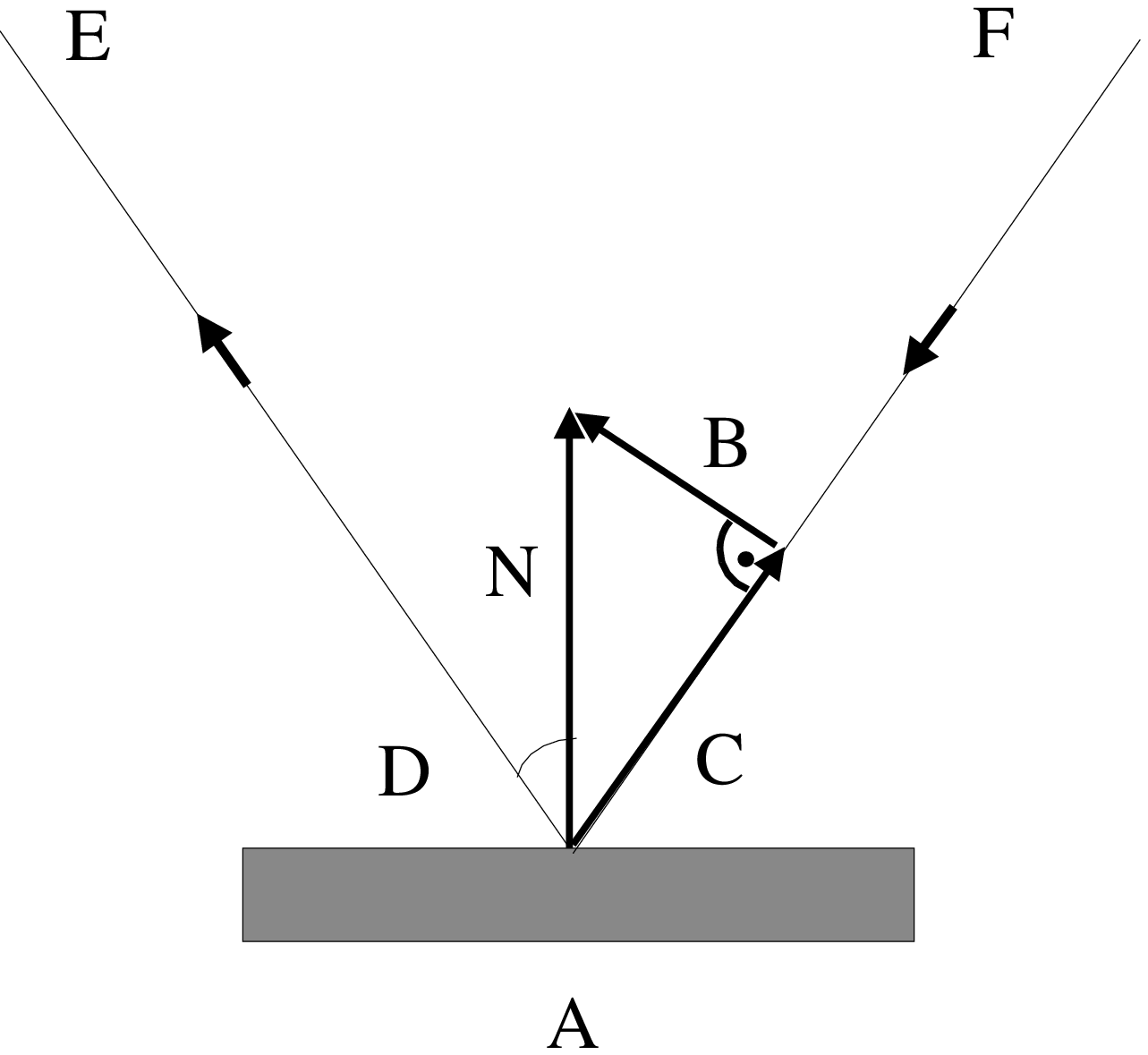}}
\end{minipage}\hfill
\begin{minipage}{8cm}
\caption[Construction of the RFs]{Construction of the RFs.\newline
The mirror normal can be decomposed into two orthogonal parts. One is
parallel to the propagation direction, $\vec{z}$. It is given by
$<\vec{z},\vec{N}> \cdot \;\vec{z}$. \newline
The other, $\vec{\beta}$, is parallel to the unknown x-axis in the
incidence plane. The x-axis direction can be calculated
by normalizing $\vec{\beta} = \vec{N} - < \vec{z} , \vec{N} > \cdot
\;\vec{z}$.\label{rfcon}}
\end{minipage}
\end{figure}
a direction is obtained, which is orthogonal to the propagation
direction, points from it to the mirror normal, and lies in the incidence
plane, the requirements for the x-axis in section \ref{mirrorrf}.

The y-axis can be calculated by:
\begin{eqnarray}
\vec{y} = \vec{x} \otimes \vec{z} \;,
\end{eqnarray}
where $\otimes$ denotes the standard vector crossproduct.
\end{paragraph}
\begin{paragraph}{Calculation of the rotation angles}
$ $\\

Two different subsequent RFs have one axis in common, that is the
propagation direction, $\vec{\bf e}_3\hspace{0.05cm}^\prime \mbox{
(RF1)} = \vec{\bf e\rm}_3 \mbox{ (RF2)}$. One needs to establish the
rotation angle around this axis, arbitrarily oriented in space, that
moves $\vec{\bf e\rm}_1\hspace{0.05cm}^\prime \mbox{ (RF1)}$ onto its
counterpart $\vec{\bf e\rm}_1 \mbox{(RF2)}$. This can be done
numerically for the calculated unit vectors of the RFs, which shall not
be disussed in detail\footnote{The condition of counter-clockwise
rotation unfortunately excludes the use of the scalar product.}. 
\end{paragraph}
\begin{paragraph}{The final matrix product}
$ $\\

Now the complete optical train can be calculated:
\begin{eqnarray}
\centerline{$\bf T\rm = \bf R\rm_{tel,pol} \cdot \bf M\rm_{defl} \cdot
\bf M\rm_{main} \cdot \bf M\rm_{entrance} \cdot \bf
R\rm_{C2,main}\cdot \bf M\rm_{C2} \cdot \bf
R\rm_{C1,C2} \cdot \bf M\rm_{C1} \cdot \bf
R\rm_{sun,C1}$}\\
\centerline{$= \bf T\rm (n,k,d/\lambda,\nu,\delta_{entr},\theta_{entr},\mbox{azimuth,time,date})$} \;.
\end{eqnarray}
\bf R \rm are rotation matrices according to eq. (\ref{mrot}) with the
corresponding angles, \bf M \rm the Mueller matrices of the optical
elements like eq. (\ref{mirrmat}) for the mirrors, and according to
eq. (\ref{retrot}) for the entrance window.
\end{paragraph}
\end{subsubsection}
\end{subsection}
\begin{subsection}{Telescope calibration data}
The principle of the telescope calibration is identical to the
polarimeter calibration. It consist of the creation of known input
Stokes vectors, and the measurement of the corresponding polarimeter
output. After the application of the inverse of the \bf X\rm-matrix the
properties of the polarimeter can be assumed to be removed from the
data. The difference is the time dependence of the \bf T\rm-matrix,
which makes a fit of constant matrix entries like for the polarimeter
impossible.\\

The main problem of the telescope calibration unit is the size needed. It has
to cover the full area of the first coelostat mirror or the entrance
window. This can not be achieved with a single sheet polarizer of
sufficient quality. Large sheet polarizers are subject to spatial
inhomogeneties due to deformations of the surface. The telescope
calibration unit therefore consists of an array of sheet polarizers. The
single sheets are placed in separate mountings and can be aligned
individually. The whole array is rotateable to create different linear
polarizations (see Fig. \ref{tel_cali}). Unfortunately the
calibration unit is too heavy to be placed freely in the beam path above
the first mirror\footnote{This would additionally require a larger
covered area, and an accurrate co-motion with the mirror.}. The possible
positions of the calibration unit are on top of the first mirror, which
also solves the problem of accurate co-motion to guarantee an input
constant in time, and on top of the entrance window after the coelostat
mirrors.\\
\begin{figure}
\begin{minipage}{8cm}
\psfrag{75}[b1][b1][4]{\Huge \hspace*{2.5cm}75 mm}
\psfrag{300}[b1][b1][4]{\Huge\hspace*{-3cm}300 mm}
\resizebox{8cm}{!}{\includegraphics{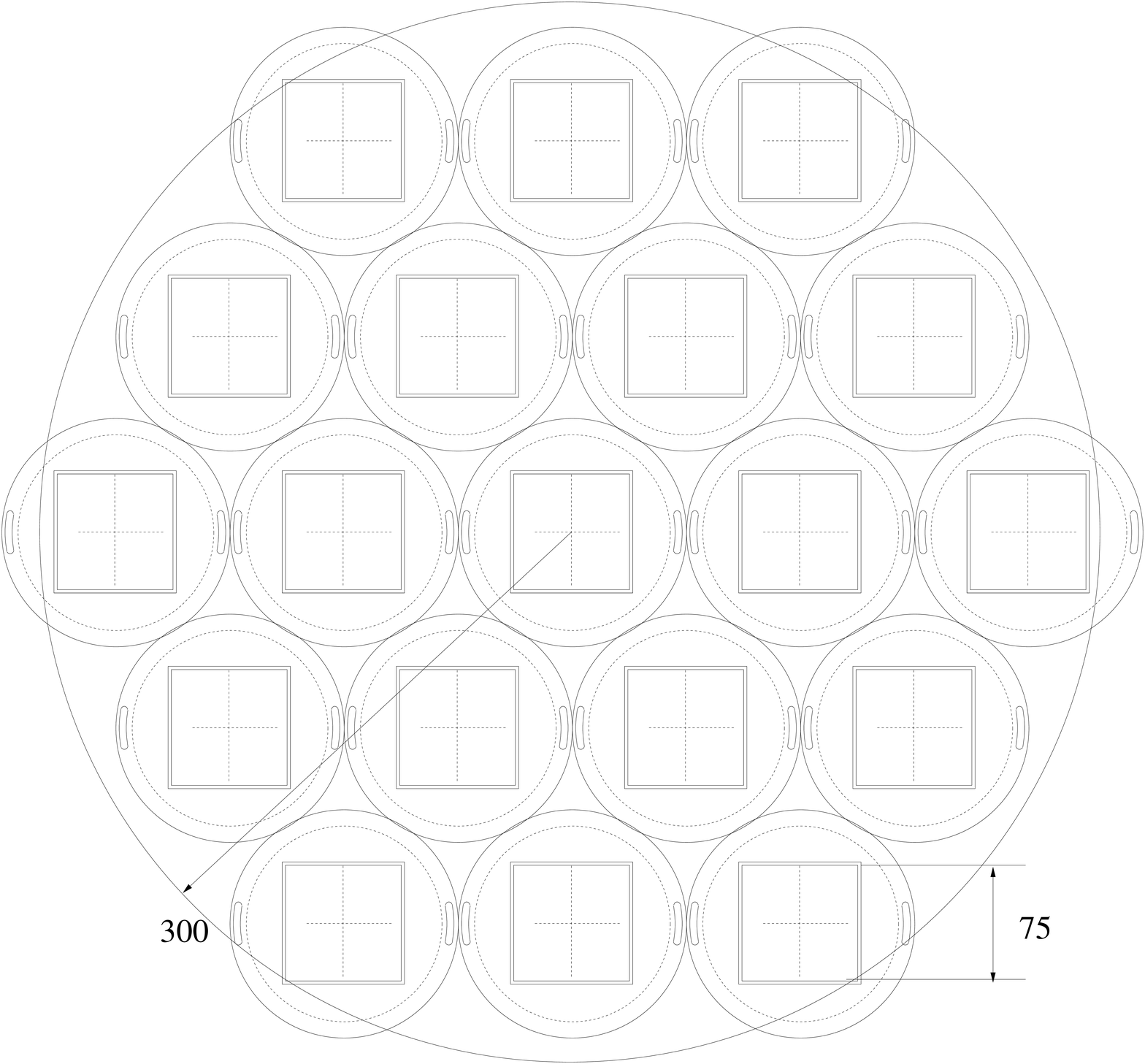}}
\end{minipage}\hfill
\begin{minipage}{7.4cm}
\caption{Section of the telescope calibration unit.\newline
The covered area corresponds to the diameter of the first mirror. The 19
quadratic sheet polarizers can be separately aligned. The array can be
rotated as a whole to create different linear polarizations. It can be
placed either on top of the first coelostat mirror C1
(Fig. \ref{polonc1img}) or on the entrance window
(Fig. \ref{polonwindow}). \label{tel_cali}}
\end{minipage}
\end{figure}
The telescope calibration unit thus allow two different sets of data to
be created. Each set will consist of a full (half) revolution of the
calibration unit with a stepsize of 10$^\circ$ (5$^\circ$). The data
sets include different optical elements of the telescope and will have
to be evaluated in correct order, but not fully separetely.
\begin{subsubsection}{Calibration unit on top of the entrance window}
\begin{figure}
\begin{minipage}{8cm}
\psfrag{B}{\Huge \hspace*{-.45cm}polarizer}
\psfrag{A}{\Huge \hspace*{-.9cm}entrance window}
\resizebox{8cm}{!}{\includegraphics{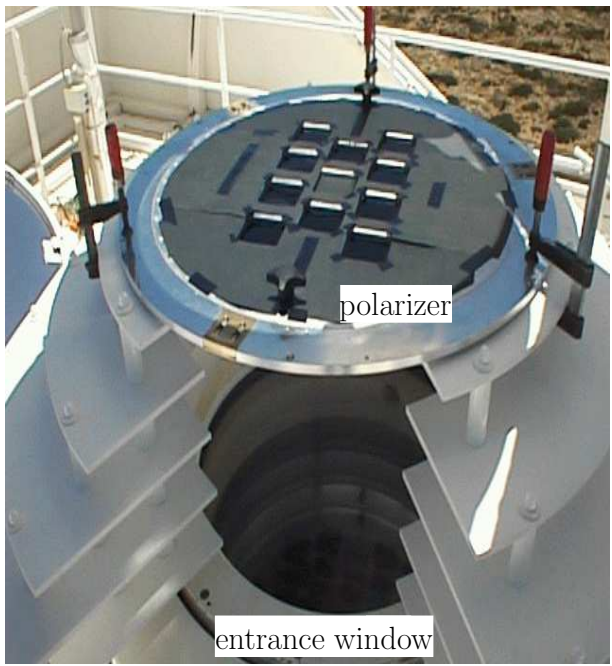}}
\end{minipage}\hfill
\begin{minipage}{7.5cm}
\caption{The array of sheet polarizers in the infrared for the TIP on
top of the entrance window. Only 10 of the 19  polarizing sheets were
delivered to that time. The whole array can be rotated to create
different linear polarizations.\label{polonwindow}}
\end{minipage}
\end{figure}
Fig. \ref{polonwindow} shows the corresponding telescope calibration unit
of the TIP with polarizer sheets in the infrared wavlength range on
that position.\\
\begin{figure}[ht]
\resizebox{8cm}{!}{\includegraphics{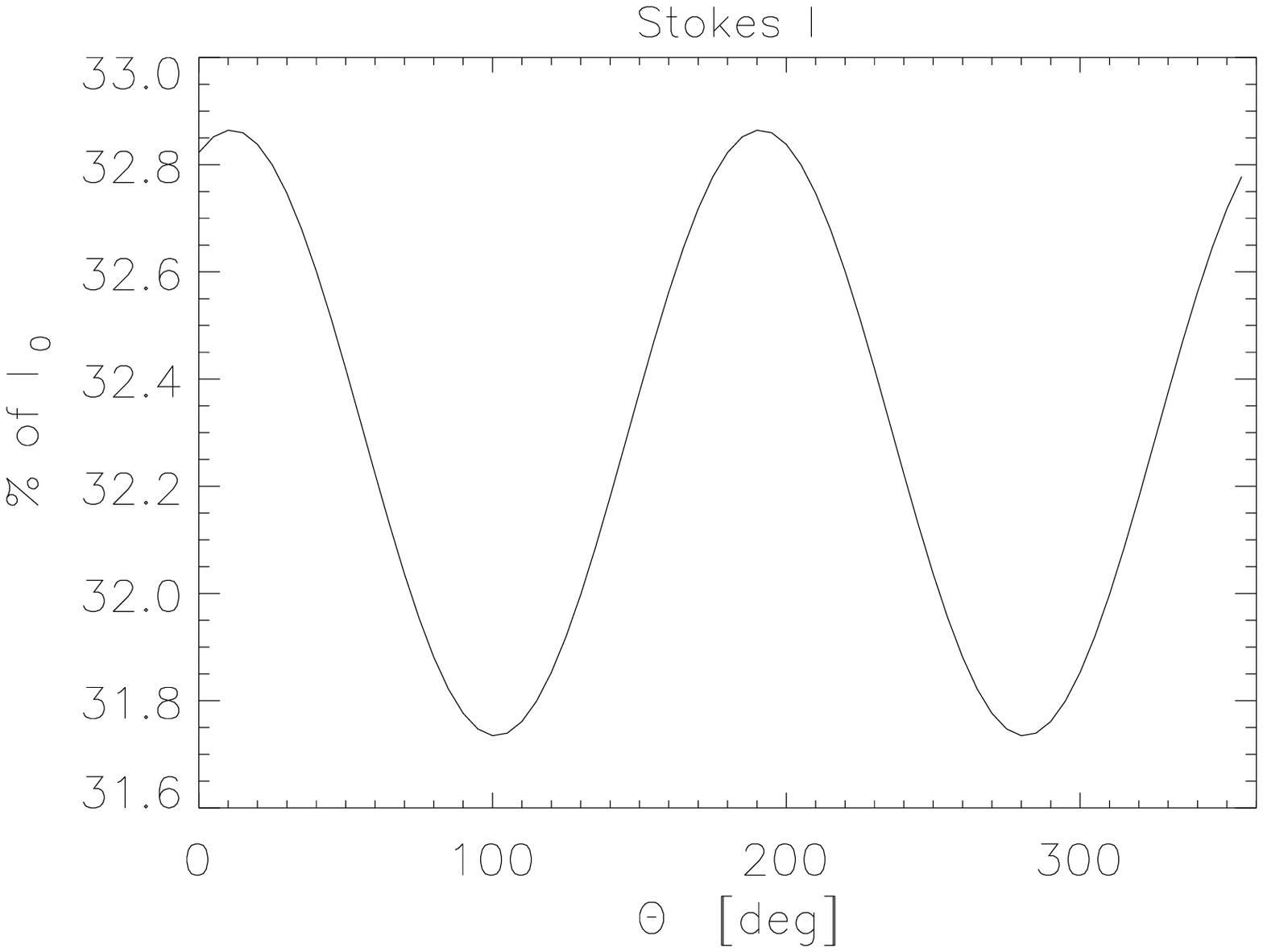}}
\resizebox{8cm}{!}{\includegraphics{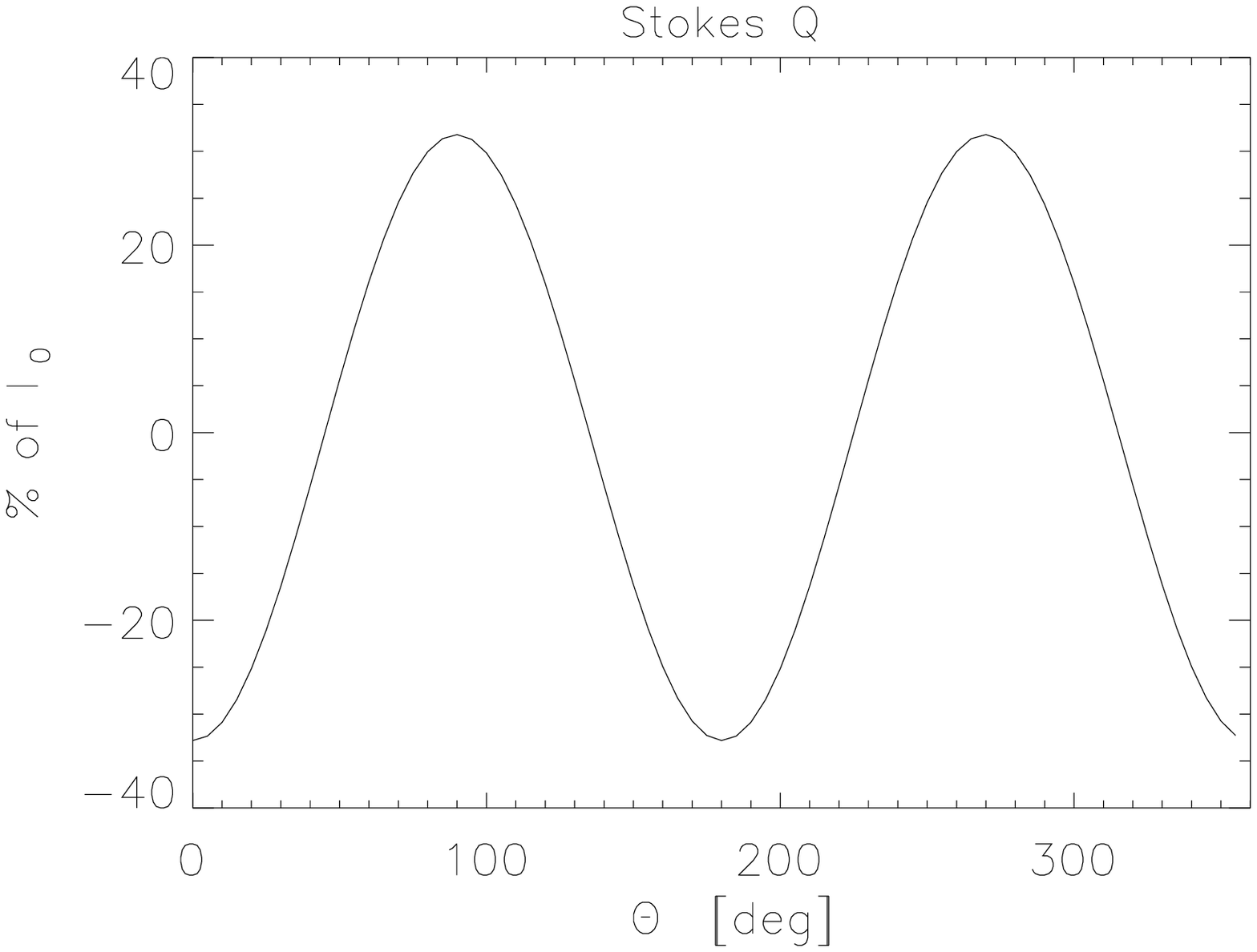}}\\
\resizebox{8cm}{!}{\includegraphics{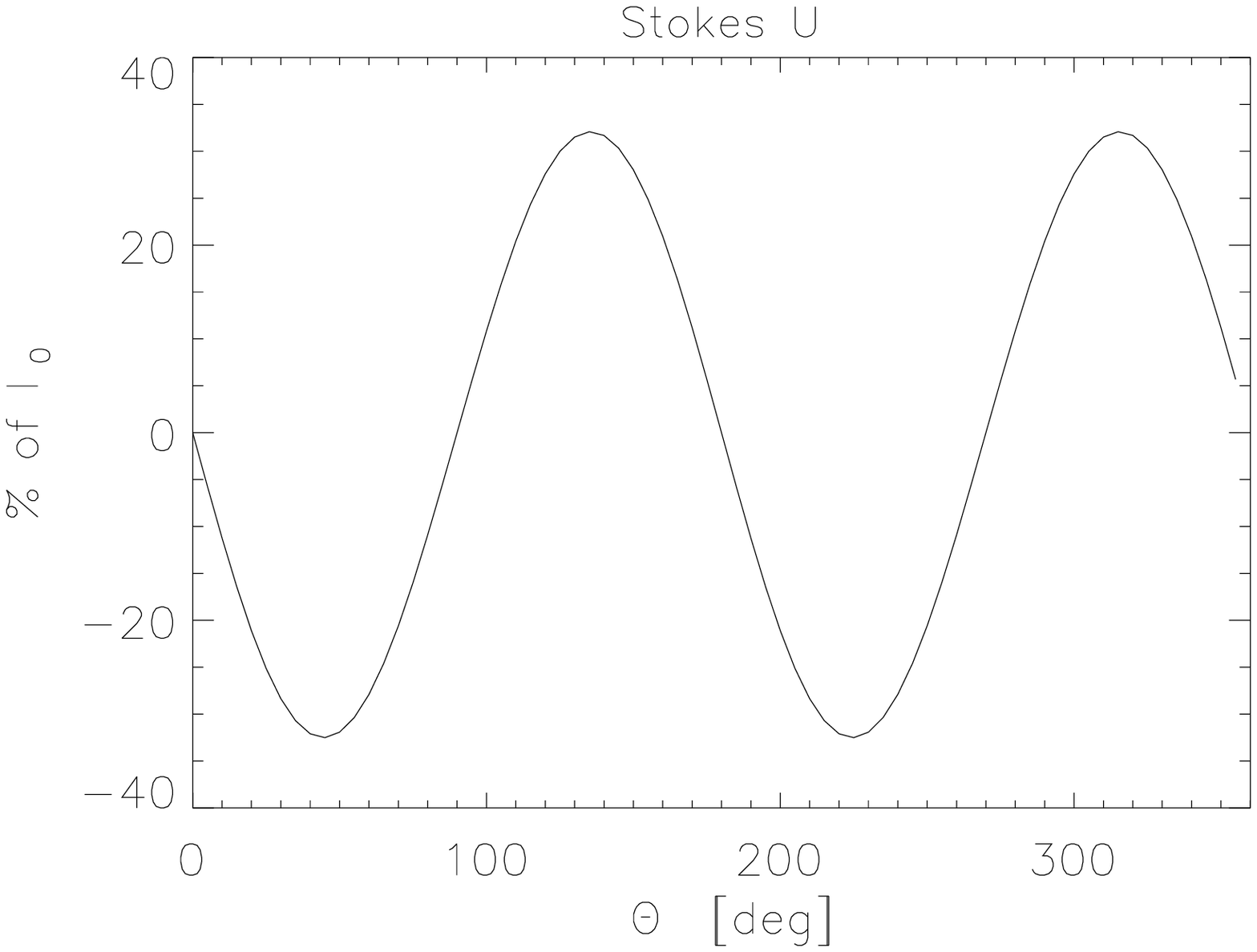}}
\resizebox{8cm}{!}{\includegraphics{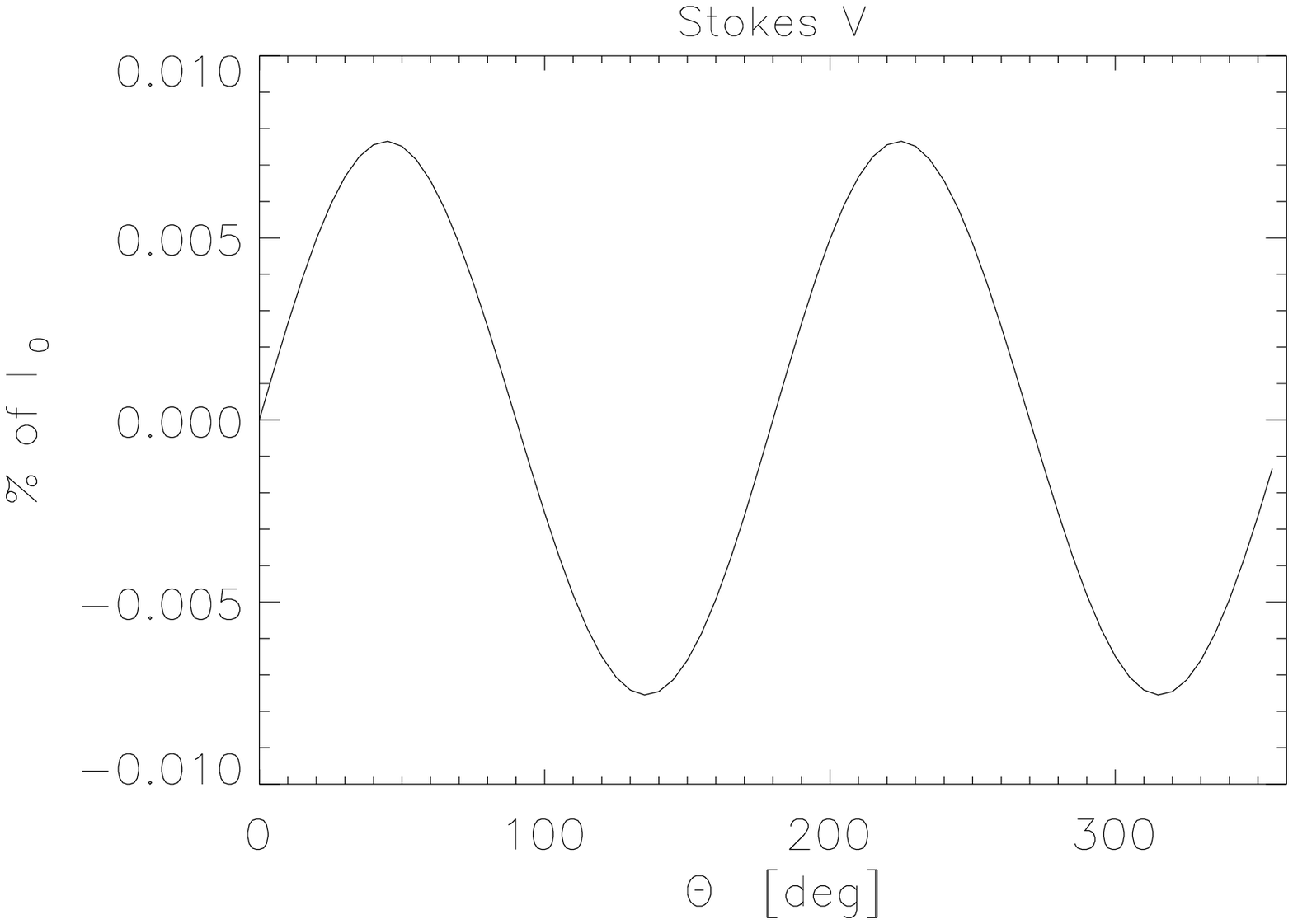}}
\caption{Stokes vector for a full rotation of the telescope calibration
unit on the entrance window in the model. Date: 21.07.02, 12:00 UT, azimuth =
0$^\circ$. Mirror parameters $(n,k,d/\lambda,\nu)$ =
(0.30,3.63,0.094,1.54), taken from an ASP telescope calibration. The
window is excluded, $\theta_{tel,pol} = 90^\circ$.\newline
The array of polarizers is placed on top of the entrance window and
rotated for an angle $\theta \in [0,360^\circ ]$ in already polarized
light. The resulting matrix $(\bf T\rm)_{pol,window}$ is applied on
unpolarized sunlight with the intensity $\rm I_0$. The transmitted intensity is
given in $\%$ of the incident intensity. See text for more
details.\label{polonwindowimg}}
\end{figure}

If the telescope calibration unit is placed on top of the entrance
window, the optical elements below are the entrance window, the main
and the deflection mirror. The zero position of the polarizer is the W
direction (=$\vec{\bf e}_1$ of the main mirror RF), the array has to be
rotated from W to N to give a counter-clockwise rotation.

Both mirrors are usually coated at the same time. Their polarimetric
properties can be assumed to be identical, and should exhibit only small
changes on longer time scales due to their evacuated environment.

With the inclusion of the entrance window as optical element and using
the description of optical thin mirrors (see section \ref{mirrorrf}
above) the following parameters are open in the reduced model:
\begin{itemize}
\item $n_{vac},k_{vac},d/\lambda$ for the mirrors. The label 'vac'
indicates that these values are supposed to be valid only for the
mirrors in the vacuum tank.
\item $\theta_{entrance}, \delta_{entrance}$ for the entrance window.
\item $\theta_{tel,pol}$ between telescope and polarimeter.
\end{itemize}
Even if it seems that the coelostat mirrors can be neglected for this
part, they still have an influence. The polarizer array is rotated in
already polarized light. This leads to intensity variations like already
discussed in section \ref{secpolcal}. The equation connecting input and
output is:
\begin{eqnarray}
\bf S\rm_{out} = I_0(t) \cdot \bf X \rm \cdot \bf T\rm_{vac} \cdot \bf
L\rm(\theta_{pol}) \cdot \begin{pmatrix} 1 \cr \alpha (t)
\cr \beta (t) \cr \gamma (t) \cr \end{pmatrix}\; .\label{tel_cal_eq}
\end{eqnarray}
$\bf T\rm_{vac}$ is the restriction of the telescope model on the
elements after the polarizer. The time dependence of $\rm I_0(t), \alpha
(t), \beta (t), \mbox{ and }\gamma (t)$ takes into account small variations in
total intensity and the polarization content resulting from the motion
of the first coelostat mirror. The values of
$\alpha, \beta,$ and $\gamma$, i.e. the polarization introduced by the
coelostat mirrors, are of order of some $\%$. If the polarizer now
is rotated, the intensity in $\bf S\rm_{out}$ may vary by the same amount,
with the total intensity $\rm I_0$ also unknown. The intensity variation
can be seen in Stokes I in Fig. \ref{polonwindowimg}. The macxima and
minima in the curve occur, if the polarizer is aligned parallel or
anti-parallel to the linear polarization produced by the two coelostat
mirrors. For an azimuth of C1 $\neq 0^\circ$ the order of the intensity
variations increases strongly.

To remove this intensity variation the following procedure may be
used\footnote{Keeping the polarizer fixed as for the polarimeter
calibration data is no option here}. The left side of
eq. (\ref{tel_cal_eq}) is normalized with the intensity after the
application of the inverse \bf X\rm-matrix:
\begin{eqnarray} 
\bf S\rm_{out}^\prime = \begin{pmatrix} 1 \cr \delta \cr \eta \cr \nu
\cr \end{pmatrix} = \frac{\bf X\rm^{-1} \cdot S\rm_{out}}{ (\bf
X\rm^{-1} \cdot \bf S\rm_{out})_0} \; ,
\end{eqnarray}
where $\delta, \eta,$ and $\nu$ is the polarization information.

The right side is constructed from an unit intensity input, and the final
input vector is also normalized: 
\begin{eqnarray}
\bf S\rm_{in} =  \frac{\bf T\rm_{vac} \cdot \bf L\rm(\theta_{pol}) \cdot
\begin{pmatrix} 1 & 0 & 0 & 0 \cr \end{pmatrix}^T}{(\bf T\rm_{vac} \cdot
\bf L\rm(\theta_{pol}) \cdot \begin{pmatrix} 1 & 0 & 0 & 0 \cr \end{pmatrix}^T)_0} \stackrel{!}{=}
\begin{pmatrix} 1 \cr \delta \cr \eta \cr \nu \cr \end{pmatrix}\; .
\end{eqnarray}
This should allow to perform a least square fit of the free parameters
$n_{vac},k_{vac},d/\lambda,\theta_{entrance}, \delta_{entrance}$, and
$\theta_{tel,pol}$. The intensity information lost by the normalization
is not needed for the fit of the model parameters, as the \bf
T\rm-matrix is intensity normalized anyway, i.e. divided by $\rm
T_{00}$. See the discussion on additional intensity variations due to
geometrical effects in the following section.
\end{subsubsection}
\begin{subsubsection}{On top of the mirror C1\label{polontop}}
\begin{figure}[ht]
\begin{minipage}{8cm}
\psfrag{A}{\Huge \hspace*{-.7cm} polar axis}
\psfrag{B}{\Huge \shortstack{\vspace*{-.1cm}C1}}
\psfrag{C}{\Huge \hspace*{-.4cm}polarizer}
\resizebox{8cm}{!}{\includegraphics{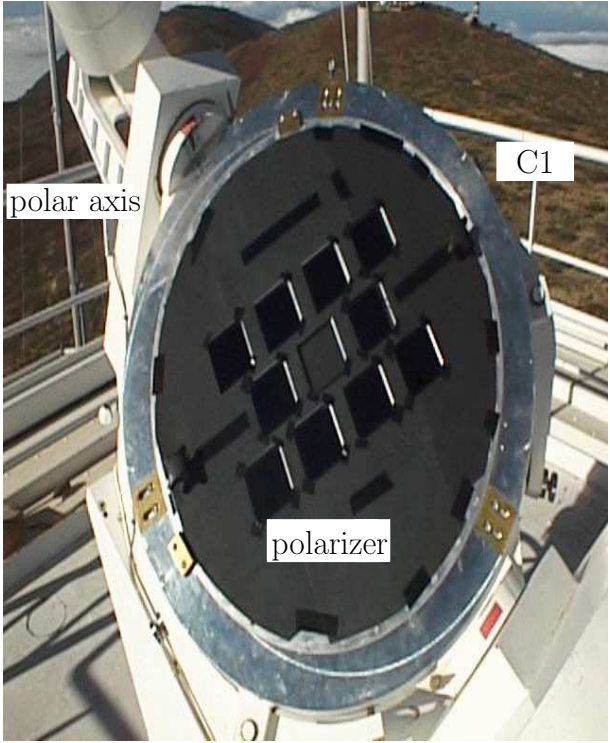}}
\end{minipage}\hfill
\begin{minipage}{7.4cm}
\caption{The array of sheet polarizers in the infrared for the TIP on
top of the first mirror. Only 10 of the 19  polarizing sheets were delivered to
that time. The whole array can be rotated on the mirror to create
different linear polarizations.\label{polonc1img}}
\end{minipage}
\end{figure}
 
In this case the light passes the polarizer twice. After the second
tranmission through the calibration unit the beam path includes all
remaining elements in the telescope model. The properties of the
entrance window and the subsequent mirrors can assumed to be known. The
open parameters now are $n_{coel}, k_{coel}$, and $d/\lambda$, where the
label 'coel' refers to the two coelostat mirrrors only. The thickness of
coating may be dropped for optical thick coatings, see the discussion in
\ref{mirrorrf}.\\

This seems to be easy to achieve, but the setup causes a mayor
problem. The polarizers of the calibration unit are assumed to be ideal,
and they are passed again after the first mirror. The only way to retain
the information on the properties of this mirror is to use here a
non-normalized \bf T\rm-matrix, contrary to the section above and the
application on observational data. It seems necessary to explicitly
derive how the information is still available, even with an ideal
polarizer.\\ 

To simplify the calculation a 'quasi-ideal'\footnote{'ideal' would additionally
require $r_x =r_y = 1$.} mirror, i.e. the assumption of a retardance of
$\delta = 180^\circ$ in eq. (\ref{mirrmat}), is used. Its Mueller matrix is:
\begin{eqnarray}
\bf M\rm_{mirror, ideal}  = \begin{pmatrix} 
\alpha & \beta& 0& 0\cr
\beta & \alpha& 0& 0\cr
0 & 0 &  - \gamma  & 0 \cr
0 & 0 & 0 & -\gamma \cr
\end{pmatrix} \;, \label{idmirror}
\end{eqnarray}
where $\alpha = (r_x+r_y)/2, \beta = (r_x-r_y)/2$, and $\gamma =
\sqrt{r_x r_y}$.

If the optical train for the rotated polarizer with the angle $\theta$
before and $-\theta$\footnote{The minus arises due to the 
counter-clockwise definition of the angle. The x-axis of the mirror is assumed
to coincide with the zero position of the array of polarizers.} after
the mirror is calculated, the transmitted intensity is:
\begin{eqnarray}
\rm I(\theta) = \frac{1}{4} \cdot \left( \alpha \cdot ( 1 + \cos ^2 \; 2
\theta) + 2 \cdot \beta \cdot \cos \;2 \theta + \gamma \cdot \sin ^2 \;2
\theta \right) \label{inteq}
\end{eqnarray}
For $r_x = r_y$ the intensity I is constant. In all other cases a
modulation of the intensity dependent on the mirror properties remains.\\

It should thus be possible to extract the properties of the first coelostat
mirror from the intensity I in the data set. Fig. \ref{polonC1}
displays a result for eq. (\ref{inteq}) predicted from the telescope
model.
\begin{figure}[ht]
\resizebox{8cm}{!}{\includegraphics{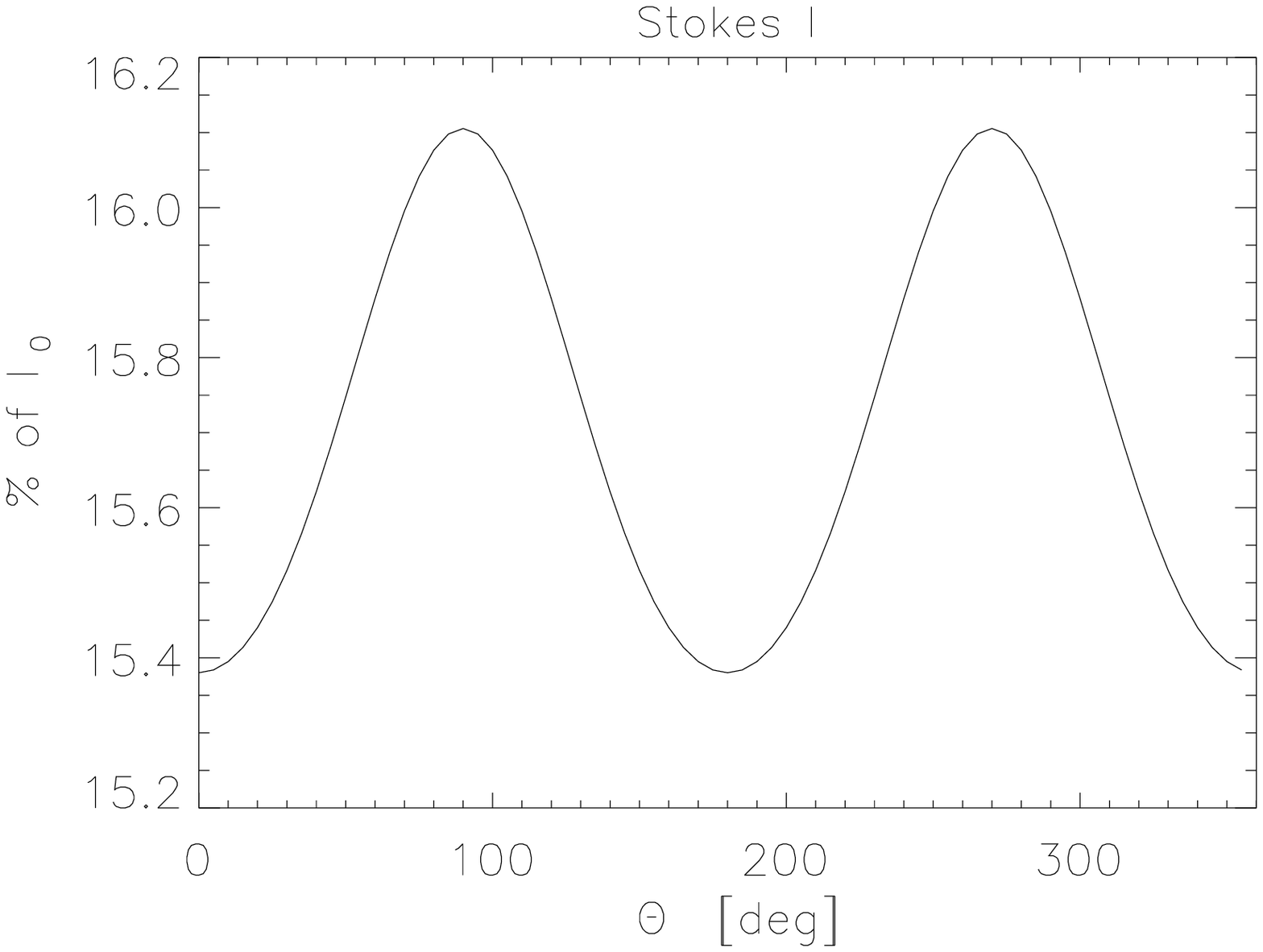}}
\resizebox{8cm}{!}{\includegraphics{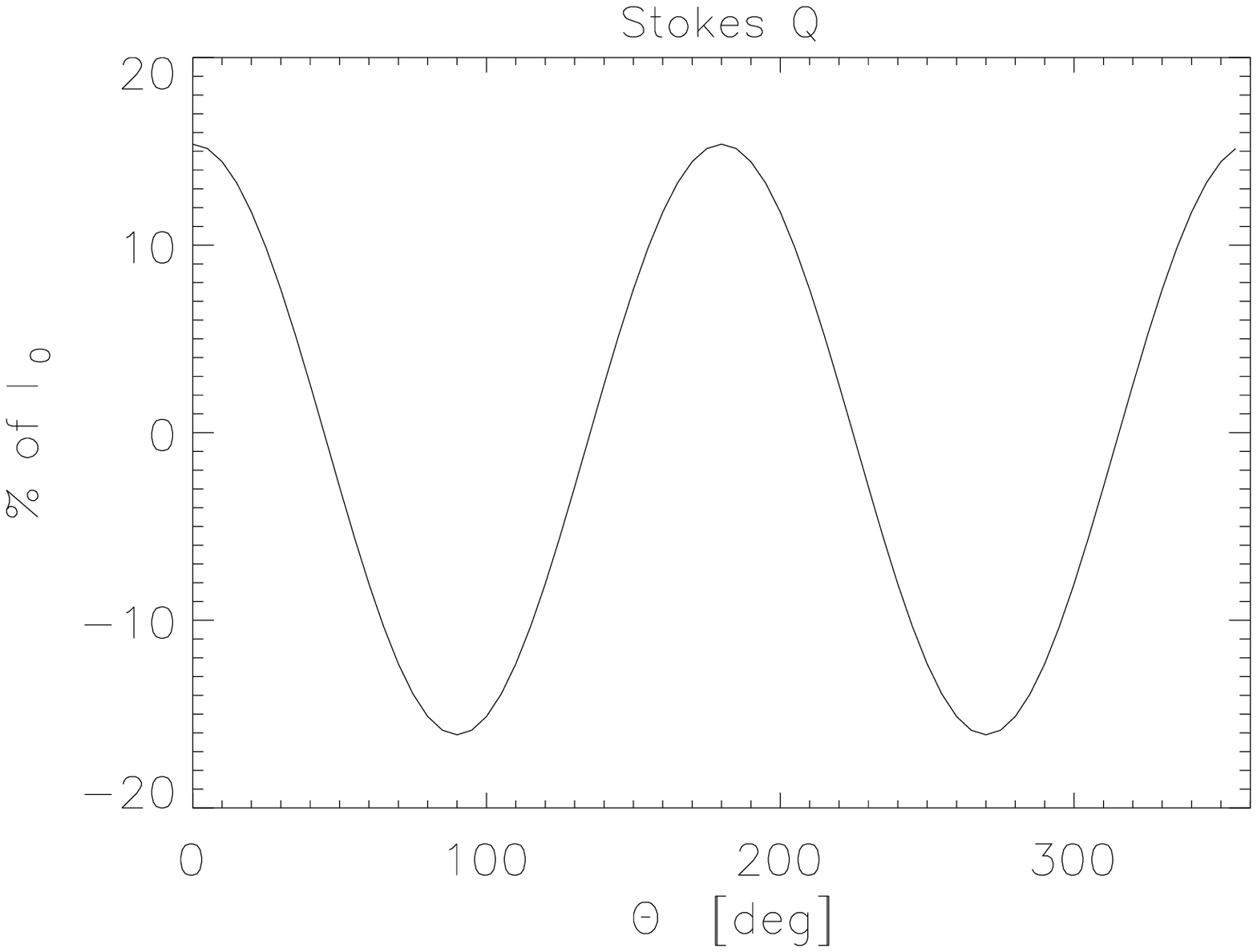}}\\
\resizebox{8cm}{!}{\includegraphics{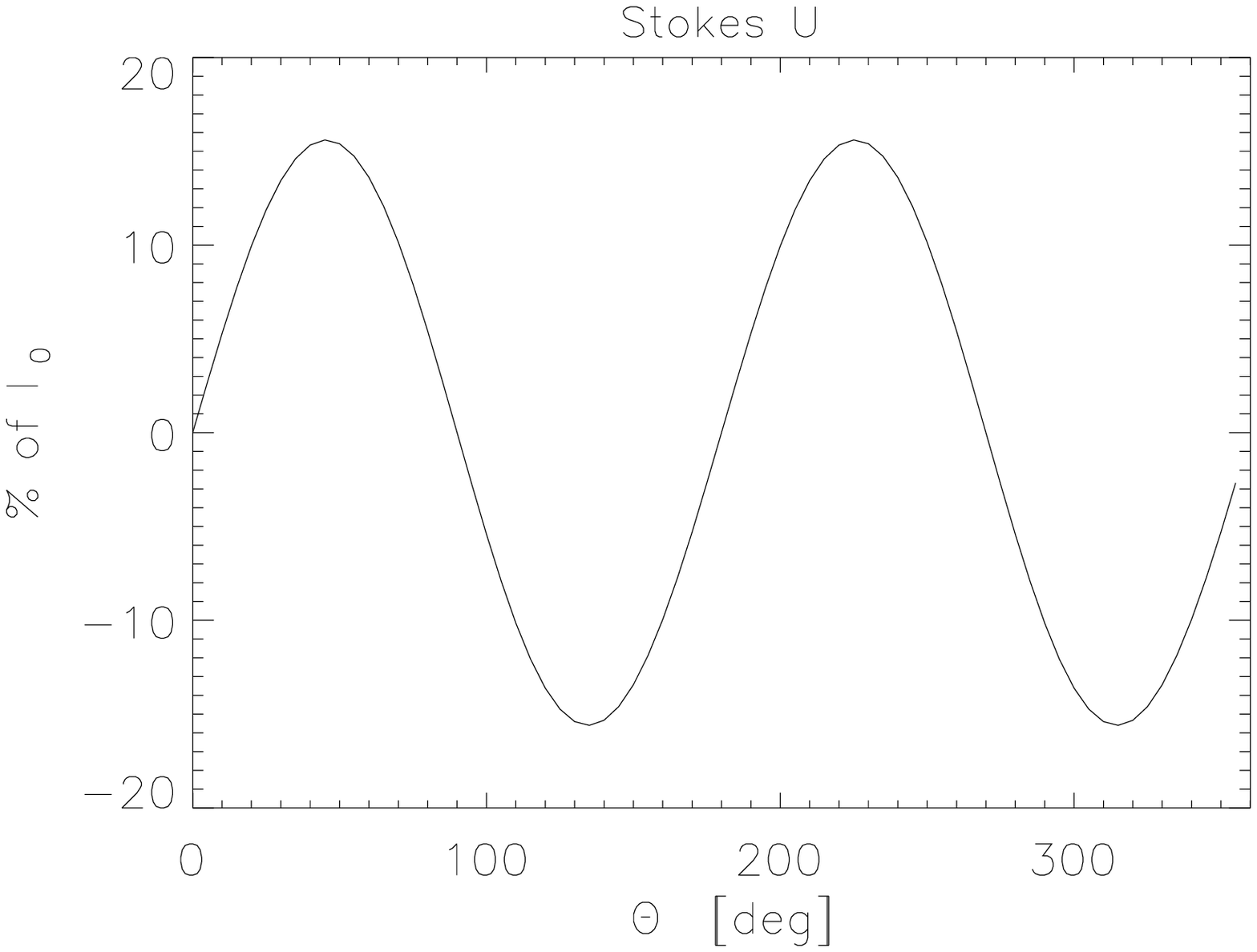}}
\resizebox{8cm}{!}{\includegraphics{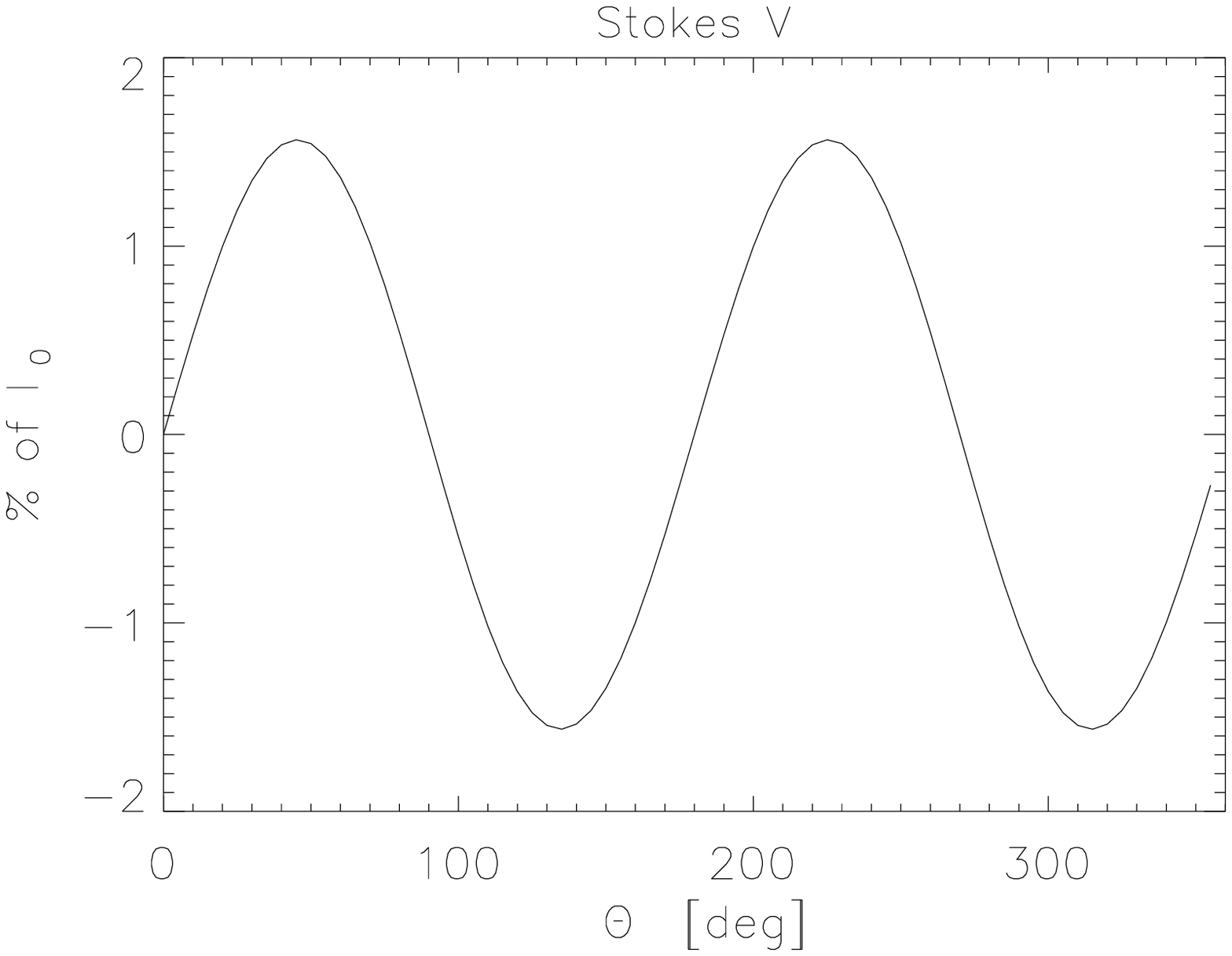}}
\caption{Stokes vector for a full rotation of the telescope calibration
unit on the first mirror C1 in the model. Date: 21.07.02, 12:00 UT, azimuth =
0$^\circ$. Mirror parameters $(n,k,d/\lambda,\nu)$ =
(0.30,3.63,0.094,1.54), taken from an ASP telescope calibration. The
window is excluded, $\theta_{tel,pol} = 90^\circ$.\newline
The array of polarizers is placed on top of C1 and rotated for an angle
$\theta \in [0,360^\circ ]$. The resulting matrix $(\bf T\rm)_{pol,C1}$
is applied on unpolarized sunlight with the intensity $\rm I_0$. The
transmitted intensity is given in $\%$ of the incident intensity. See
text for more details.\label{polonC1}}
\end{figure}
The non-normalized matrix $(\bf T\rm)_{pol,C1}$ with the inclusion of
the polarizer array on C1 is applied on unpolarized sunlight with the
intensity $\rm I_0$. The physical mirror paramters were taken from an
ASP telescope calibration and can be found in the caption. The variables in
eq. (\ref{inteq}) here are $\alpha = 0.897, \beta = -0.0079,$ and
$\gamma = 0.893$, calculated from
eqs. (\ref{mirr1})-(\ref{mirr2})\footnote{The calculated retardance of the
first mirror is $\delta = 175.16^\circ$, which motivates the neglection of the
off-diagonal terms $\propto \sin \delta$ in the lower right corner of
the mirror matrix.}. The total transmitted intensity drops to about 16
$\%$ of the incident intensity, which will require longer integration
times. Note that the position of the extrema in Stokes I (=extrema of
eq. (\ref{inteq})) are not directly related to the extrema of the polarizations
(dependent on angle $\theta_{pol}$).\\

From observational data taken with a calibration unit of similar design
in the infrared range (see Fig. \ref{polonc1img}) an additional
complication seems to be present. The array of polarizers does not cover
the full area of the mirror. There are strong hints that in the rotation
of the array an intensity modulation appears due to a cross-mounting holding
a folding mirror in the vacuum tank. This mirror reflects light out of
the beam path to the telescope guiding system. A rotation of the
structure of Fig. \ref{tel_cali} over this cross-like pattern of
non-neglectible extension leads to a variation in intensity of an order
similar to the one due to eq. (\ref{inteq}). It may be necessary to
remove this purely geometrical effect by a calculation of the area
shaded for each angle $\theta$ to obtain the information important for
the polarimetric properties.
\end{subsubsection}
\begin{subsubsection}{Additional possibilities for the telescope calibration}
There are some other methods to establish the instrumental polarization
of the telescope. They mainly differ in the way used to create the known
polarization input of the incoming beam.
\begin{itemize}
\item The usage of the presumably unpolarized sun light from
a non-magnetic region at disc center, corresponding to a Stokes vector
of (I,0,0,0). This has some disadvantages:
\begin{itemize}
\item The application of the telescope matrix on unpolarized light only
uses the first column. No direct control of the entries of the
cross-talk between polarization states is possible.
\item The window properties will most likely not be well determined, as
they are more important for the polarized states than intensity.
\end{itemize}
This input of unpolarized light should only be used in addition to other
data in the determination of the instrumental polarization, but is well
suited as a control value for predictions of the telescope model.
\item The usage of a sunspot with a magnetic field aligned parallel to
the LOS. The magnetic field lines are assumed to be
vertically upwards, with regard to the solar surface, in the center of
the umbra of a sunspot. Depending on the position of the spot on the
solar disk a point may contain field lines parallel to the LOS. The
emerging light of this point then is polarized only circularly,
corresponding to a Stokes vector of (I,0,0,V). This has the only
drawback to find a suited sunspot near disk center at
the day of the calibration measurements. As this coincidence can not be
planned, this possibility should also be seen as an addition.
\item The usage of mirror parameters measured in a laboratory from
samples. These samples always must be kept under the same
environment conditions as the telescope mirrors. This method seems to
have too many error sources to be used. The results for the
mirror parameters at the Dunn VTT strongly deviate from literature
values for aluminium coatings, which were measured in optical
laboratories. The mirrors are part of a complicated optical
setup, where not all polarization-sensitive elements in the beam path
may have been explicitly included in the telescope model. Most likely
this will lead to effective values for the physical parameters only weakly
connected to, for example, the actual refraction index of a single mirror.
\end{itemize}
\end{subsubsection}
\end{subsection}
\end{section}
\end{chapter}
\begin{chapter}{Observations with the ASP: evaluation of polarimetric data\label{polarimetric}}
\begin{minipage}{12cm}
''\it Forget about the ASP for a moment...\rm''\\
M.Collados Vera (2002, in a discussion)\\
\end{minipage}\\
$ $\\
The evaluation of two data sets from the ASP was performed prior to the
diploma thesis as preparation to the calibration of POLIS. It was
intended as a preliminary examination of the quality of the data, and to
get used to polametric data. All interpretations and conclusions drawn
stem from the limited knowledge of the author on the discussed
subjects\footnote{And may be wrong.}. Again here actual data from POLIS
was planned to be used, but this was impossible due to the delayed setup.\\

The section shall demonstrate the properties and some evaluation
possibilities of polarimetric data. Some of the effects mentioned in section
\ref{solmag} can be displayed.
\begin{section}{Measurement data: size and content}
The data sets available were a scan of 239 scan steps over a magnetic
active region with emergent flux, and a single sunspot with 120 scan
steps.
\begin{figure}[ht]
\psfrag{A}{\huge \hspace*{-.8cm}$\shortstack[c]{$\frac{U}{I_c}$ \\\vspace*{-.2cm}$[\%]$} $ }
\psfrag{B}{\huge \hspace*{-.5cm}$\shortstack[c]{$\frac{V}{I_c}$ \\\vspace*{-.2cm}$[\%]$}$}
\psfrag{C}{\huge \hspace*{-.5cm}$\shortstack[c]{$\frac{Q}{I_c}$ \\\vspace*{-.6cm}$[\%]$}$}
\psfrag{D}{\huge \hspace*{-.5cm}$\%$}
\psfrag{I}{\huge \hspace*{-.5cm}I$(\lambda)$}
\psfrag{U}{\huge \hspace*{-.5cm}U$(\lambda)$}
\psfrag{Q}{\huge \hspace*{-.5cm}Q$(\lambda)$}
\psfrag{V}{\huge \hspace*{-.5cm}V$(\lambda)$}
\psfrag{E}{\Large \hspace*{-.4cm}$\lambda \;[ nm]$}
\psfrag{F}{\Large \hspace*{-.4cm}$\lambda \; [nm]$}
\resizebox{16cm}{!}{\includegraphics{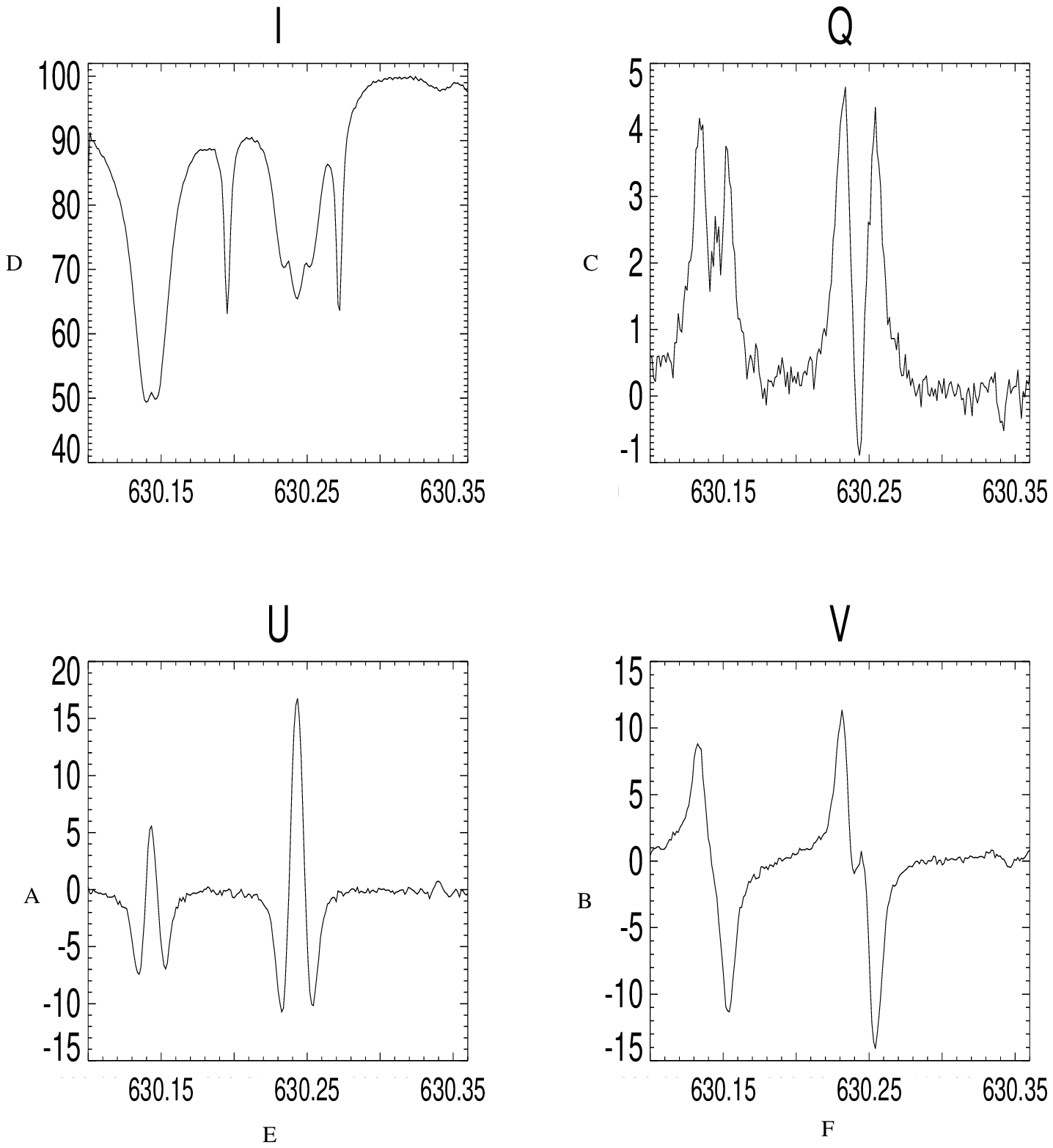}}
\caption[Stokes spectra at 630 nm]{Stokes spectra $(I,Q,U,V)\;
(\lambda)$ of the wavelength region around 630 nm. All profiles are
normalized by the continuum intensity $I_c$ calculated from
the spectral range around 630.30 nm.\newline
In Stokes I four spectral lines can be identified. The
sharp lines at 630.20 nm and 630.27 nm are telluric $O_2$-lines formed
in the atmosphere of the earth. They can be used for a calibration of
velocities, as they show no systematic Dopplershift due to any systematic
motion. The two FeI lines at 630.15 and 630.25 nm are of solar
origin. The line with the greater Land\'e factor shows a visible
splitting.\newline
The linear polarizations Q and U correspond to the case depicted in
Fig. \ref{classzeeman}, bottom right. \newline The observation direction is not
perpendicular to the magnetic field, so also a vertical component
is present as proven by the signal in Stokes V. These profiles were
taken from the data set of an active region and show a great amount of
linear polarization, while in the sun spot data set the magnetic field
is mainly vertical.\label{stokesprof}}
\end{figure}
The polarimetric data measured by the ASP is the intensity and the
polarization content for the wavelength range from about 630.0 nm to
630.4 nm (cp. table \ref{asp_prop}). Fig. \ref{stokesprof} gives the
profiles in one row of a data image from the active region. There are
four spectral lines included in the intensity spectrum of Stokes I:
\begin{itemize}
\item two telluric $O_2$-lines, which are formed in the atmosphere of
the earth at their rest wavelength. They can be used as reference for
absolute velocity calibrations or to measure the spectral dispersion per
detector pixel.
\item two FeI lines of solar origin. These may show Dopplershifts due to
relative motions, between earth and sun, the solar rotation and finally
motions of material in the photosphere of the sun. The displacement from their
rest wavelength gives the LOS velocity. In the profiles of
Fig. \ref{stokesprof} the magnetic field strength is sufficient to
produce a visible splitting of the line with the greater Land\'e factor
in the intensity profile. The intensities of the linear
polarizations Stokes Q and U give information on the horizontal field
component, while the signal in Stokes V is proportional to the LOS
component.
\end{itemize}
More information on the spectral lines can be found in table \ref{lines}.
Along the slit height corresponding to 74 arcsecs on the solar image 229
rows of (I,Q,U,V) profiles are available for each scan step. A reduction
to two-dimensional maps of selected parameters is shown in for example
Fig. \ref{poldata} and \ref{sunspot}.
\clearpage
\end{section}
\begin{section}{Evaluation of the Stokes V signal\label{stokesveval}}
\begin{figure}
\begin{minipage}{8cm}
\psfrag{A}{\Huge \hspace*{-.1cm}$\lambda_{zcr}$}
\psfrag{B}{\Huge \hspace*{-1.3cm}$\shortstack[c]{\vspace*{-.5cm}$\lambda
\; [nm]$}$} 
\psfrag{C}[B1][B1][1.4]{\Huge
\hspace*{-1.2cm}$\shortstack[c]{$\frac{V}{I_c}$ \vspace*{0.2cm}\\\vspace*{-1.cm}$[\%]$}$} 
\psfrag{D}{\Huge \hspace*{-1.2cm}a$_{blue}$}
\psfrag{E}{\Huge \hspace*{-.5cm}a$_{red}$}
\psfrag{F}{\Huge \hspace*{-.6cm}A$_{blue}$}
\psfrag{G}{\Huge \hspace*{-.4cm}A$_{red}$}
\psfrag{H}{\Huge \hspace*{-.5cm}0}
\psfrag{I}{\Huge \hspace*{-.8cm}-20}
\psfrag{J}{\Huge \hspace*{-.8cm}-40}
\psfrag{K}{\Huge \hspace*{-.6cm}20}
\psfrag{L}{\Huge \hspace*{-.7cm}40}
\resizebox{8cm}{!}{\includegraphics{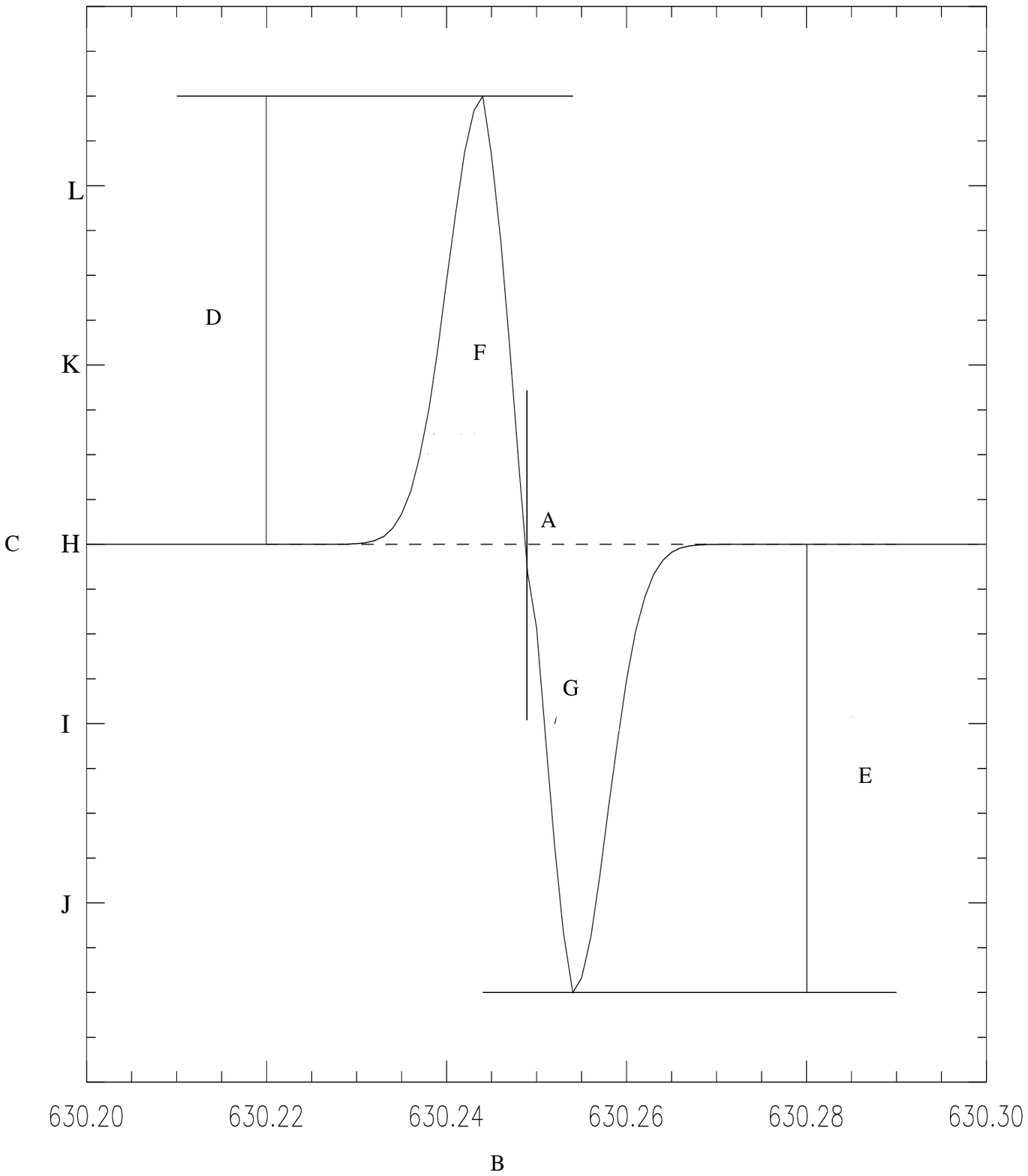}}
\end{minipage}
\begin{minipage}{8cm}
\caption[Stokes V profile]{Artificial Stokes-V profile for the FeI line
at 630.25 nm.\newline
The amplitudes $a_{blue/red}$ can be used for the construction of
magnetograms. The positions of minmum and maximum exchange when the
magnetic field lines are of oppposite direction. Subtraction of the
areas under the single peaks, $A_{blue/red}$, gives the net circular
polarization (NCP). The wavelength of the zero-crossing, $\lambda_{zcr}$,
is the position of the $\pi$-component, which is not influenced by the
magnetic field. The calculation of the zero-crossing velocity from
$\lambda_{zcr}$ gives the velocity of material inside the magnetic
field. This can be compared to the velocity derived from the
intensity profile, containing mostly the non-magnetic
environment.\label{stokesvprofil}}
\end{minipage}
\end{figure}
An accurate measurement is easiest for the Stokes V signal for two reasons:
the magnetic fields on the sun are usually more vertical than horizontal with
regard to the solar surface, thus they are more or less oriented
parallel to the LOS. Therefore the signal in V is usually the
strongest. Secondly, the polarization effects of mirrors or other
optical devices in the beam path are smaller for circular
polarization. Before the use of vector polarimeters so-called
magnetographs (cp. the introduction to chapter \ref{vecpola}) were
mainly in operation for solar magnetometry, which measure the amount of
circular polarization only.\\

Fig. \ref{stokesvprofil} gives an idealized Stokes V profile for a
magnetic sensitive spectral line. It is useful to define the following
quantities for an evaluation:
\begin{itemize}
\item $a_{red / blue}$ are the maximal amplitudes of the signal at the
position of the left or right circular component. A magnetic field with
the oppposite direction exchanges the position of the peaks, i.e. then
the minimum would preceed. The amplitude can be used for the
construction of magnetograms like in Fig. (\ref{poldata}) or
(\ref{sunspot}). They give an estimation of vertical strength and
the direction, sometimes called polarity, of the field.
\item $A_{red / blue}$ is the area under the single peaks. Their
subtraction results in the net circular polarization (NCP).
\item The zero-crossing wavelength $\lambda_{zcr}$ is the position
of the transition ($\pi$-component) of the spectral triplet, that is not
influenced by the magnetic field (cp. Fig. \ref{zeemanbild}).
\end{itemize}
\begin{subsection}{The zero-crossing velocity}
The last item, the zero-crossing wavelength, is important to establish
the LOS velocity of material in the magnetic field, $v_{zcr}$. This may
be different to the velocity derived from the intensity line profile,
which contains contributions from non-magnetic areas. Examples of
velocity maps are displayed in Fig. \ref{dopplergramm} for the sunspot
data. The pattern of velocities towards and away from the
observer, roughly anti-symmetric to the direction to the center of the
solar disk (not drawn), is caused by the Evershed effect. This is
assumed to be a radial outflow from the sunspot center, where the
different viewing angle on the spot changes the measured LOS
velocity\footnote{see Westendorp et al. (2001), \cite{weste}, or
Schlichenmaier and Schmidt (2000, \cite{schmidt1} and \cite{schmidt2})
for a detailed examination of LOS-velocities in a sunspot}.
\begin{figure}
\psfrag{A}{\large -1.5 km/s}
\psfrag{B}{\large \hspace*{-1.9cm}+1.5 km/s}
\psfrag{C}{\large -1.5 km/s}
\psfrag{D}{\large \hspace*{-1.9cm}+1.5 km/s}
\begin{minipage}{8cm}
\includegraphics[width=7cm,height=8.8cm]{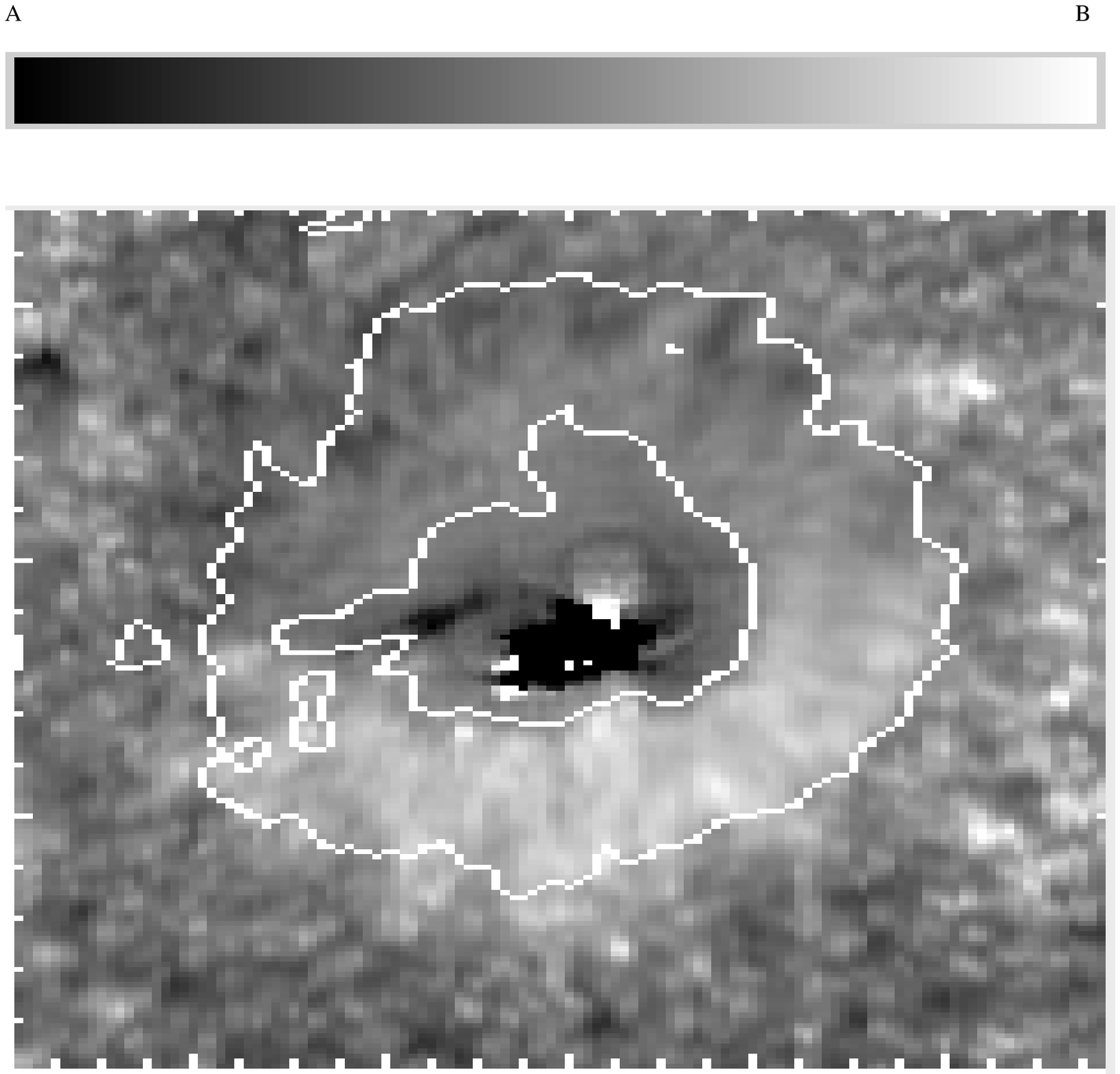}
\end{minipage}\hfill
\begin{minipage}{8cm}
\includegraphics[width=7cm,height=8.8cm]{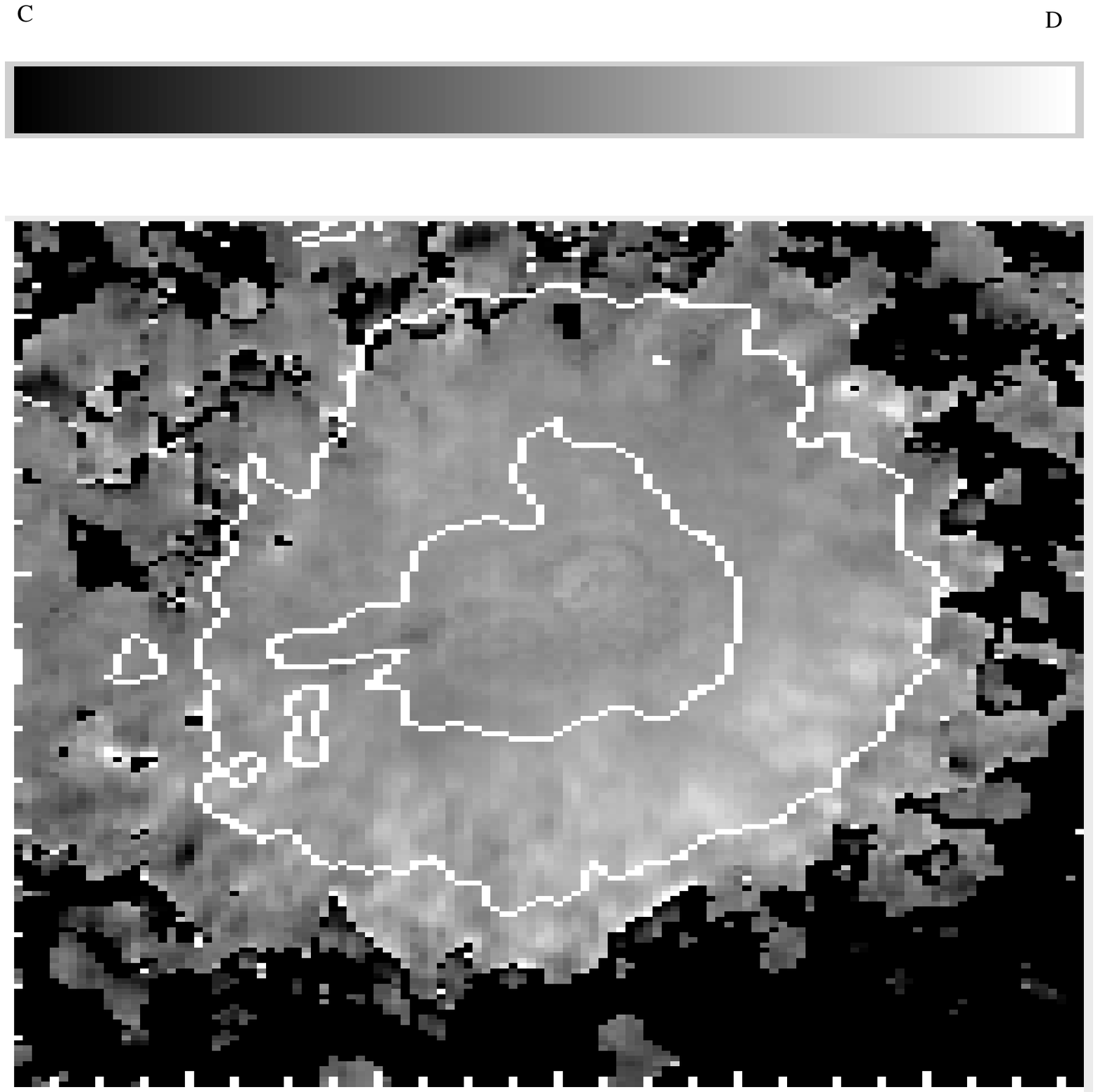}
\end{minipage}
\caption[Dopplergram and zero-crossing map of the sunspot]{Dopplergram
(\it left\rm) and zero-crossing map (\it right\rm) of the sun spot from
Fig. \ref{sunspot}. The Doppler shift caused by the relative movement of
earth and sun and the sun rotation has been subtracted.
\newline
Outside the spot single granulation cells can be identified, in which
hot material is rising, then cooling and sinking down at their
borders. The anti-symmetrical pattern of motions towards (lower
hemisphere) and away from the observer (upper hemisphere) in the spot is
caused by the Evershed effect. The radial outflow of material from the
spot center has an effective LOS component depending on the azimuthal
position in the spot. For the area in the central umbra, where the
line profile splits up, the evaluation method with a parabola fit is not
suited, as the artifacts in the left panel display.\label{dopplergramm}
The zerocrossing map of the velocity in the magnetic
field shows the same general structure. Here the central umbra with
strong Stokes V signals presents no difficulties for the calculation of
the zero-crossing wavelength. In the uniform black area outside the spot
no zero-crossing can be established (insufficient V signal). The magnetic
field extends over the boundary of the penumbra in the continuum (outer
white line). The limit for the  evaluation of a Stokes V profile was a
minimal polarization content of 1 $\%$, which is exceeded in the outer
regions as well.}
\end{figure}
The diagnostic importance of the zero-crossing velocity is shown in
Fig. \ref{veldiff}. The difference between the LOS velocity from the
intensity profile, $v_{dop}$, and $v_{zcr}$ displays some features:
\begin{itemize}
\item Single magnetic elements outside the spot may show up as well as down
flows relative to their environment.
\item The Evershed effect can still be seen in the two images, which
give the difference of the velocities, $v_{dop}-v_{zcr}$, for $v_{zcr} >
v_{dop}$ and $v_{zcr} < v_{dop}$ separately. Above the spot the 
filling factor, i.e. the fraction of area covered by magnetic fields,
can assumed to be 100 $\%$. Therefore no velocity differences should be
seen at all, as all material is inside magnetic fields. Single
unresolved, concentrated magnetic structures with a small filling factor
inside a detector element would contribute to the Stokes V signal, but only
little to the intensity due to the small area covered. If these
structures have a significantly differing mass velocity, they may cause
an additional shift of the zero-crossing wavelength relative to the
intensity line core. The clearly visible remaining velocity differences
therefore hint to unresolved magnetic elements in the penumbra.
\item The rectangle in the left image highlights a substructure inside
the penumbra. In an extension of a brightening in the continuum
intensity a radially oriented region shows $v_{zcr}$ to be greater than
the doppler velocity by about 1 km/s. Such a feature can be explained by
the model of single flux tubes embedded in the background field of a sun
spot (Schlichenmaier, \cite{schliche}, 1997). This model predicts nearly
horizontal flux tubes in the outer penumbra, inside which flows of up to
15 km/s can appear. If one assumes the remaining velocity difference
 to be caused by a horizontal flow in a single flux tube, and a viewing
angle of about 10$^\circ$, the horizontal speed would need to be here
app. 6 km/s greater than in its environment.
\end{itemize}
\begin{figure}
\psfrag{E}{\large -1.5 km/s}
\psfrag{F}{\large \hspace*{-1.9cm} 0 km/s}
\psfrag{G}{\large 0 km/s}
\psfrag{H}{\large \hspace*{-1.9cm} 1 km/s}
\includegraphics[width=7cm,height=8.8cm]{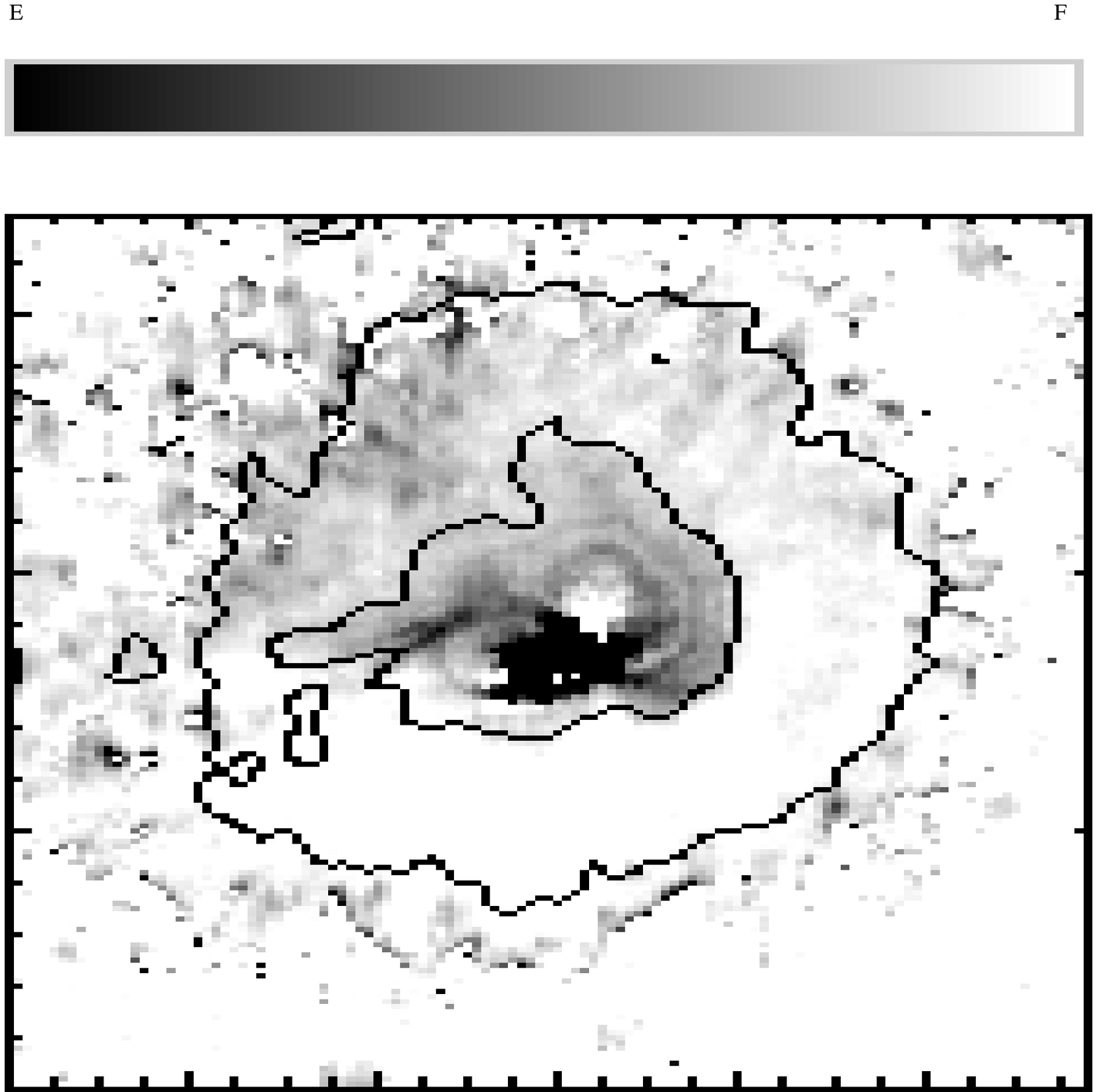}
\hfill
\includegraphics[width=7cm,height=8.8cm]{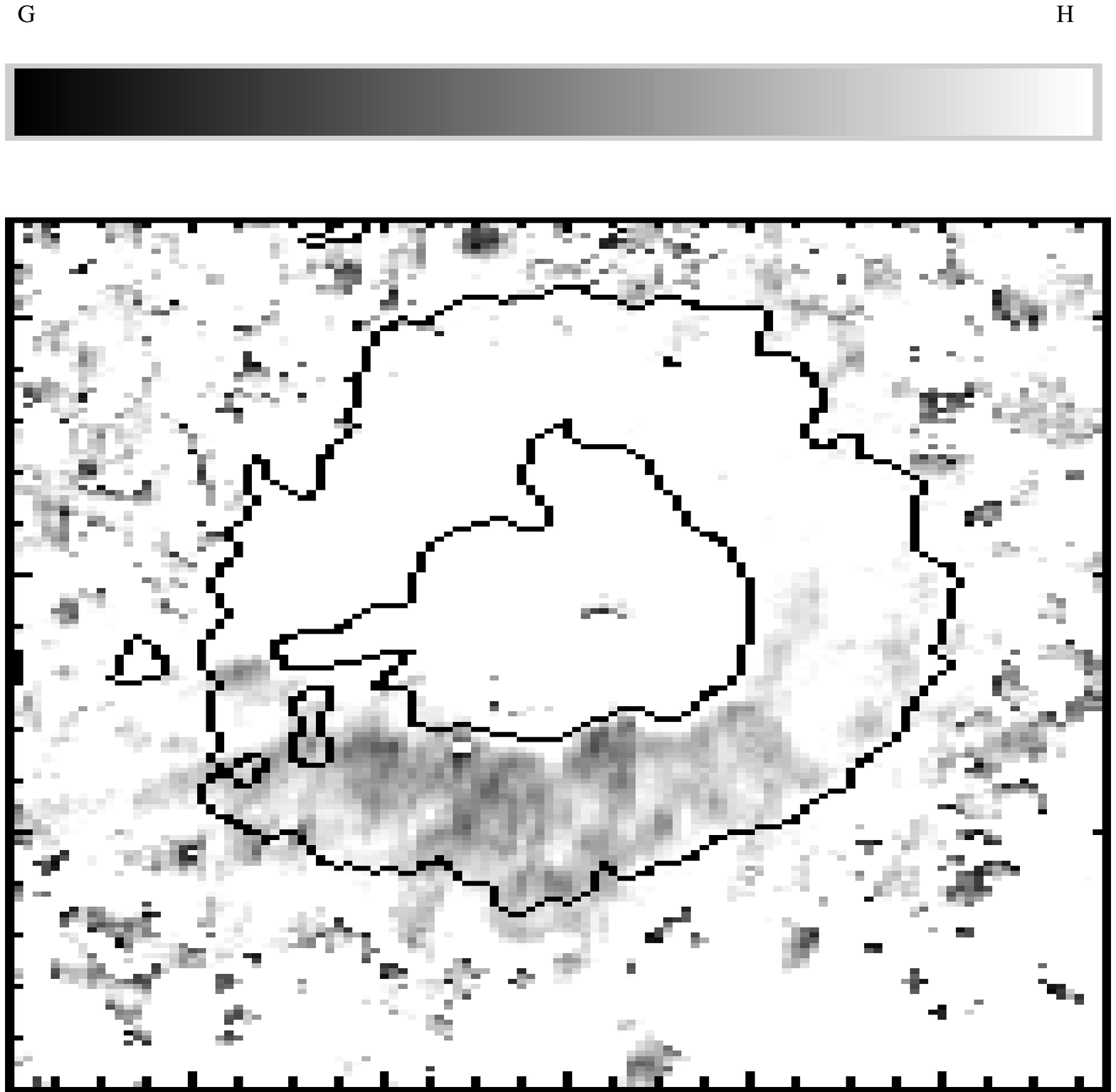}
\caption[Velocity difference $v_{dop}-v_{zcr}$]{The difference between
dopplergram and zero-crossing velocity from Fig. \ref{dopplergramm},
(\it left\rm) $v_{zcr} > v_{dop}$, (\it right\rm) $v_{zcr} < v_{dop}$.\newline
Outside the spot the velocity in isolated magnetic elements shows up as
well as down flows relative to the non-magnetic environment. The Evershed
effect remains even after the subtraction of both images. The white
rectangle marks a feature in the outer penumbra, where in an extension
of a brightening in the intensity a radial oriented region with a
velocity difference of about 1 km/s can be seen.\label{veldiff}}
\end{figure}
\end{subsection}
\begin{subsection}{Net circular polarization}
The area under the single wings of the Stokes V profile, $A_{blue}$ and
$A_{red}$, should be equal for a static atmosphere in local thermal
equilibrium. Fig. \ref{ncpmap} displays, that this is very often not
the case in the sunspot. In the upper hemisphere the complete penumbra
shows net circular polarization (NCP), i.e. an integration
along the wavelength dimension leads to a residual value not equal to zero. The
NCP value changes abruptly at the border of penumbra and umbra from the
continuum image, in difference to the magnetogram, where the transition
is smooth. This behaviour can be also be explained by the model of
moving flux tubes on the background field of the sunspot
(Schlichenmaier, \cite{schliche}, 1997, and Schlichenmaier et al.,
1998). The existence of more horizontal, isolated flux tubes in
the outer penumbra, in which the plasma moves radially outwards with a
speed of about 15 km/s, leads to gradients in either the velocity and
the magnetic field strength along the LOS. If gradients in both
environment parameters are present, the produced Stokes V signal has a
non-vanishing NCP. Inside the umbra no isolated elements appear.

The particular shape of the NCP, being roughly symmetric to an axis
from the top to the bottom of the image, is predicted by the thesis of
M\"uller, \cite{mueller}, 2001. In this thesis synthetic NCP maps of
axialsymmetric sunspots for different spectral lines were examined,
which were constructed by placing the final results of the moving tube
simulations into a three-dimensional background field. The calculation
of the NCP of the Stokes V profile along the LOS around the spot
resulted in a symmetric shape for the iron line at 630.25 nm. For
another iron line at 1564 nm the NCP should be anti-symmetric. This
could be confirmed by observations with the Tenerife Infrared Polarimeter
(TIP) (Schlichenmaier et al., 2001).
\begin{figure}
\begin{minipage}{8cm}
\psfrag{A}{\large -12 pm}
\psfrag{B}{\large \hspace*{-1.5cm}+12 pm}
\includegraphics[width=6.7cm,height=8.8cm]{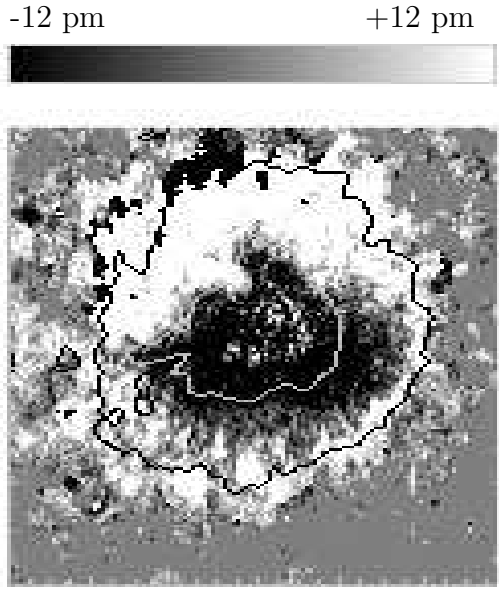}
\end{minipage}
\begin{minipage}{8cm}
\caption[NCP map of the sunspot]{Map of the net circular polarization
(NCP). The value is calculated as in Schlichenmaier et al. (2001), NCP = $\int
V(\lambda) d\lambda$, with V in $\%$ of $I_c$.\newline
The NCP gives the difference between the areas $A_{red}$ and $A_{blue}$,
if the V signal is zero outside the magnetic sensitive spectral line. If
the NCP is not equal to zero for the whole profile, the Stokes V signal
is not fully anti-symmetric. The NCP apruptly changes at the border
bewtween umbra and penumbra in the upper hemisphere. This indicates quite
different environment conditions, mostly in material velocities, as the
magnetogram in Fig. \ref{sunspot} shows a smooth transition.\label{ncpmap}}
\end{minipage}
\end{figure}
\end{subsection}
$ $\\
To give one additional example of polarimetric data, Fig. \ref{poldata}
shows some maps of the other data set taken with the ASP. This active
region is dominated by the increased appearance of opposite
polarities, i.e. magnetic fields with opposite directions, opposite to
the sun spot data with mainly a single polarity. A time series
was taken for the area in the center of the image, where the dopplergram
shows the disturbation of the granulation pattern by the magnetic
fields, to study the evolution of this emerging magnetic flux. A
complete evaluation of the data set requires an inversion. An extended
discussion of the data can be found in the report on the
'Hauptpraktikum',\cite{beck}.\\
\end{section}
$ $\\
\fbox{\fbox{\parbox{16cm}{Summary:\newline
Polarimetric data of the Stokes vector allows to calculate the
vector magnetic field, i.e. field strength and direction. The
zero-crossing wavelength gives the velocity of material inside magnetic
elements, which may differ from the Doppler velocity of the intensity
profile. The NCP hints to gradients in velocity and field strength. It
can be used to test theoretical models. The environment conditions in
umbra and penumbra of sunspots are different. Last, but not least: every
interpretation of polarimetric data rests on the assumption, that the
effects are of solar origin only. The calibration of the data should be
better than the order of the observed effect, especially for the correct
interpretation of weak polarization signals.}}}
\begin{figure}
\begin{minipage}{8cm}
\psfrag{C}{\large 25$\% \cdot I_{max}$}
\psfrag{D}{\large \hspace*{-2cm}100$\% \cdot I_{max}$}
\includegraphics[width=8cm,height=5.8cm]{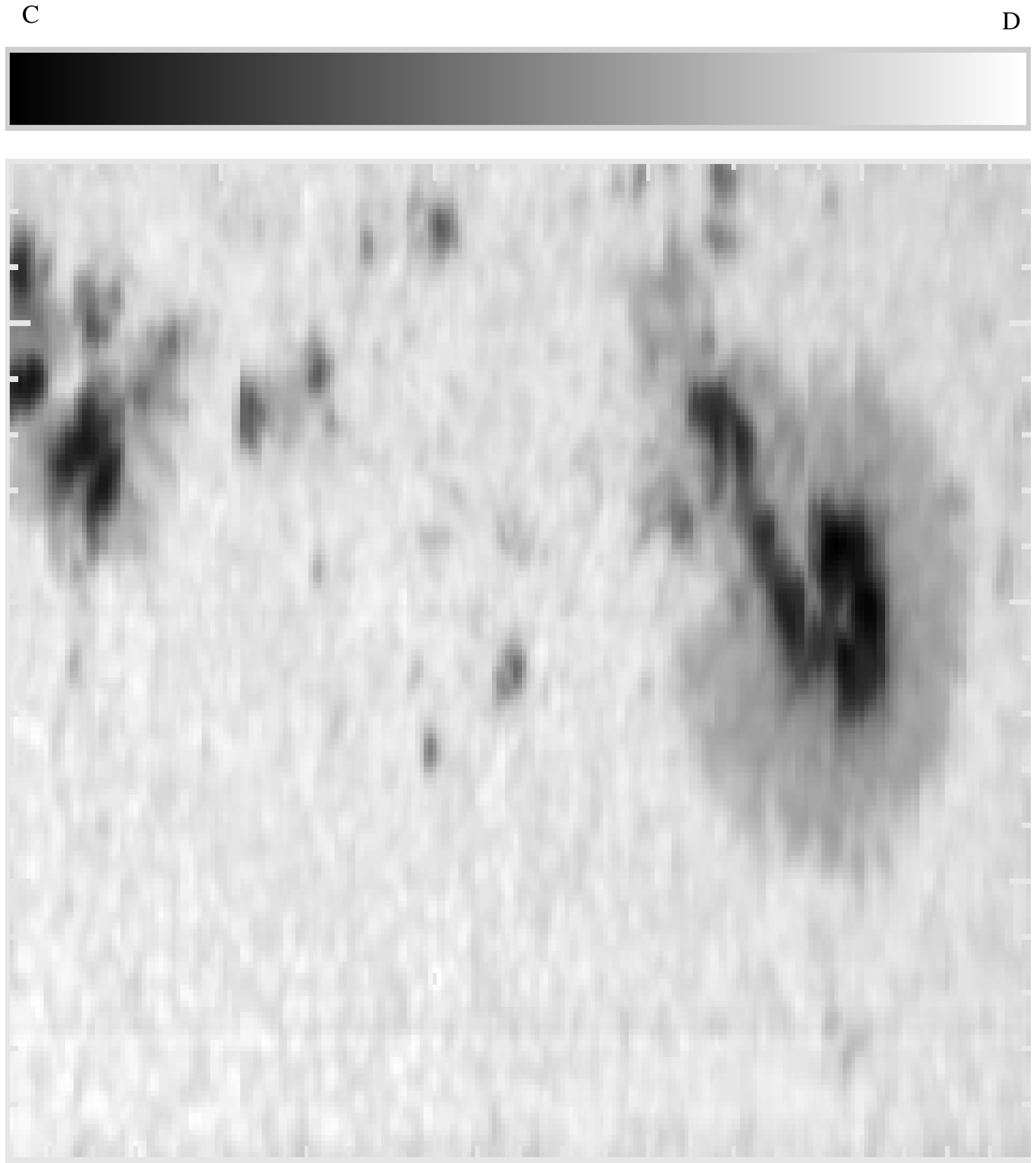}
\end{minipage}
\begin{minipage}{8cm}
\caption[Different maps of the active region]{Data set of the ASP of an
active region, taken on 29.09.00 (courtesy W.Schmidt), preliminary
evaluation by the author. The image area is 90'' x 63'', consisting of
240 steps of 0.375'' stepwidth. The spectral line is FeI at 630.25
nm.\vspace*{1cm}\newline 
The continuum intensity drops to 25 $\%$ of the maximum intensity in the
sunspots and pores. The spot itself can be divided in dark umbra and
brighter penumbra. In comparison with Fig. \ref{sunspot} the minimal
intensity is higher, while the Stokes-V amplitude is lower.\label{poldata}}
\end{minipage}\\
\begin{minipage}{8cm}
\psfrag{E}{\large -17$\% \cdot I_{mean}$ }
\psfrag{F}{\large \hspace*{-2cm}+16$\% \cdot I_{mean}$}
\includegraphics[width=8cm,height=5.8cm]{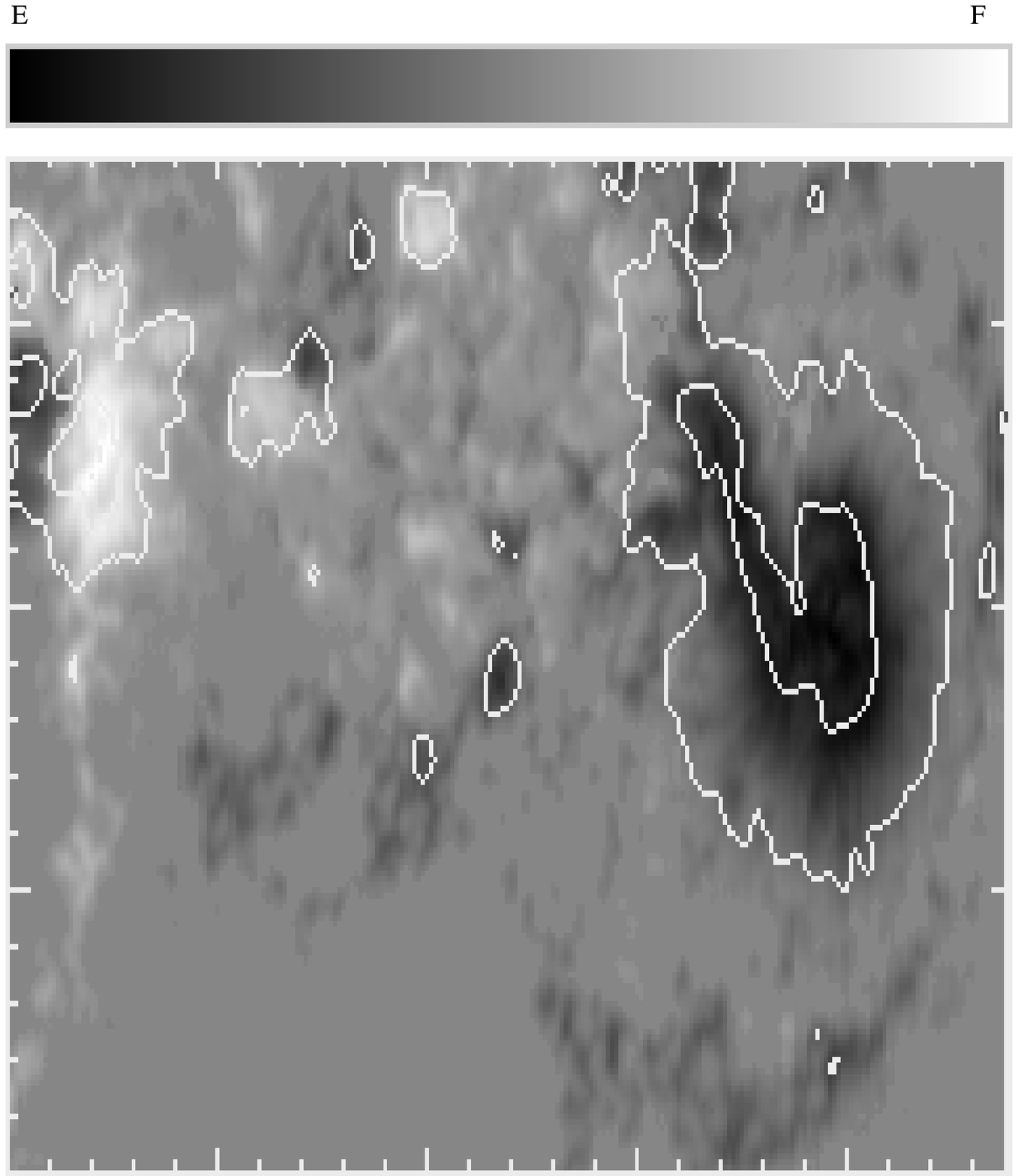}
\end{minipage}
\begin{minipage}{8cm}
\normalsize The signed amplitude of the Stokes-V signal in $\%$ of the mean
intensity with superimposed borders of the umbra and penumbra.\\
The amplitude is proportional to the magnetic field strength, while its sign
distinguishes the field direction to or away from the observer. Some of
the pores and the sunspot at the left border of the area can be seen to
be of opposite polarities. The region in the center of the image shows
additional magnetic activity, which is not visible in the continuum
intensity above. 
\end{minipage}\\
\begin{minipage}{8cm}
\psfrag{A}{\large +1.5 km/s}
\psfrag{B}{\large \hspace*{-1.6cm}-1.5 km/s}
\includegraphics[width=8cm,height=5.8cm]{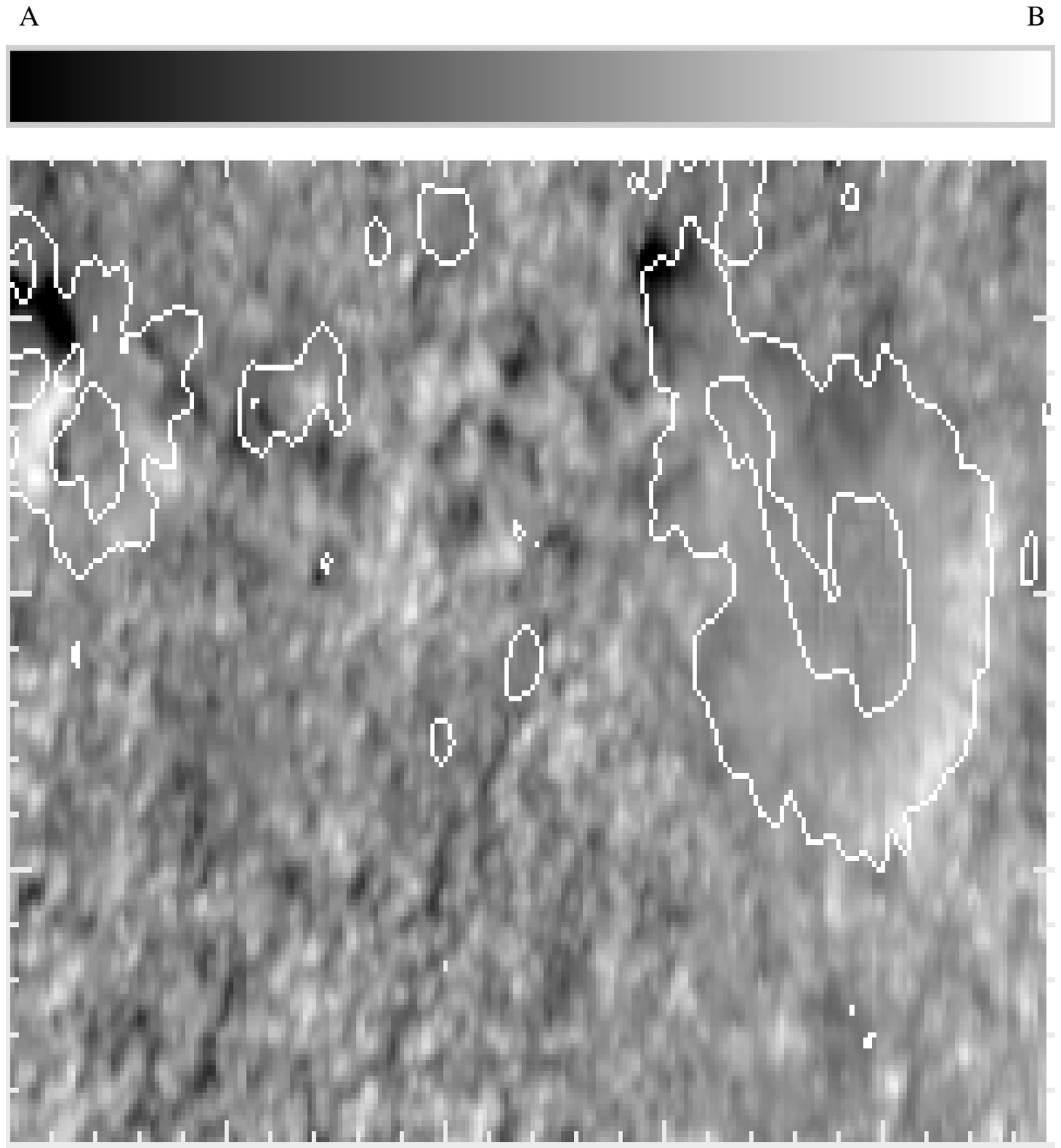}
\end{minipage}
\begin{minipage}{8cm}
A Dopplergramm of the LOS velocities relative to the solar
surface with superimposed borders of the umbra and penumbra.\\
The velocity pattern outside the strong magnetic fields reflects the
movements of the granulation, hot material rising and cooler
sinking. The \it Evershed flow \rm can be seen in the greater
sunspot. The granulation pattern between the two spots is slightly
distorted, presumably due to the presence of magnetic field in that area.
\end{minipage}\\
$ $\\
$ $\\
\begin{minipage}{8cm}
\includegraphics[width=7.05cm,height=4.7cm]{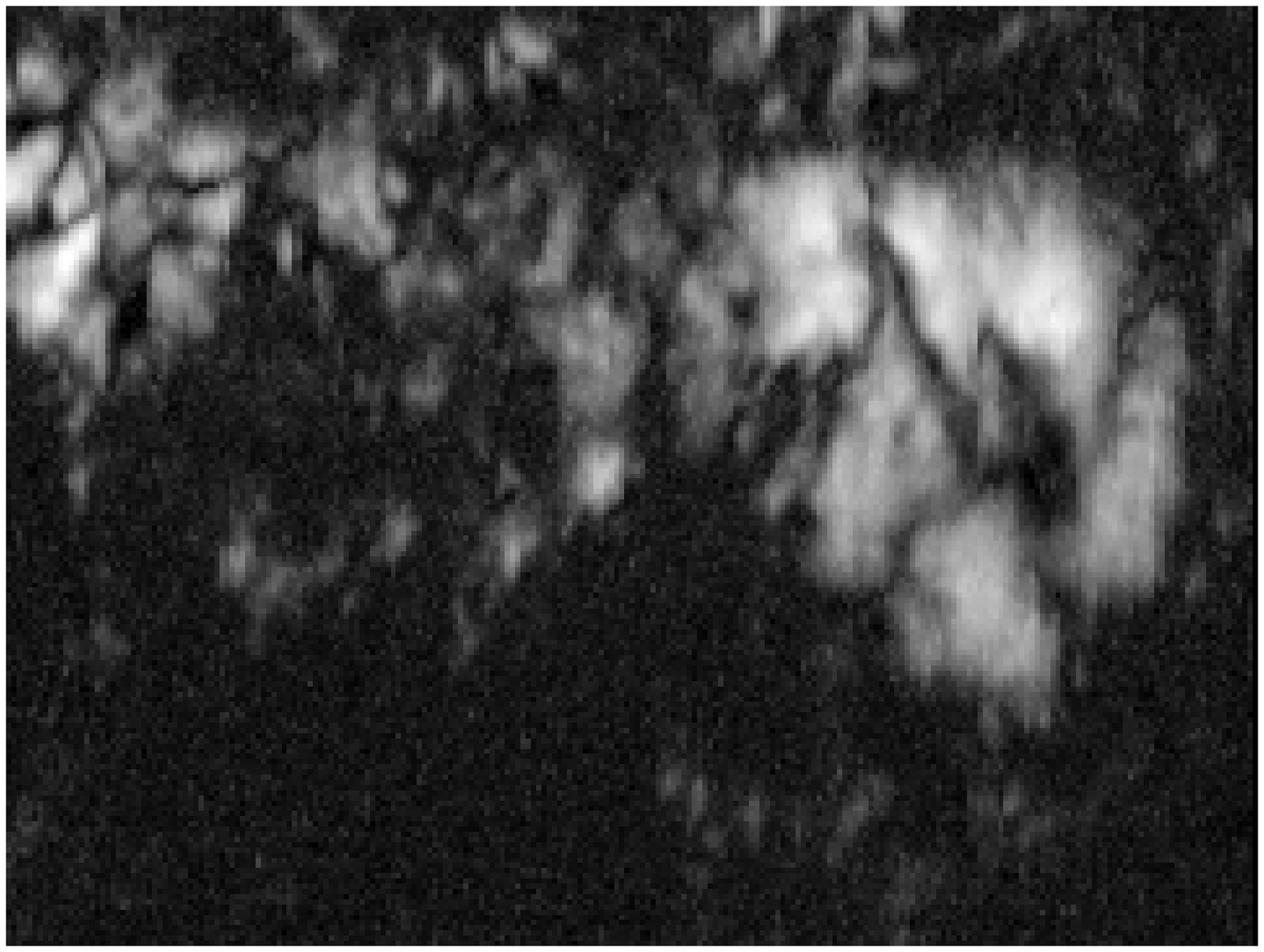}
\end{minipage}
\begin{minipage}{8cm}
A logarithmic map displaying $\log (\int | \mbox{Q}(\lambda) |
d\lambda )$.\\
The Stokes parameter Q measures linear polarization,
which corresponds to more horizontal magnetic fields. These are 
strongest in the area between the two pores of different polarity at the
left border. This can be explained by field lines connecting the
oppposite polarities in a loop, where at the top of the loop the field is
horizontal. The cross-like structure in the greater sunspot is caused by
the axialsymmetric field of a spot, where the angle between
LOS and magnetic field lines is the same for different positions around
the spot center.
\end{minipage}
\end{figure}
\end{chapter}

\addcontentsline{toc}{chapter}{Bibliography}
\begin{appendix}
\begin{chapter}{The Mueller matrices\label{muellermat}}
In the Stokes formalism the effects of optical active elements can be
formulated as matrices \bf M \rm $\in \mathbbm{R}^{4x4}$, which are
applied on the Stokes vector \bf S \rm $\in \mathbbm{R}^4$ as defined in
equation (\ref{stokesdef}). The matrix elements are determined from the
eletrical field components before ($E_x, E_y$) and after ($E_x^\prime,
E_y^\prime$) the transmission through the element from the equation
\begin{eqnarray}
 \bf S\rm^\prime = \bf M \rm \cdot \bf S\rm \label{gleichung} \;\; .
\end{eqnarray}
For this calculations an equivalent definition of the Stokes parameters
with complex field amplitudes is very useful. It is given by
substituting equation (\ref{efield}) with
\begin{eqnarray}
E_x ( t )&=& E_{x,0} \cdot \exp \left[ \; i\;(\omega t +
\delta_x)\;\right]\\
E_y ( t )&=& E_{y,0} \cdot \exp \left[ \; i\;(\omega t +
\delta_y)\;\right]\nonumber
\end{eqnarray}
and changing the defintion of the Stokes parameters to 
\begin{eqnarray}
\label{stokesdefkomplex}
S_0 & = & E_x E_x^* + E_y E_y^*\\
S_1 & = & E_x E_y^* - E_y E_x^* \nonumber\\
S_2 & = & E_x E_y^* + E_y E_x^* \nonumber\\ 
S_3 & = & i (\; E_x E_y^* - E_y E_x^* \;) \nonumber \;\;.
\end{eqnarray}
An explicit calculation and also the proof of equivalency can be found
in \cite{collett}. Here shall only the results for the different optical
elements being used be presented.
\begin{section}*{The polarizer}
A polarizer is an optical element with a transmission dependent on the
oscillation direction of the electrical field and can be described by
\begin{eqnarray}
\label{efieldpol}
E^\prime_x = p_x \cdot E_x \nonumber \;\;\\
E^\prime_y = p_y \cdot E_y \nonumber \;,
\end{eqnarray}
with $0 \le p_x, p_y \le 1$.\\
Inserting the electrical field amplitudes into equation
(\ref{gleichung}) one obtains the Mueller matrix of a linear polarizer as
\begin{eqnarray} \bf M \rm _{pol} = \frac{1}{2}
\begin{pmatrix} p_x^2 + p_y^2 & p_x^2 - p_y^2 & 0 & 0 \cr p_x^2 - p_y^2
& p_x^2 + p_y^2 & 0 & 0 \cr 0 & 0 & 2 \cdot p_x\cdot p_y & 0 \cr 0 & 0 & 0 & 2
\cdot p_x\cdot p_y \cr 
\end{pmatrix} \; .
\end{eqnarray}
\end{section}
\begin{section}*{The retarder}
A retarder has different propagation velocities for the two directions
(x,y). It changes the relative phase between the field
components. Here is
\begin{eqnarray}
E^\prime_x = e^{+ i \Phi /2} \cdot E_x \;\;\nonumber\\
E^\prime_y = e^{- i \Phi /2} \cdot E_y \nonumber\; ,
\end{eqnarray}
and the Mueller matrix resulting is
\begin{eqnarray} \bf M \rm_{ret} = \begin{pmatrix} 1 & 0 & 0 & 0 \cr 0 & 1 &
0 & 0 \cr 0 & 0 & \;\;\; \cos \Phi  & \sin \Phi  \cr 0 & 0 & -\sin \Phi 
&  \cos \Phi  \cr 
\end{pmatrix} \; .
\end{eqnarray}
\end{section}
\begin{section}*{The rotator }
A rotator rotates (as its name says) the field components by a fixed
angle $\theta$. Here is 
\begin{eqnarray}
E_x^\prime = \;\;\, E_x \cdot \cos \theta  + E_y \cdot \sin \theta\;\;\nonumber\\
E_y^\prime = -E_x \cdot \sin \theta  + E_y \cdot \cos \theta \; ,\nonumber
\end{eqnarray}
and the matrix is
\begin{eqnarray} \bf M \rm _{rot}(2 \theta) \rm = 
\begin{pmatrix} 1 & 0 & 0 & 0 \cr 0 & \;\;\;\cos 2\;\theta  & \sin
2\;\theta  & 0 \cr 0 & -\sin 2\;\theta  & \cos 2\;\theta 
& 0 \cr 0 & 0 & 0 & 1 \cr 
\end{pmatrix} \; .\label{mrot}
\end{eqnarray}
\end{section}
\begin{section}*{Rotated optical elements}
The Mueller matrix of the rotator is especially important as it also
allows one to determine the effects of rotated optical elements, which
often occur in optical designs. The rotation can be included by two
rotator matrices around the optical element itself by
\begin{eqnarray}
\bf M\rm_{elem}(\theta) = \bf M\rm_{rot}( - 2 \theta)\cdot \bf
M\rm_{elem}\cdot \bf M\rm_{rot} ( 2 \theta ) \nonumber \; .
\end{eqnarray}
One example is the rotating retarder used by either the ASP or POLIS :
\begin{eqnarray}
\bf M \rm _{ret}(\theta) = \begin{pmatrix} 1 & 0 & 0 & 0 \cr 0 & c^2
+s^2d & sc(1-d) & -se \cr 0 & sc(1-d) & s^2 + c^2d & ce \cr 0 & se & -ce
& d \cr
\end{pmatrix} \; ,\label{retrot}
\end{eqnarray}
with $c = \cos 2 \theta , s = \sin 2 \theta , d = \cos \delta$ and $e =
\sin \delta$. $\delta$ is the retardance of the modulator and
$\theta$ the angle between its fast axis and the x-direction, as always
counter-clockwise when looking at the light source\footnote{I really
like that definition.}.
\end{section}
\begin{section}*{Polarizing beam splitter}
The Mueller matrix of the polarizing beam splitter is\footnote{No derivation
given.}
\begin{eqnarray}
\bf M \rm^\pm _{beam splitter} = \frac{r^\pm}{2}\begin{pmatrix} 1 & \pm 1 & 0
& 0 \cr 1 & \pm 1 & 0 & 0 \cr 0 & 0 & 0 & 0 \cr 0 & 0 & 0 & 0 \cr
\end{pmatrix} \; .\label{beamsplit}
\end{eqnarray}
The matrix multiplication with the rotated retarder above, application on a
Stokes vector (I,Q,U,V) and the selection of the first vector entry
gives the intensities of the two beams labelled + respectively - as
\begin{eqnarray}
I^\pm (\theta,\delta) = \frac{r^\pm}{2} \left\{I \pm Q \cdot ( c^2_{
2\;\theta} + s^2_{2\;\theta}\; c_\delta) \pm U \cdot
s_{2\;\theta}\; c_{2\;\theta}(1-c_\delta) \mp V \cdot
s_{2\;\theta}\;s_\delta\right\}\label{iplusminusa} \; ,
\end{eqnarray}
on which the measurement scheme relies.
\end{section}
\end{chapter}
\begin{chapter}{Alignment of the polarimeter calibration unit
\label{aligncalib}}
In the polarimeter calibration unit two elements have to be aligned
correctly, i.e. the direction of the transmission axis of the linear
polarizer, and the direction of the fast axis of the retarder have to be
established. The measurement setup used is shown in
Fig. \ref{alignsetup}. In addition to the optical elements under
examination a second polarizer as analyzer, a light source, and an
intensity detector, in this case a photo diode, are needed. The usage of
a wavelength filter for 630 nm is optional to increase the resemblance
to the conditions of operation in Tenerife. The field stop reduces the
stray light.
\begin{figure}[ht]
\caption{Measurement setup for the alignment of the polarimeter
calibration unit.\newline
The optical train consists of an artifical light source, a field stop,
a wavelength filter for 630 nm (optional), the polarimeter calibration
unit, a second polarizer as analyzer, and a photo diode as intensity
detector. The elements of the calibration unit can separately be rotated
clockwise, when looking at the light source, for
180$^\circ$.\label{alignsetup} }
\begin{center}
\psfrag{A}{\large lamp}
\psfrag{G}{\large \hspace*{-1cm}photodiode}
\psfrag{H}{\large \hspace*{-.5cm}polarizer}
\psfrag{B}{\large \hspace*{-.5cm}field stop}
\psfrag{E}{\large \hspace*{-.5cm}retarder}
\psfrag{D}{\large \hspace*{-2.5cm}calibration unit}
\psfrag{F}{\large \hspace*{-1cm}analyzer}
\psfrag{C}{\large filter}
\resizebox{12cm}{!}{\includegraphics{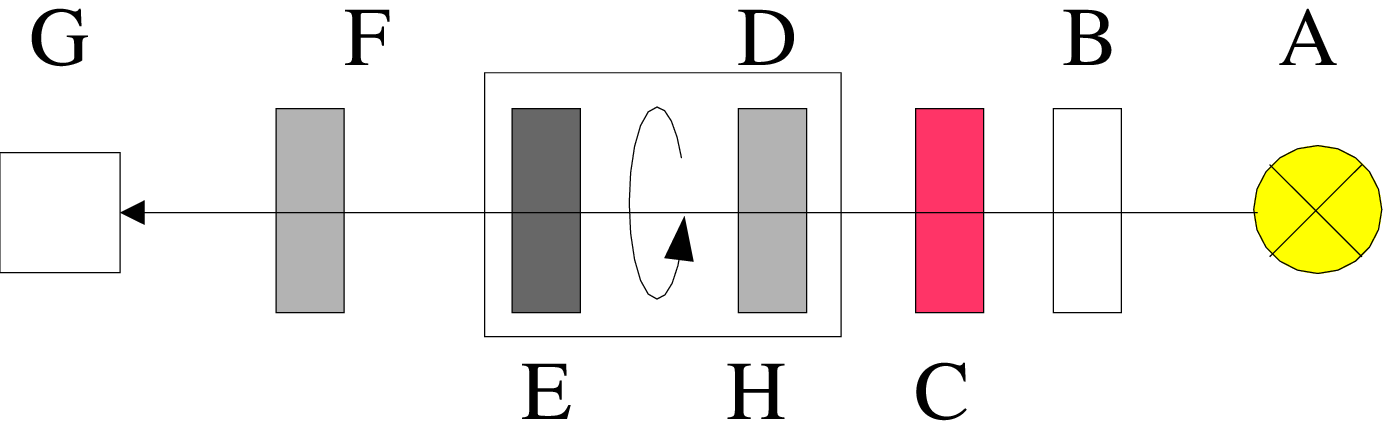}}
\end{center}
\end{figure}

The axis of the polarizer is established with the use of the second
polarizer, the analyzer. The method introduced in Fig. \ref{rotate_pol}
works without any axes known beforehand.  Note that all position
angles derived may be off by 90$^\circ$, i.e. the axes marked on the
polarizers may be the blocking instead of the transmission axes, and the
retarder may get the slow instead of the fast axis marked. This can be
simply cross-checked with a light source of known polarization, for
example reflection under the Brewster angle.
\begin{figure}
\begin{minipage}{8cm}
\psfrag{A}{\Huge \hspace*{-1.8cm}polarizer}
\psfrag{B}{\Huge \hspace*{-.8cm}analyzer}
\psfrag{C}{\Huge axis}
\psfrag{E}{\Huge to photodiode}
\psfrag{D}{\Huge $\alpha$}
\psfrag{F}{\Huge $2\, \alpha$}
\psfrag{G}{\Huge (1)}
\psfrag{H}{\Huge (2)}
\psfrag{I}{\Huge }
\psfrag{J}{\Huge \hspace*{-.8cm}$180^\circ$}
\resizebox{8cm}{!}{\includegraphics{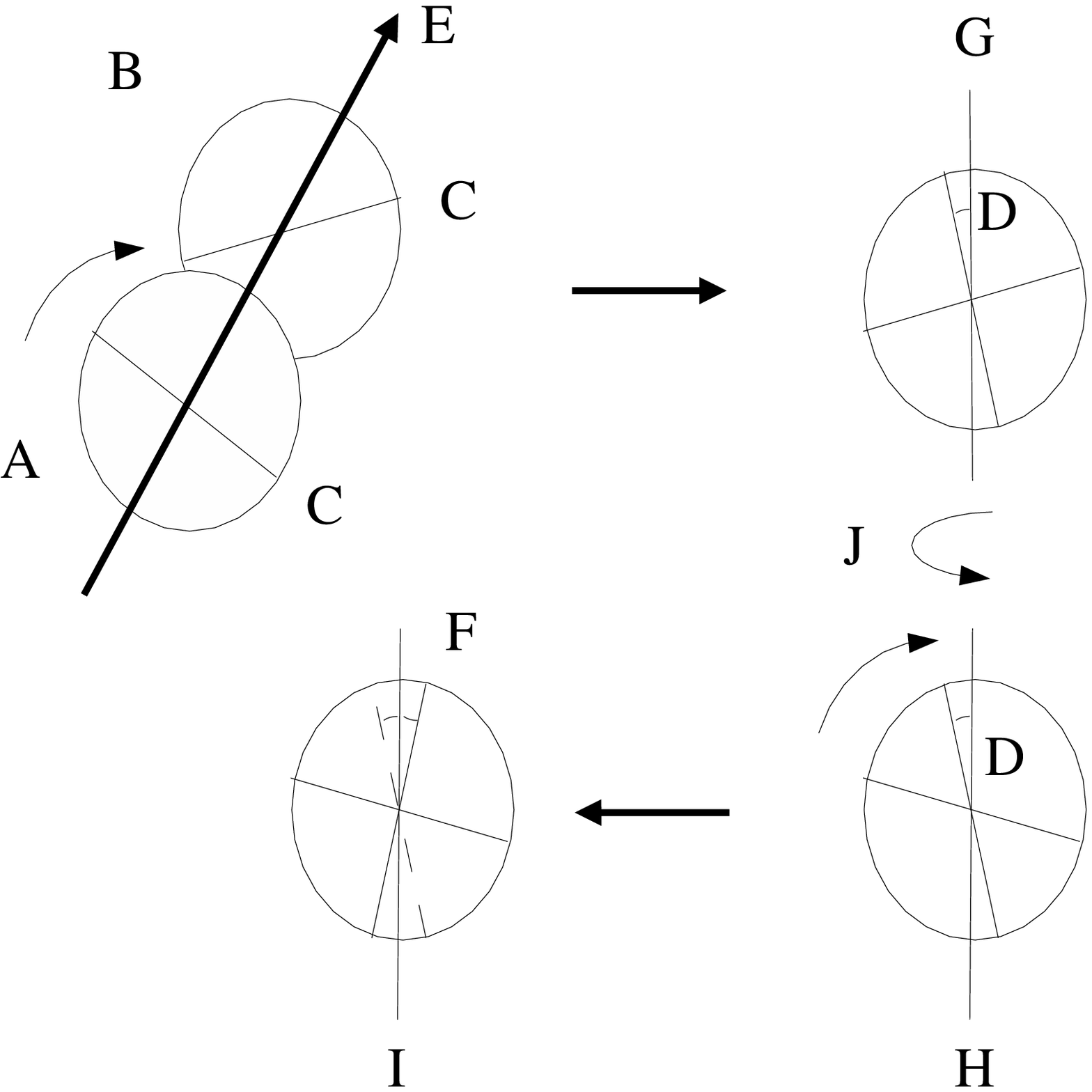}}
\end{minipage}
\begin{minipage}{8cm}
\centerline{\large (1)}
\resizebox{8cm}{!}{\includegraphics{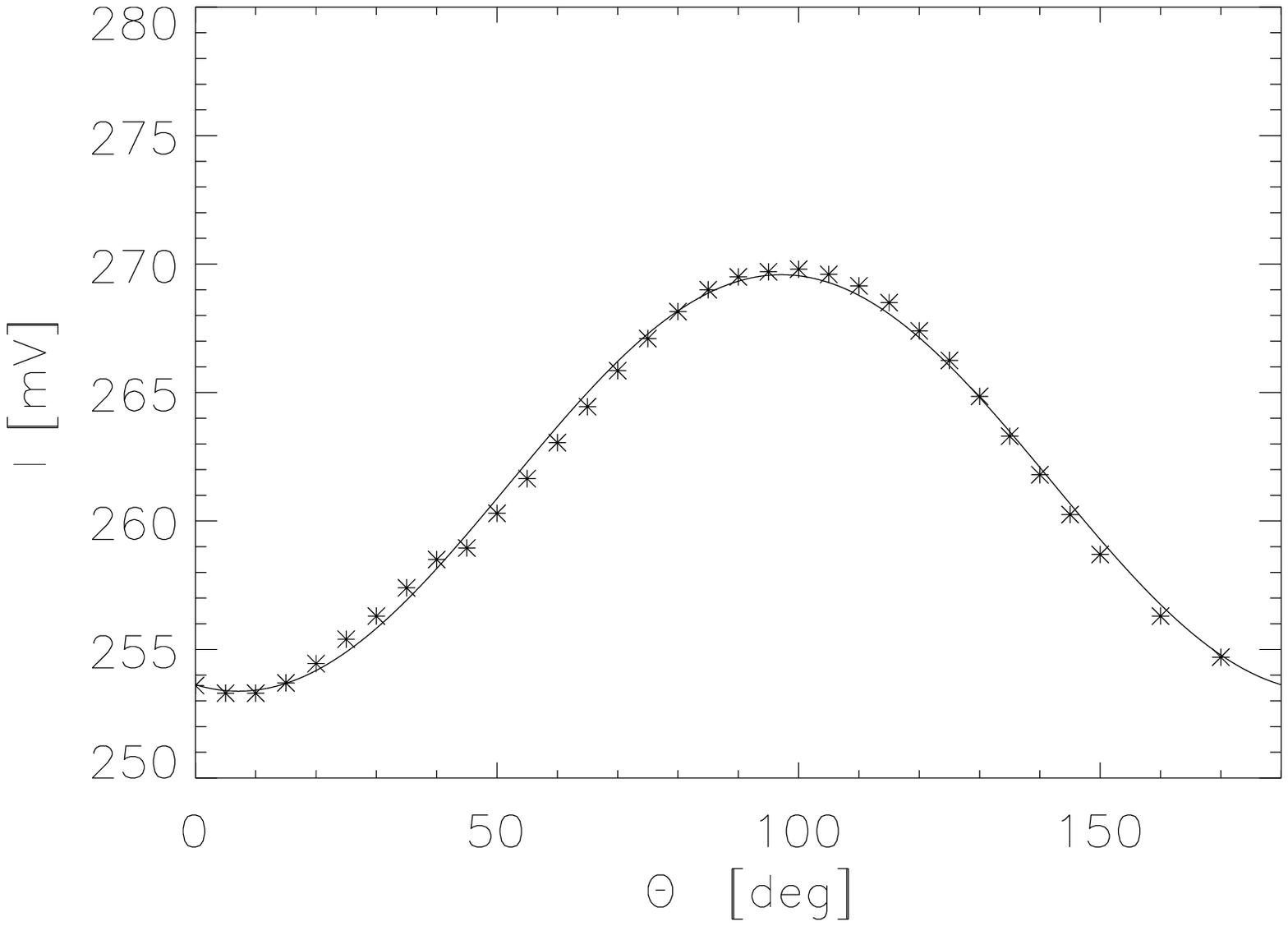}}\\
\centerline{\large (2)}\\
\resizebox{8cm}{!}{\includegraphics{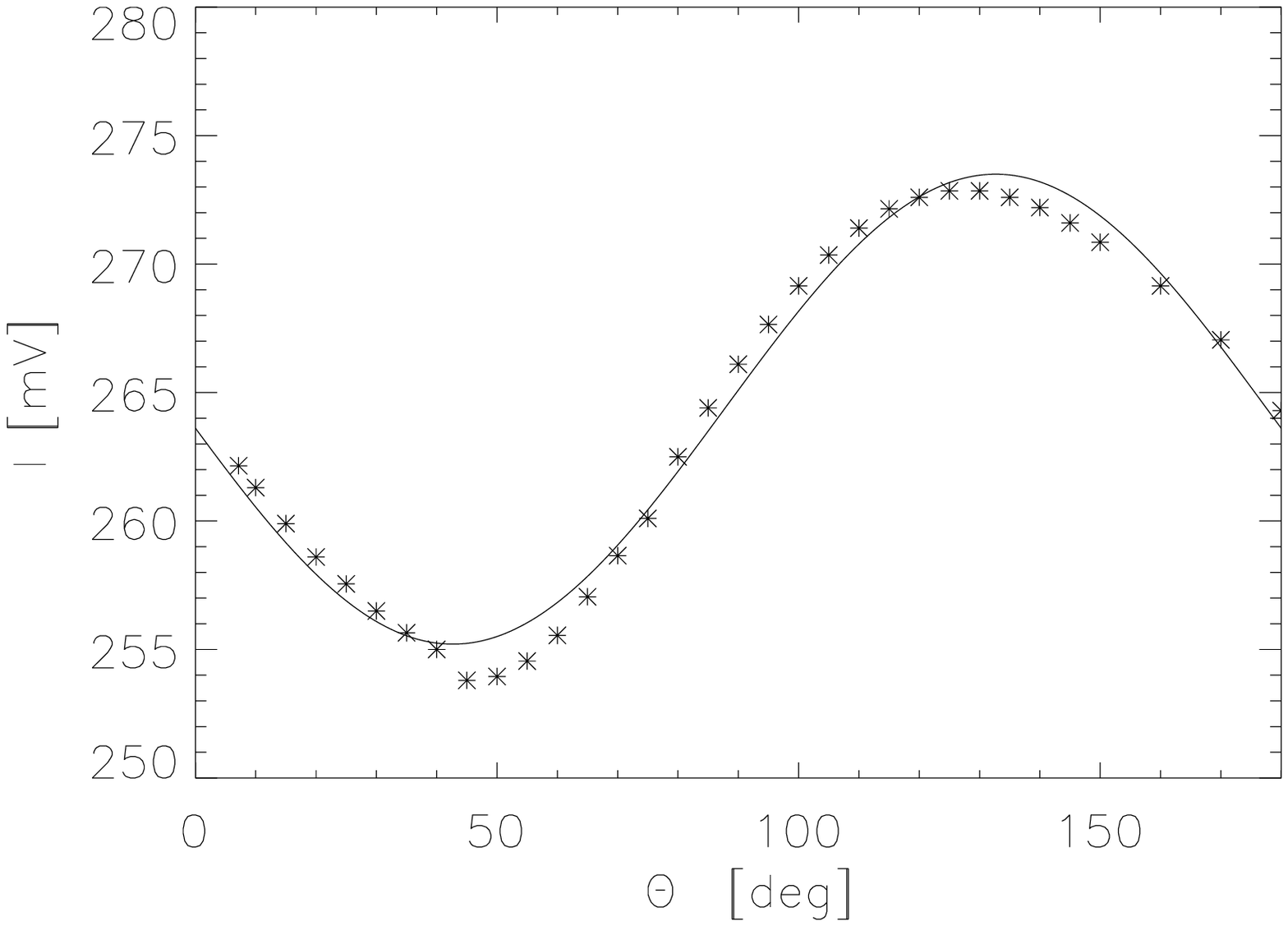}}
\end{minipage}
\caption{The transmission axes of the polarizer and the analyzer have 
arbitrary, and unknown directions. If the polarizer is rotated around
the propagation direction, the resulting curve of transmitted intensity,
(1), shows a minimum, when polarizer and analyzer are crossed. In the
crossed position, $\theta_{pol,min} \sim 7.3^\circ$, the axis of the
polarizer forms an unknown angle, $\alpha$, with the vertical.\newline
Now the analyzer is rotated by 180$^\circ$ around the vertical
axis. This causes its transmission axis to jump by $2\, \alpha$. The
polarizer is rotated again for 180$^\circ$, resulting in curve
(2). \newline
The minmimal intensity now occurs at the angle $\theta_{pol,min} + 2\,
\alpha \sim 40.3^\circ$. The value was established from a least square
fit to the intensities for $\theta_{pol} > 45^\circ$ only. The motor
seems to have positioning problems around 45$^\circ$, but it is the spare unit
hopefullly never to be used.\newline
Half the difference between the angles of minimal intensity is the angle
to the vertical, $\alpha$. Thus the transmission axis of the polarizer
is 16.5$^\circ$ off the vertical in the first minimal position, i.e. it
is at -23.8$^\circ$ for $\theta_{pol}$ = 0. The analyzer can then also be
marked. \label{rotate_pol}}
\end{figure}
\begin{figure}[ht]
\begin{minipage}{8cm}
\begin{center}
\psfrag{A}{\Huge polarizer}
\psfrag{C}{\Huge retarder}
\psfrag{B}{\Huge analyzer}
\psfrag{E}{\Huge to photodiode}
\resizebox{4cm}{!}{\includegraphics{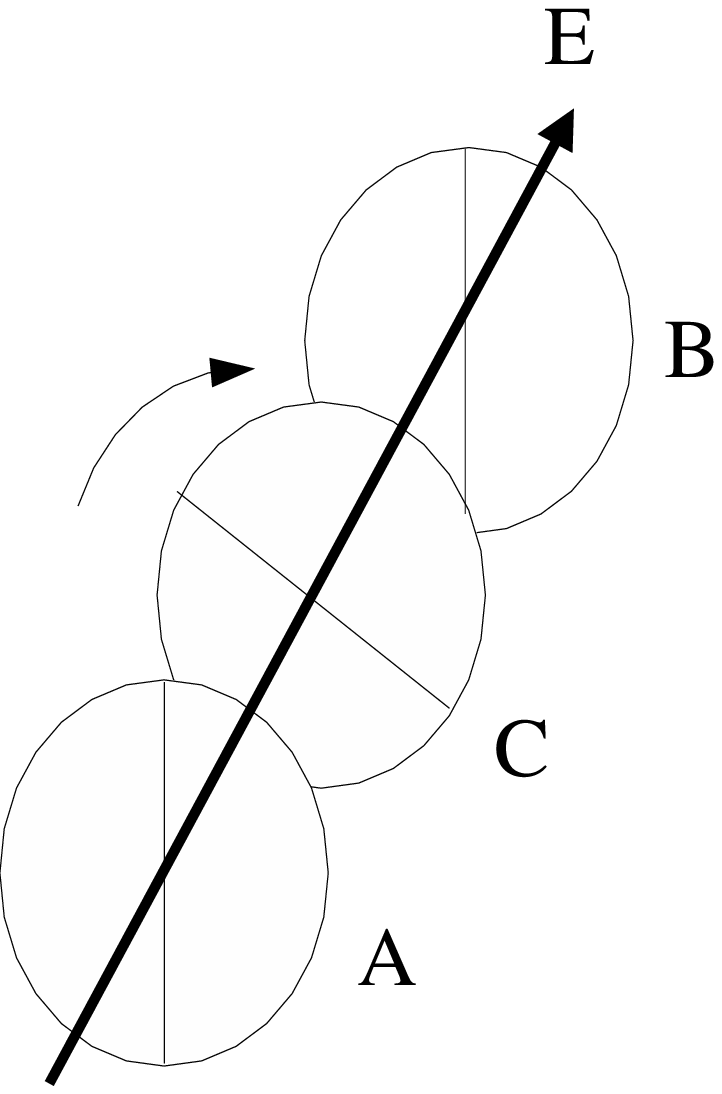}}
\end{center}
\end{minipage}
\begin{minipage}{8cm}
\resizebox{8cm}{!}{\includegraphics{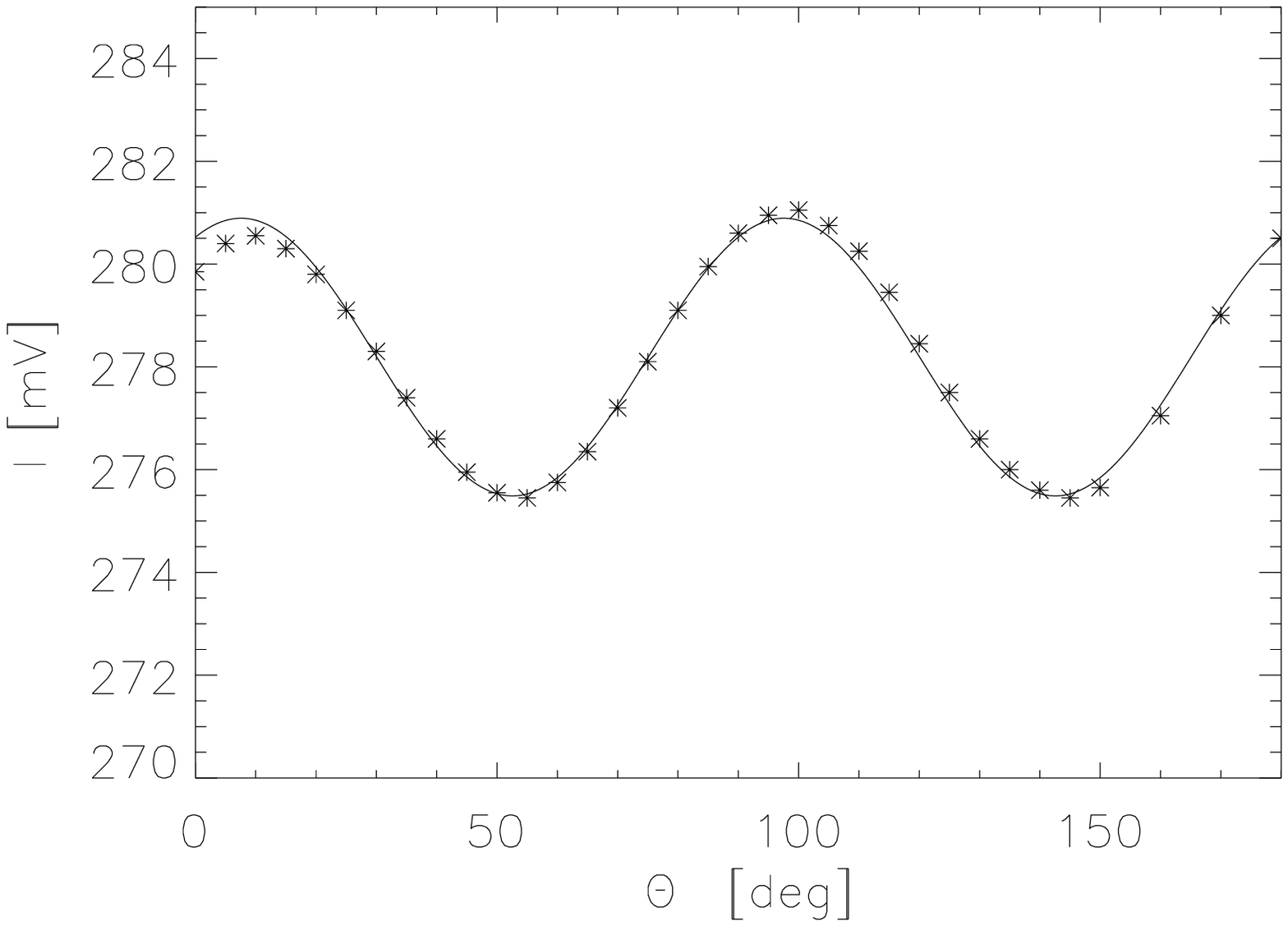}}
\end{minipage}
\caption{The axis of the retarder is established by inserting it between
the polarizers, now both vertically oriented. The intensity curve for a
rotation of the retarder shows maxima, when the fast axis is parallel to
the transmission axes of the polarizers. The first maximum appears for
$\theta_{ret} = 7.92^\circ$. The axis of the retarder therefore is off the
vertical by -$7.92^\circ$ at $\theta_{ret} = 0$.}
\end{figure}
\end{chapter}
\begin{chapter}{The analytical telescope model: discussion and
comparison\label{compltel}}
Section \ref{telmodel_theory} describes a way for the construction of
the telescope model that relies on some numerical procedures. The
developed procedure can be applied to optical setups other than a
coelostat as well, if the beam path is known. For the restriction on the
polarimetric properties of a coelostat it is possible to find an
analytical solution. Section \ref{cap_eq} presents the formulae of this
solution, section \ref{thickmirr} the mirror matrix for optical thick
coatings, and section \ref{comparison} a discussion on the plausibility
of results from the model, and a comparison between the analytical and
the numerical method.
\begin{section}{The analytical solution of the telescope model\label{cap_eq}}
The paper by Capitani et al, \cite{capitani}, derives analytically the
rotation and incidence angles needed for the description of the
instrumental polarization of a coelostat. Only the resulting equations
for the important parameters $i_1, i_2, \theta_1, \theta_2$ will be
presented here.
\begin{subsection}{Position of the second coelostat mirror}
This is identically to the discussion in section \ref{beampath}. The
position of C2 is fixed by its azimuth $A_{C2}$, and the horizontal angular
height, $h_{C2}$. For the VTT Tenerife this in fact is easier to establish
due to the motion of the first mirror C1 on a circle, while for the
solar tower in Arcetri described by Capitani the mirror moves on two
orthogonal set of rails (Cap., eqs. (13a)-(13d)). This complicates the
calculation of the azimuth due to another degree of freedom.
\end{subsection}
\begin{subsection}{Analytical equations}
The equations derived by Capitani, listed in the required order of calculation,
now are:
\begin{eqnarray}
\sin H_{C2} &=& \frac{ \cos h_{C2} \sin A_{C2}}{\cos \delta_{sun}}\\
\cos H_{C2} &=& \frac{ \sin h_{C2} + \sin \phi \sin \delta_{sun}}{\cos
\phi \cos \delta_{sun}}\\
H &=& \frac{1}{2} \left( H_{C2} - H_{sun} \right)\\
\sin 2 \theta_1 &=& \frac{2 \sin \delta_{sun} \sin H \cos H}{\sin ^2
\delta_{sun} \cos ^2 H + \sin ^2 H}\\
\cos 2 \theta_1 &=& \frac{\sin ^2 \delta_{sun} \cos ^2 H \cos H- \sin ^2
H}{\sin ^2 \delta_{sun} \cos ^2 H + \sin ^2 H} \;\;\rightarrow \theta_1\\
\sin \theta_1 \sin i_1 &=& - \sin H
\end{eqnarray}
\begin{eqnarray}
\cos \theta_1 \sin i_1 &=& -\sin \delta_{sun} \cos H\\
\cos i_1 &=& \cos \delta_{sun} \cos H \;\;\rightarrow i_1\\
i_2 &=& \frac{1}{2} \left( 90^\circ - h_{C2} \right)\\
A &=& A_{C2} - A_{sun}\\
\sin \theta_2 &=& - \frac{\cos h_{sun} \sin A}{\sin 2 i_1}\\
\cos \theta_2 &=& \frac{\sin h_{C2} \cos 2 i_1 - \sin h_{sun}}{\cos
h_{C2} \sin 2 i_1}\;\;\rightarrow \theta_2
\end{eqnarray}
These equations were mainly used to check the results of the numerical
routines, but of course they allow the construction of the telescope
model by themselves. For the agreement between the two methods see
section \ref{comparison} below.
\end{subsection}
\end{section}
\begin{section}{Mirror matrix for optical thick
coatings\label{thickmirr}}
The mirror matrix for the description of optical thick coatings is in
principle identical to eq. \ref{mirrmat}, but the equations for the
calculation of the entries simplify. The matrix is:
\begin{eqnarray}
\bf M\rm_{mirror} = \frac{r^2_\bot}{2}\begin{pmatrix} 
X^2 + 1  &X^2 - 1 & 0& 0\cr
X^2 - 1 &X^2 + 1 & 0& 0\cr
0 & 0 & \;\;\; 2\cdot X \cdot \cos \delta & 2\cdot X \cdot \sin \delta \cr
0 & 0 & -2\cdot X \cdot \sin \delta & 2\cdot X \cdot \cos \delta\cr
\end{pmatrix} \;, \label{mirrmatthick}
\end{eqnarray}
with $X^2 = \frac{r_\parallel}{r_\bot}$.

The entries can be calculated with
\begin{eqnarray}
f^2 = \frac{1}{2} \left( n^2 - k^2 - \sin ^2 i + \sqrt{(n^2 - k^2 -
\sin^ 2 i)^2 + 4 n^2 k^2 }\right)\\
g^2 = \frac{1}{2} \left( k^2 - n^2 + \sin ^2 i + \sqrt{(n^2 - k^2 -
\sin^ 2 i)^2 + 4 n^2 k^2 }\right)
\end{eqnarray}
by
\begin{eqnarray}
X^2 &=& \frac{f^2 + g^2 - 2\cdot f \cdot \sin i \tan i + \sin ^2 i \tan ^2
i}{f^2 + g^2 + 2\cdot f \cdot \sin i \tan i + \sin ^2 i \tan ^2 i}\;,
\mbox{ and}\\
\tan \delta &=& \frac{2\cdot g \cdot \sin i \tan i}{\sin ^2 i \tan
^2-(f^2+g^2)} \; .
\end{eqnarray}
$i$ is the incidence angle on the mirror, $n$ and $k$ refraction and
extinction coefficient. The thickness of the coating and the properties
of the substrate do not enter in the calculation.
\end{section}
\begin{section}{Discussion and comparison of model results \label{comparison}}
The best test of the telescope model is the comparison with actual
measurement data from the telescope, but some consistency checks are
possible with the model alone.
\begin{figure}
\resizebox{8cm}{!}{\includegraphics{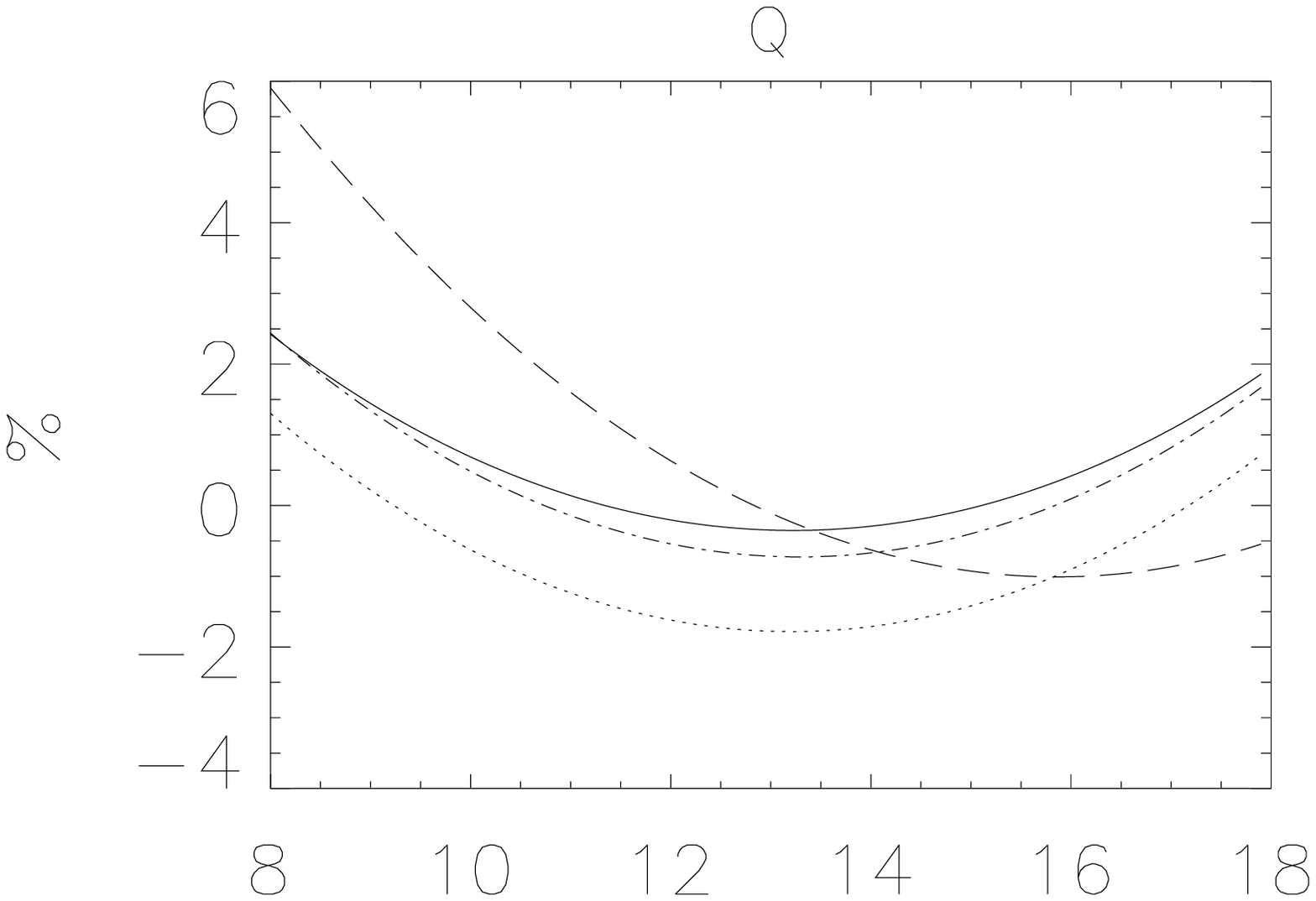}}
\resizebox{8cm}{!}{\includegraphics{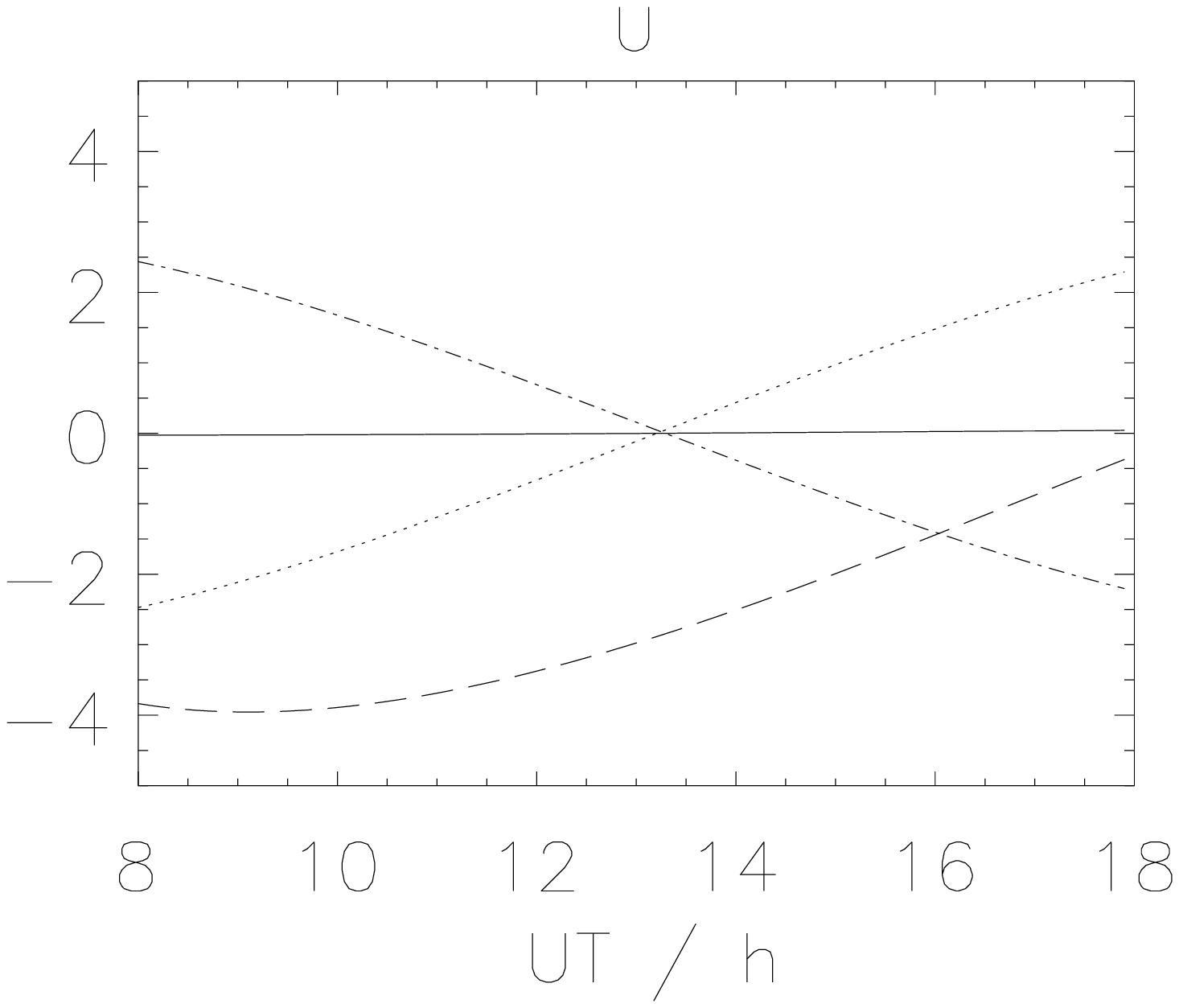}}\\
\resizebox{8cm}{!}{\includegraphics{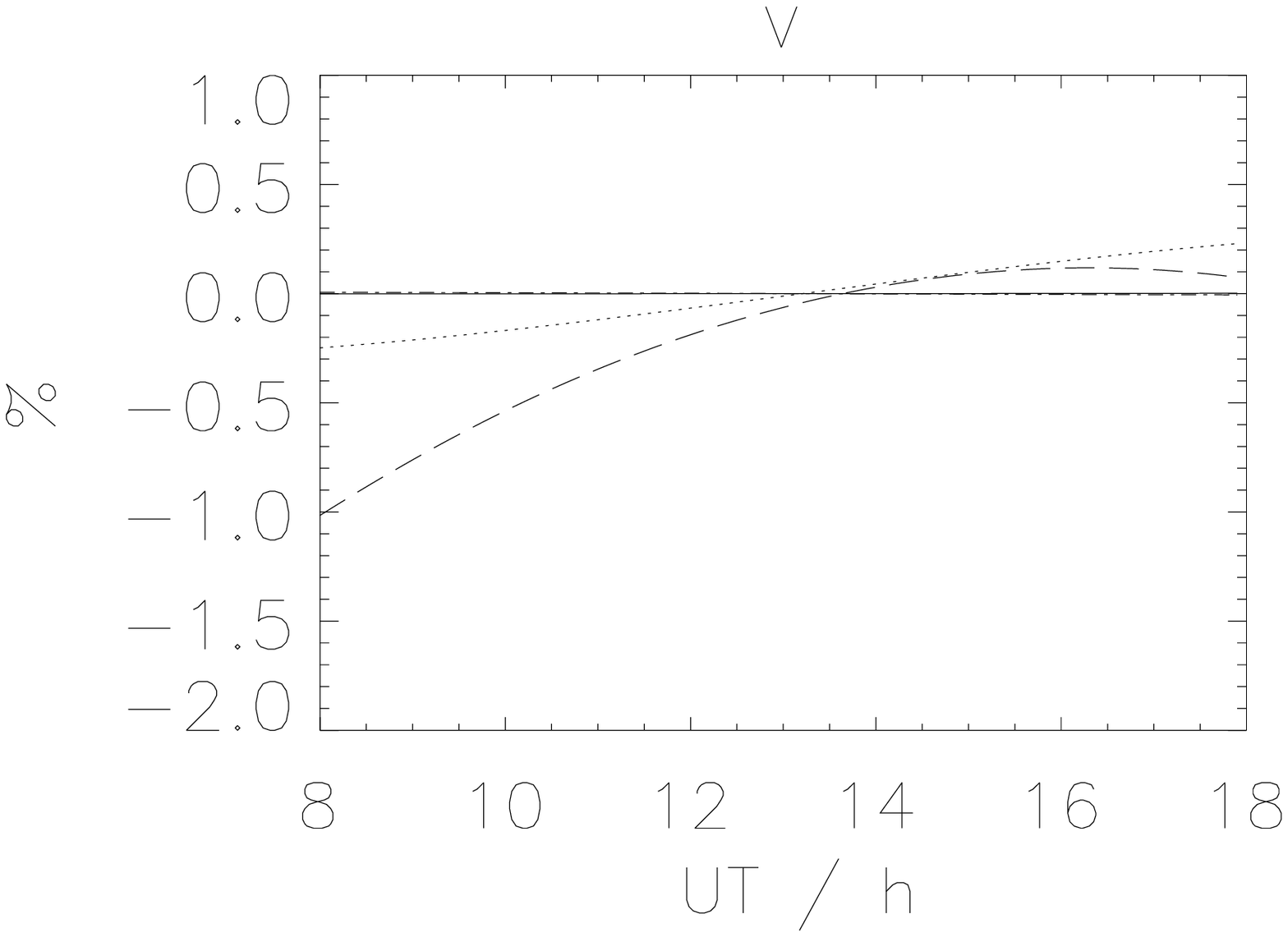}}
\hfill
\begin{minipage}{8cm}
\vspace*{-5cm}
\caption[Telescope output vector]{Telescope output vector for
unpolarized input in the model.\newline
The polarization content in $\%$ of Stokes I for 21.03.02 (\it
solid\rm), 21.07.02 (\it dotted\rm) and 21.01.02 (\it dash dotted\rm),
azimuth $0^\circ$. The fourth curve (\it long dashes\rm) is for 21.07.02,
azimuth $40^\circ$. The polarization remains well below 10 $\%$ of the
total intensity, but if one moves the first mirror from
the zero position, the telescope polarization strongly
increases.\label{stokesout}}
\end{minipage}
\end{figure}
\begin{itemize}
\item image rotation of the coelostat:\\
If the first mirror C1 is not on the zero azimuth position, the 
coelostat will perform a rotation of the sun image of
\begin{eqnarray}
p  = - \mbox{asin} \left( \frac{\cos \phi\cdot \sin(\mbox{azimuth})
}{\cos \delta_{sun}} \right) + \mbox{azimuth}\;.\label{imrot}
\end{eqnarray}
This can be checked with the model in the following way:
\begin{itemize}
\item the definition of an angle of the linear polarization by
\begin{eqnarray}
\tan \alpha = \frac{Q}{U}\;,
\end{eqnarray}
which can be calculated for $\bf S\rm_{in}$, and also $\bf S\rm_{out} = \bf
T\rm\cdot\bf S\rm_{in}$.
\item the usage of ideal mirrors, i.e. the mirror matrices are set to
diag(1,1,-1,-1). This corresponds to mirror reflectivities of 100
$\%$ regardless of the oscillation direction of the incoming light and
no cross talk terms $Q \leftrightarrow U$.
\end{itemize}
The rotation of the defined polarization axis can be shown by the
difference $\alpha_{in}-\alpha_{out}$. This has to be compared to the
value of $p$ predicted by Eq. (\ref{imrot}). An example is displayed in
Fig. \ref{imagerot} for the date 21.07.02 and an 
azimuth setting of C1 of 50$^\circ$ for 8:00-18:00 UT. The remaining
differences between theoretical value and the output from the modellized
coelostat are of order of 0.005$^\circ$ and arise from the used
numerical procedures.
\begin{figure}
\begin{minipage}{9cm}
\resizebox{9cm}{!}{\includegraphics{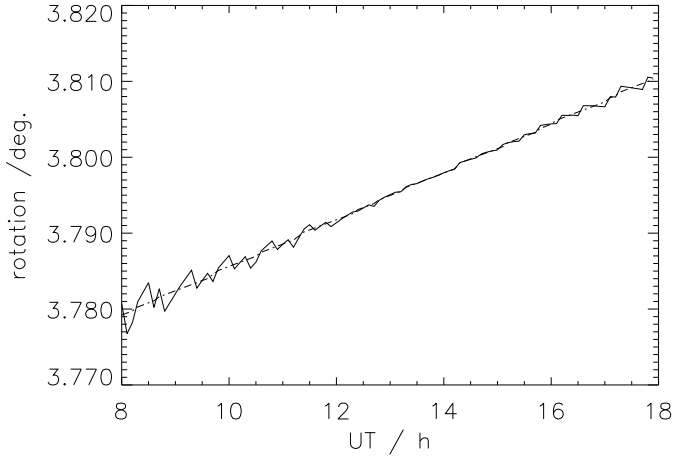}}
\end{minipage}
\begin{minipage}{7cm}
\caption[Image rotation]{Image rotation produced by the modellized
coelostat in degrees. \newline
Shown is the difference of the angles $\alpha_{in}-\alpha_{out}$ for the date
21.07.02 and an azimuth setting of C1 of 50$^\circ$ for 8:00-18:00 UT
(solid line). The dash-dotted line is the prediction of
Eq. (\ref{imrot}) for the rotation angle. The increase of the rotation
angle is caused by the changing declination of the sun.\label{imagerot}}
\end{minipage}
\end{figure}
\end{itemize}
\begin{itemize}
\item comparison with the results in Cap. et al., \cite{capitani}:
\begin{itemize}
\item symmetry properties: the matrix entries plotted in
Fig. (\ref{telmatr}) for three dates show either a symmetric or
antisymmetric course with regard to noon. The resulting pattern is in
accordance with the properties given in Cap., \cite{capitani}, eq. (14). 
\item telescope output for unpolarized input: Fig. \ref{stokesout}
displays the Stokes vector, which leaves the telescope optics according
to the model for an unpolarized input. Comparing with Cap.,
\cite{capitani}, Fig. (9 a-c), which show measurements at the Donati
Solar tower in Arcetri, one can see a similar range of values for the
polarization content.\\
The long-dashed curve is for an azimuth position of 40$^\circ$ on
21.07.02 . The polarization created in the telescope increases, if the
first mirror is moved away from the zero azimuth position. 
\end{itemize}
\end{itemize}
\begin{figure}
\resizebox{9cm}{!}{\includegraphics{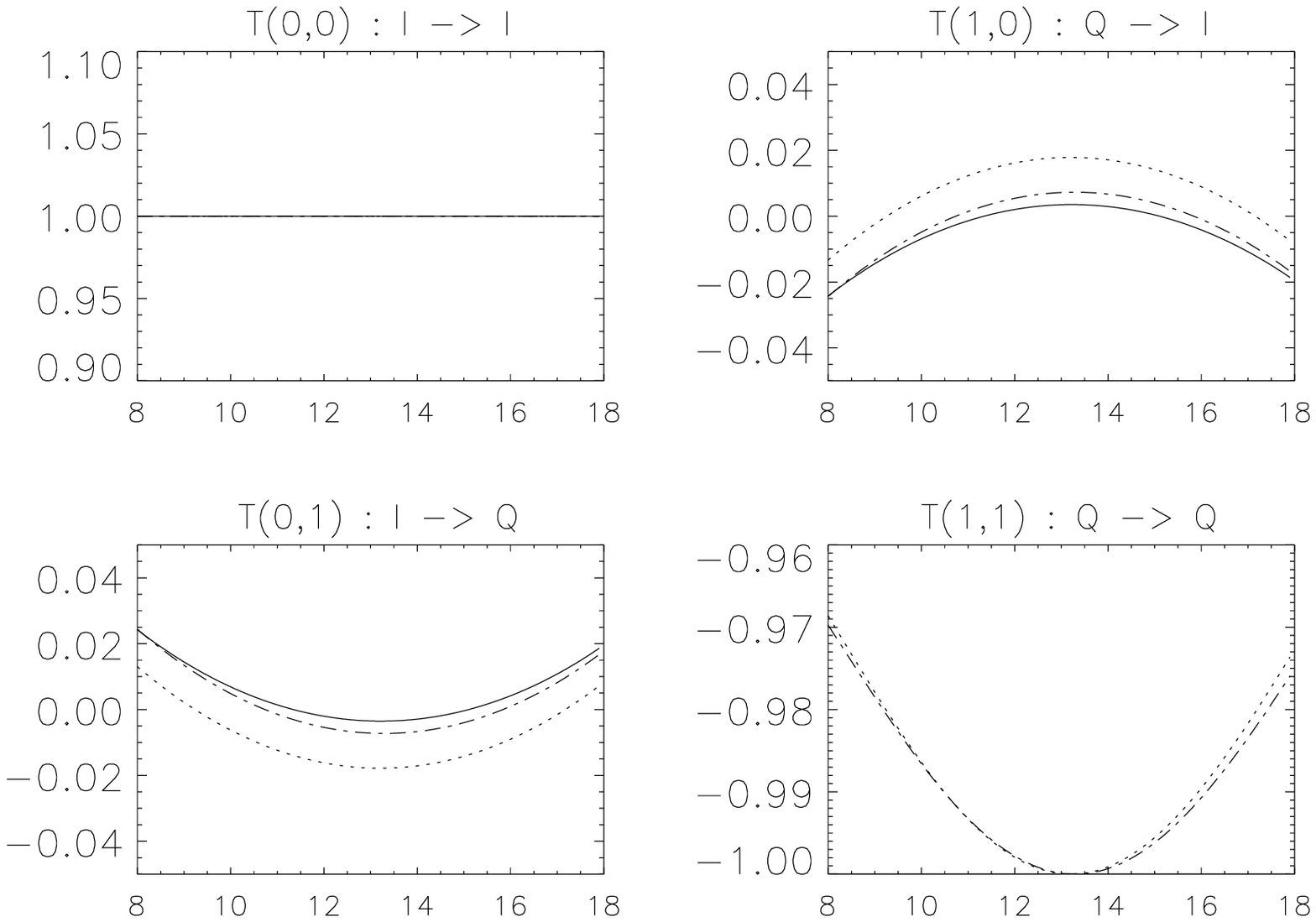}}
\resizebox{9cm}{!}{\includegraphics{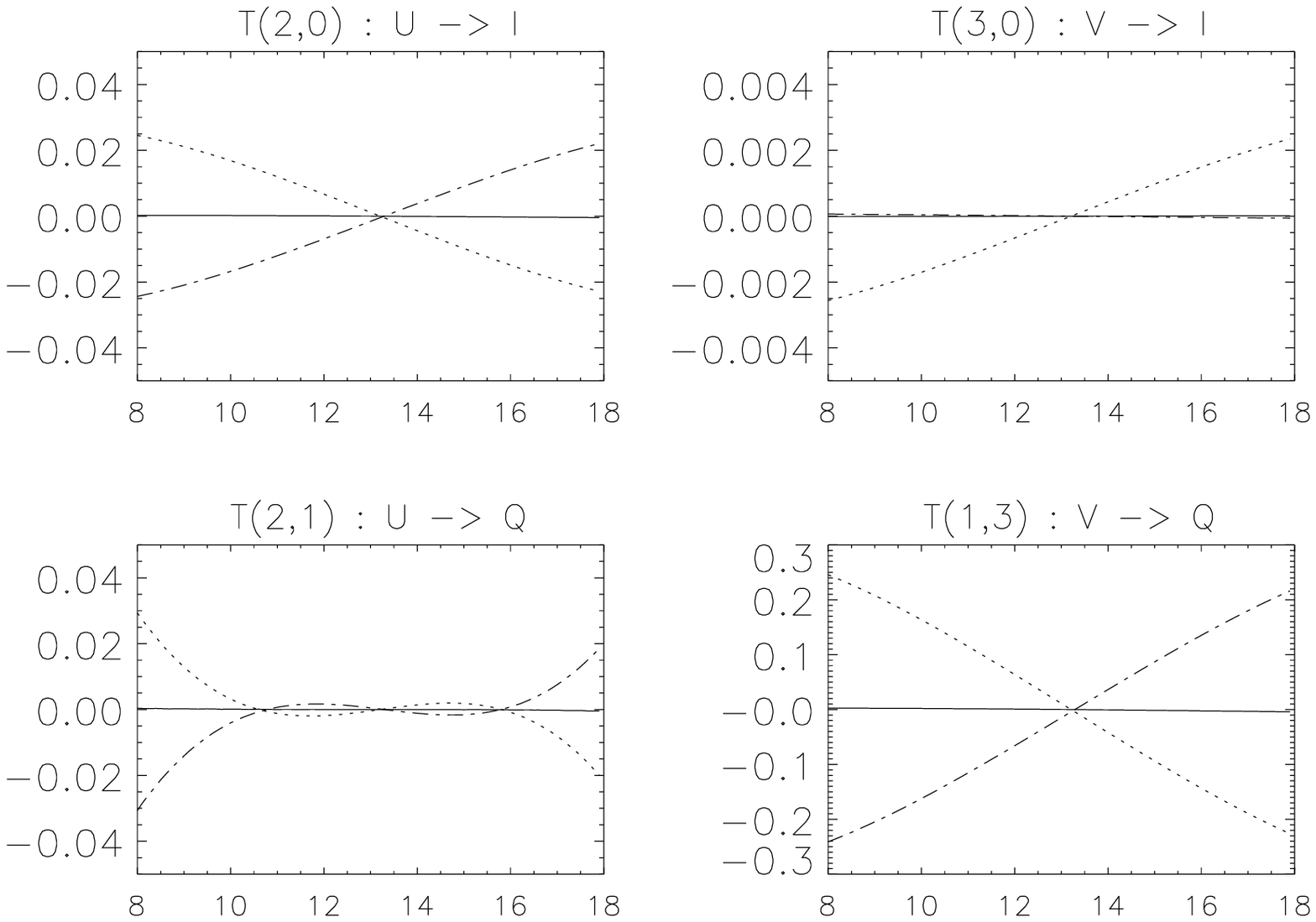}}\\
\resizebox{9cm}{!}{\includegraphics{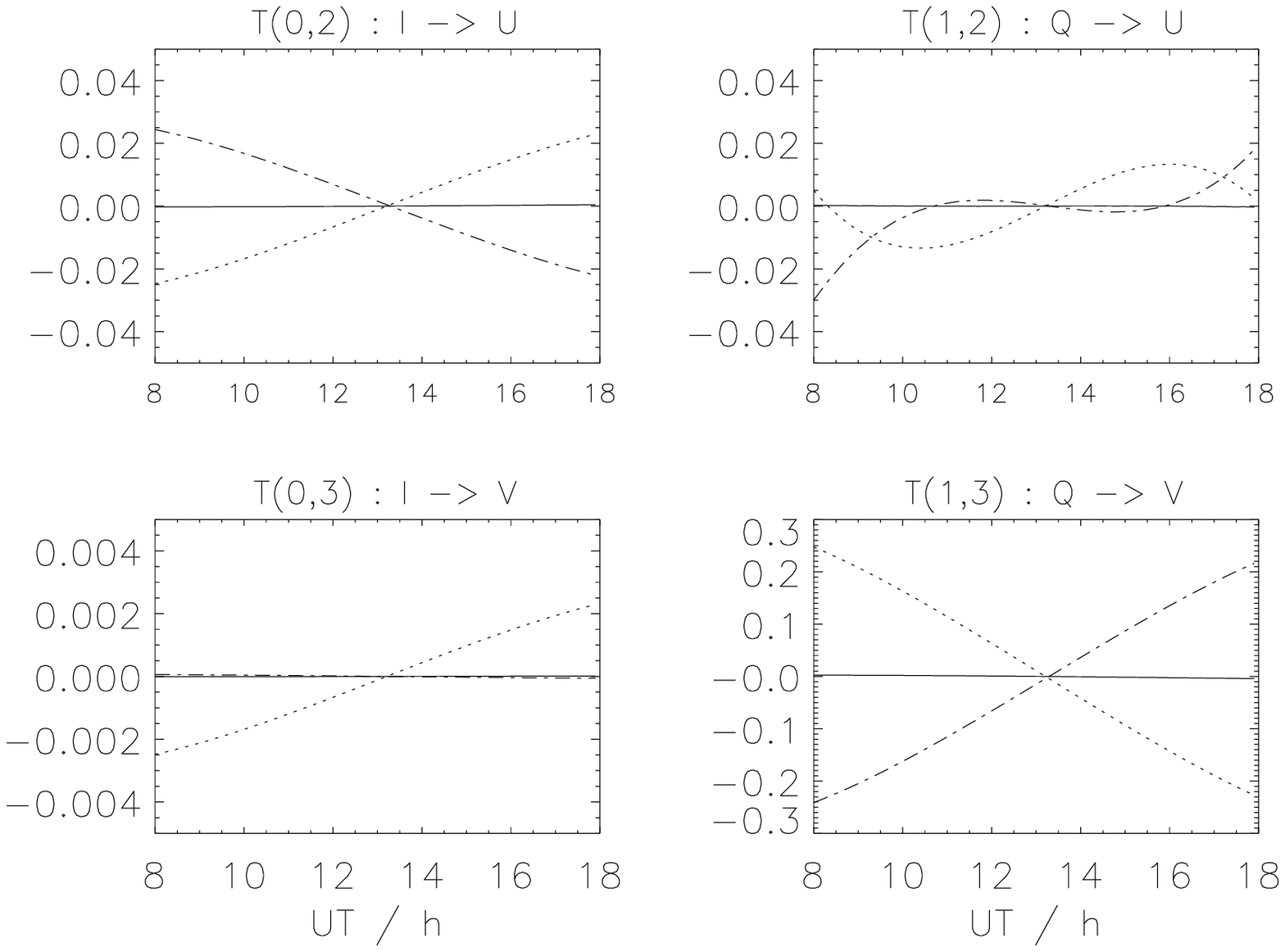}}
\resizebox{9cm}{!}{\includegraphics{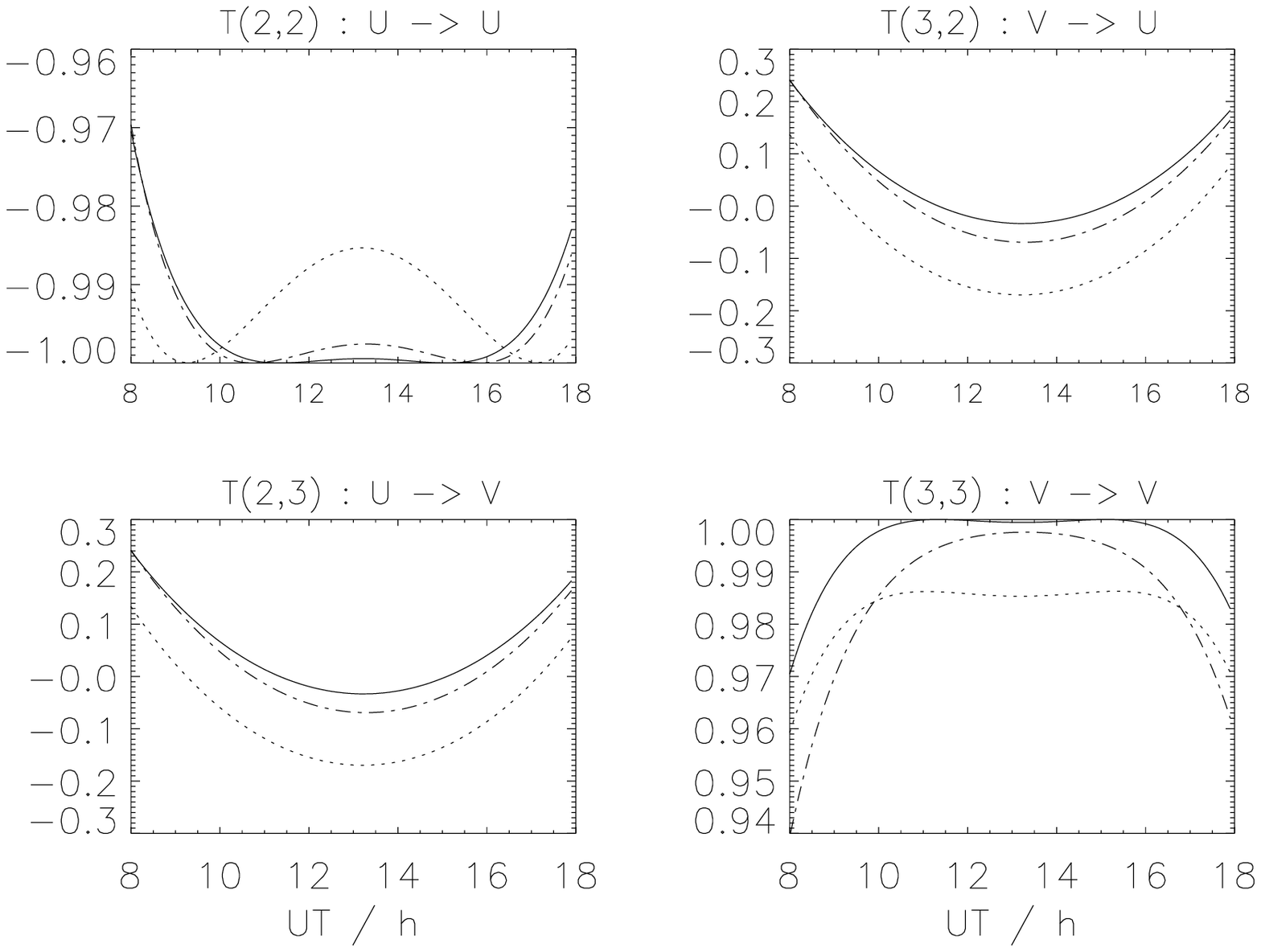}}
\caption[Telescope matrix]{Normalized telescope matrix for three dates.\newline
The 16 entries of the telescope matrix resulting from the model are
plotted for 8:00-18:00 UT. The azimuth of C1 was always set to
0$^\circ$. (\it solid\rm) 21.03.02, (\it dotted\rm) 21.07.02, (\it 
dash-dotted\rm) 21.01.02 . The matrix entries are either symmetric or
antisymmetric to noon. A comparison of absolute values is more
difficult, as this depends on the parameter set used for the physical
properties of the mirrors. The cross talk between different
polarizations has contributions from geometrical effects (image
rotation) and the mirror properties.\label{telmatr}}
\end{figure}
\end{section}
\end{chapter}
\begin{chapter}{Technical characteristics\label{instruchar}}
\begin{section}{The grating\label{grating}}
The optical design of the grating is a combination of an Echelle and a
reflective Littrow configuration. As the entrance pupil with the central
obscuration of the telescope is imaged onto the grating, the central
part of the grating is not illuminated. This area is used for a 
small folding mirror that feeds the light beam to the collimator (cp.
Fig. \ref{polis_section}). The grating is slightly tilted about a
horizontal axis and the imaging mirror is located above the
collimator. The final image plane is formed above the grating. The
configuration has only small reflection angles and thus a high optical
quality across a large field of view.
\begin{table}[ht]
\begin{tabular}{|c|c|c|} \hline
\multicolumn{3}{|c|}{\rule[-2mm]{0mm}{6mm} \bf Grating}\cr \hline
79 grooves / mm & blaze angle: 63.4$^\circ$& dispersion: 12 pm/mm @ 400 nm \cr\hline
\multicolumn{3}{|c|}{\rule[-2mm]{0mm}{6mm} \bf CCD}\cr \hline
652$\times$488 pixels & $ 12 \mu \times 12 \mu$ pixel size& frame rate:
max. 40 Hz \cr \hline
\multicolumn{3}{|c|}{\rule[-2mm]{0mm}{6mm} \bf Modulator}\cr \hline
$\delta = 3/8 \lambda$ @ 630 nm & $\delta = 5/8 \lambda$ @ 400 nm &
zero-order \cr \hline
\multicolumn{3}{|c|}{\rule[-2mm]{0mm}{6mm} \bf Interference filters}\cr \hline
5 nm FWHM @ 630 nm & 1 nm FWHM @ 400 nm & \cr\hline
\multicolumn{3}{|c|}{\rule[-2mm]{0mm}{6mm} \bf Demodulation}\cr \hline
15 Hz frame rate & synchronized with retarder rotation & \cr\hline
\multicolumn{3}{|c|}{\rule[-2mm]{0mm}{6mm} \bf Computer control}\cr \hline
\multicolumn{3}{|c|}{\rule[-2mm]{0mm}{6mm} PC-based with Windows NT}\cr\hline
\end{tabular}
\caption{Technical characteristics.\label{instrutable}}
\end{table}
\end{section}
\end{chapter}
\begin{chapter}{Many thanks to}
In no special order:\\
$ $\\
\bf Wolfgang Schmidt \rm for the topic, the assistance during the work, the
scientific background, etc.\\
$ $\\
\bf My parents \rm for their unlimited sponsorship all the time\\
$ $\\
\bf Manolo Collados \rm for his great patience, and a very important visit to
Tenerife\\
$ $\\
\bf Bruce Lites \rm for his great patience, and the help in the
calibration of the ASP data sets\\
$ $\\
\bf Michael Sigwarth \rm for exactly the same\\
$ $\\
\bf Kai Langhans \rm for the introduction into IDL\\
$ $\\
\bf Thomas Kentischer \rm for the work on POLIS, the opportunity to
smoke inside, and his great patience\\
$ $\\
\bf Tayeb Aiouaz \rm for a great time, and much fun with negative
pressure\\
$ $\\
\bf Helmold Schleicher, Jo Bruls \rm for the new calculations of formation
heights - when only asked if the old values were correct\\
$ $\\
\bf Carsten Baur, Stefan Henninger, Michael Engler \rm for the weekly
'Skatrunde'\\
$ $\\
\bf Everybode else I forgot\rm
\end{chapter}
\end{appendix}
\begin{chapter}[Zusammenfassung (in german)]{\vspace*{-.5cm}Zusammenfassung (in german)\vspace*{-.5cm}}
In dieser Arbeit wird die Kalibration des neuen Spektropolarimeters
POLIS beschrieben. Das Ger\"at resultiert aus einer Kooperation zwischen
dem Kiepenheuer Institut f\"ur Sonnenphysik (KIS), Freiburg, und dem
High Altitude Observatory (HAO), Boulder (USA). Es soll am deutschen
Vakuum-Turm-Teleskop (VTT) des KIS in Teneriffa betrieben werden.\\

Das Ger\"at bestimmt den Polarisationszustand des Sonnenlichtes in zwei
verschiedenen Wellenl\"angen- bereichen bei 400 und 630 nm. Zur Messung
wird eine rotierende Verz\"ogerungsplatte verwendet, die den
Polarisationszustand des einfallenden Lichtes moduliert. Durch einen
polarisierenden Strahlteiler wird die modulierte Polarisation in eine
Intensit\"atsmodulation umgewandelt. Die Demodulation zur Bestimmung des
Stokes-Vektors erfolgt \"uber ein gewichtetes Integrationsschema.\\

Um eine ausreichende polarimetrische Genauigkeit von 0.1 $\%$ der
Kontinuumsintensit\"at zu erreichen, muss das eigentliche Messger\"at
kalibriert werden, um seine Antwort auf verschiedene
Polarisationszust\"ande festzustellen. Weiterhin muss die instrumentelle
Polarisation des Teleskops bestimmt werden, die den Polarisationszustand des
einfallenden Sonnenlichtes ver\"andert.

Zur Kalibration des Polarimeters steht eine spezielle
Kalibrationseinheit zur Verf\"ugung, die aus einem Linear-Polarisator
und einer Verz\"ogerungsplatte besteht. Die Kalibrationseinheit wird
sich innerhalb des Vakuumtanks des Teleskop befinden, wo der Strahldurchmesser 
des Lichtb\"undels bereits auf einen Durchmesser von 6 cm verringert
ist. Mit dieser Einheit lassen sich durch Rotation der beiden optischen
Elemente verschiedene bekannte Polarisationszust\"ande
erzeugen, die im Kalibrations-Datensatz gemessen werden. Aus einem
Vergleich zwischen dem erzeugten Input und dem gemessenen Output l\"asst
sich die Polarimeter-Antwort-Funktion \bf X \rm durch einen Least-Square-Fit
bestimmen. Mit dieser 4$\times$4-Matrix kann man aus dem eigentlichen
Messwert den zutreffenden Stokes-Vektor der Polarisation am Ort der
Kalibrationseinheit rekonstruieren.

Die vollst\"andige Kalibration erfordert mehrere zus\"atzliche
Datens\"atze. Um die Detektoreigenschaften der verwendeten CCD-Kameras
zu bestimmen, werden Dunkelstrom- und Flatfield-Daten gewonnen. Aus
diesen beiden Datens\"atzen l\"asst sich die Antwort der einzelnen
Detektorpixel auf einfallende Intensit\"at ermitteln. Diese
Detektoreigenschaften werden in Gaintables f\"ur die jeweiligen
Detektorbereiche gespeichert. Der eigentliche Kalibrations-Datensatz zur
Bestimmung der Polarimeter-Antwort wird danach mit den Gaintables
korrigiert, um die Detektoreigenschaften zu entfernen.

Die korrigierten Daten werden \"uber die Wellenl\"angen gemittelt, da
der polarimetrische Gehalt im Kalibrationsdatensatz konstant sein
sollte. Entlang des Spaltes k\"onnen jedoch durch Verunreinigungen oder
r\"aumliche Inhomogenit\"aten der Kalibrationseinheit Schwankungen
auftreten. Deshalb wird die Antwort-Funktion an vier Positionen entlang
des Spaltes durch eine Matrix-Inversion des linearen Problems $\bf
S\rm_{out} = \bf X\rm \cdot \bf S\rm_{in}$ bestimmt.

Um die Eigenschaften der Kalibrationseinheit zu erfassen, werden
zus\"atzliche freie Parameter verwendet, wie z.B. die tats\"achliche
Verz\"ogerung $\delta$ der Wellenplatte. Die Anwendung der inversen
Polarimeter-Antwort-Funktion \bf X\rm$^{-1}$ auf Messdaten ergibt den
tats\"achlichen Stokes-Vektor des Lichtes am Ort der
Kalibrationseinheit.\\

Der Strahlengang im Teleskop vor der Kalibrationseinheit umfasst vier
Spiegel und das Eintrittsfenster des Vakuumtanks. Um die
polarimetrischen Eigenschaften dieses komplexen optischen Systems zu
bestimmen, werden die einzelnen Komponenten durch Mueller-Matrizen
modelliert. Bei den Spiegeln ben\"otigt man die spezifische Geometrie
der Reflexion zur Berechnung der Spiegelmatrix. Die Mueller-Matrix des
Spiegels ist jeweils nur in speziellen Koordinatensystemen g\"ultig. Zur
Berechnung des Gesamtsystems muss man daher sowohl die Spiegelmatrizen
als auch die Koordinatentransformationen bestimmen.

Das Eintrittsfenster steht unter grossem Druck und kann deswegen
spannungsinduzierte Doppelbrechung zeigen. Es wird als
Verz\"ogerungsplatte mit einer beliebigen, aber festen, Ausrichtung der
Achse modelliert.

Das Teleskopmodell ergibt die polarimetrischen Eigenschaften des Systems
durch Berechnung des jeweiligen Strahlenganges als Funktion folgender
Parameter:
\begin{itemize}
\item Spiegel: Brechungsindex n, Absorptionskoeffizient k. Die beiden
Coelostatenspiegel werden als gleich angenommen, ebenso Haupt- und
Umlenkspiegel im Vakuumtank (insgesamt 4 Parameter)
\item Fenster: Verz\"ogerung, Ausrichtung der Achse (2)
\item Ausrichtung der Kalibrationseinheit: zus\"atzliche Rotation am
Ende des Strahlenganges (1).
\end{itemize}
Diese 7 Parameter werden aus Teleskop-Kalibrationsdaten bestimmt. Zu
diesem Zweck steht eine drehbare Fassung zur Verf\"ugung, in der
Linear-Polarisatoren angebracht sind. Die Fassung kann auf dem ersten
Coelostatenspiegel oder dem Eintrittsfenster befestigt werden. Durch
eine Rotation um 180$^\circ$ lassen sich verschiedene linear
polarisierte Zust\"ande erzeugen. Der Polarisationszustand nach dem
Teleskop wird mit dem kalibrierten Polarimeter bestimmt. Danach werden
die Parameter im Teleskopmodell ebenfall mittels eines iterativen
Least-Square-Fits an die Messdaten angepasst.

Nach erfolgter Kalibration von Polarimeter und Teleskop l\"asst sich
mittels
\begin{eqnarray}
\bf S\rm_{solar} = \bf T\rm^{-1} \cdot \bf X\rm^{-1} \cdot \bf S\rm_{gemessen}
\end{eqnarray}
der Stokes-Vektor des eingefallenen Sonnenlichtes aus dem Messergebnis
bestimmen.\\

Im letzten Kapitel der vorliegenden Arbeit wird ein Datensatz eines
bereits existierenden Vektorpolarimeters untersucht. Selbst bei dieser
nur vorl\"aufigen Auswertung lassen sich Ph\"anomene wie der
Evershed-Effekt in Sonnenflecken wiederfinden.
\end{chapter}
\begin{chapter}*{Glossary\label{glossary}}
\addcontentsline{toc}{chapter}{Glossary}
\bf chromosphere \rm\\
\hspace*{1.5cm} the chromosphere is blue, until it turns grey at the
borders of reality\\
\bf convection / convective zone \rm\\
\hspace*{1.5cm} energy transport mainly by mass movement\\
\bf corona \rm\\
\hspace*{1.5cm} the corona is the part of the sun most intensively revered
in Bavaria by such immortal songs as 'Corona Bavariae, deep unter the
stary night...'\\
\bf differential rotation \rm\\
\hspace*{1.5cm} the sun as a ball of gas does not behave as a solid
body, so the rotation speed on the surface\\
\hspace*{1.5cm} is different depending on solar latitude. The values
range from 26 to 30 days on the poles\\
\bf Doppler effect \rm\\
\hspace*{1.5cm}  The principal effect is a change of frequency if a source
or an observer of radiation are\\
\hspace*{1.5cm} moving relative to each other along
their line-of-sight (LOS). In solar physics the Doppler\\
\hspace*{1.5cm} effect is used
to established the material movements on the sun by measuring the
wavelength\\
\hspace*{1.5cm} of \it absorption \rm and \it emission lines\rm, as the
relative movement of earth and sun is known\\
\bf flux tube \rm\\
\hspace*{1.5cm} a magnetic flux tube is assumed to be a highly evacuated
bundle of magnetic field lines\\
\hspace*{1.5cm} held stable against the gas pressure from
outside by the internal magnetic pressure\\
\hspace*{1.5cm} of the increased magnetic field. The field strength
rises to values of kG \\
\bf granulation \rm\\
\hspace*{1.5cm} the convective energy transport in the outer atmosphere
leads to more or less stable convection\\
\hspace*{1.5cm} cells. The pattern of the
borders of these cells is called the \it granulation\rm, while a single
cell\\
\hspace*{1.5cm} is
called a \it granule\rm. The material is supposed to rise up in the middle and
flow down at the\\
\hspace*{1.5cm} borders\\
\bf line\rm\\
\hspace*{1.5cm}\bf emission $\sim$ \rm : light of a certain wavelength
emitted by an electron transition between\\
\hspace*{1.5cm} atomic energy levels, $E = h \cdot \nu = h \cdot \lambda/c$\\
\hspace*{1.5cm}\bf absorption $\sim$ \rm : lowered intensity at a certain
wavelength in a continuum spectrum caused\\
\hspace*{1.5cm} by absorption and undirected reemission, in the sun the
outer atmosphere produces the\\
\hspace*{1.5cm} \it Fraunhofer lines \rm in this way\\
\bf magnetic\rm\\
\hspace*{1.5cm} $\sim$ \bf field \rm : the second component of the
electromagnetic field, appearing in the sun on greater\\
\hspace*{1.5cm} scale in the form of \it sun spots \rm and smaller \it flux
tubes\rm, supposed to be produced by\\
\hspace*{1.5cm} the \it solar dynamo \rm\\
\hspace*{1.5cm} $\sim$ \bf loops \rm : loops of magnetic field lines as
suggested mostly in images taken in the\\
\hspace*{1.5cm} H$_\alpha$ wavelength\\
\hspace*{1.5cm} $\sim$ \bf reconnection \rm : the annihilation of
magnetic field lines with antiparallel directions,\\
\hspace*{1.5cm} supposed to take part
on small scales and producing the energy visible in \it flares \rm or
other short\\
\hspace*{1.5cm} time features, where a great amount of energy is set free\\
\bf photosphere \rm\\
\hspace*{1.5cm} the photosphere is green for the most part, at least
where it's not red\\
\bf plasma \rm\\
\hspace*{1.5cm} ionized atoms lead to a gas of electrons and
ions, which is highly conductive, a situation\\
\hspace*{1.5cm} prevalent almost everywhere in the sun\\ 
\bf polarimetry \rm\\
\hspace*{1.5cm} in general the measurement of the polarization state of
light. Today polarimetry has a great\\
\hspace*{1.5cm} number of applications in science
(magnetic fields, saccharimetry, biology,...) and\\
\hspace*{1.5cm} technics (stress
tests, quality control of surfaces,...). In solar physics one can divide
into\\
\hspace*{1.5cm} vector polarimetry with measurement of the full \it Stokes
vector \rm and 'simple' polarimetry,\\
\hspace*{1.5cm} using only one polarization state\\
\bf pore \rm\\
\hspace*{1.5cm} a small sunspot without any visible penumbra\\
\bf quiet sun \rm\\
\hspace*{1.5cm} region on the sun with no solar activity, i.e. no
sunspots or flares, but only the diffuse small\\
\hspace*{1.5cm} scale magnetic background\\
\bf radiative core \rm\\
\hspace*{1.5cm} in the inner regions of the sun the energy is
transported mainly by radiation, whereas after\\
\hspace*{1.5cm} a small transient zone the outer atmosphere is \it
convective \rm\\
\bf resolution \rm\\
\hspace*{1.5cm} \bf spatial $\sim$ \rm : the distance which two for
example intensity features must have at least\\
\hspace*{1.5cm} on the sun to be detectable seperately\\
\hspace*{1.5cm} \bf spectral $\sim$ \rm : the minimal distance in
wavelength which can be separated\\
\bf solar cycle \rm\\
\hspace*{1.5cm} the changing of the polarity of the solar magnetic
activity with a periode of about 11 years,\\
\hspace*{1.5cm} coupled with the intensity of
activity as can be seen in for example the number of sun spots.\\
\hspace*{1.5cm} A theoretical description of the \it solar dynamo \rm
would have to include an explanantion\\
\hspace*{1.5cm} of this fact\\
\bf solar dynamo \rm\\
\hspace*{1.5cm} the magnetic fields on the sun are suppossed to be
generated by the solar dynamo.\\
\hspace*{1.5cm} It works on the mass movements of the
ionized \it plasma\rm, which constitutes a current and so\\
\hspace*{1.5cm}  gives rise to
a \it magnetic field\rm. The main factors are the \it convective \rm
movement caused by\\
\hspace*{1.5cm} the radial temperature differences and the \it
differential rotation\rm. The exact working principle\\
\hspace*{1.5cm} and for example
the \it solar cycle \rm are still far from being well understood\\
\bf sun spot \rm\\
\hspace*{1.5cm} larger area on the surface of the sun with lower
temperature than the surroundings. \\
\hspace*{1.5cm} The drop is temperature is caused by
the \it magnetic field \rm, which suppresses \it convection\rm.\\
\hspace*{1.5cm} Greater sun spots can be divided in the dark \it umbra
\rm and the brighter \it penumbra\rm.\\
\hspace*{1.5cm} Smaller spots without penumbrae are called \it pores \rm\\
\bf Stokes vector \rm\\
\hspace*{1.5cm} the Stokes formalism developed by G.Stokes about 1850
gives an easy way to describe the\\
\hspace*{1.5cm} polarization properties of light. It
uses the four Stokes parameters to include all necessary\\
\hspace*{1.5cm} information on
the polarization state, which most times are grouped into the Stokes
vector\\
\hspace*{1.5cm} for convenience\\
\end{chapter}
\end{document}